\documentclass[10pt]{ijnamb}
\def\currentvolume{1}
\def\currentissue{1}
\def\currentyear{2013}
\def\month{August}
\pagespan{1}{18}
\copyrightinfo{2013}{} 

\usepackage{graphicx}
\usepackage{subfigure}
\usepackage{float}
\usepackage{amsmath}
\usepackage{color}
\usepackage{multirow}
\usepackage{amssymb}
\usepackage{cite}

\begin{document}

\title[Proper orthogonal decomposition closure models ]
{Proper orthogonal decomposition closure models \\ for fluid flows: Burgers equation}


\author[O. San  \and T. Iliescu ]{Omer San \and Traian Iliescu}
\address{
  Department of Mathematics and Interdisciplinary Center for Applied Mathematics,
  Virginia Tech,
  Blacksburg, VA 24061, USA
}
\email{omersan@vt.edu \and iliescu@vt.edu}






\abstract{
This paper puts forth several closure models for the proper orthogonal decomposition (POD) reduced order modeling of fluid flows.
These new closure models, together with other standard closure models, are investigated in the numerical simulation of the Burgers equation.
This simplified setting represents just the first step in the investigation of the new closure models.
It allows a thorough assessment of the performance of the new models, including a parameter sensitivity study.
Two challenging test problems displaying moving shock waves are chosen in the numerical investigation.
The closure models and a standard Galerkin POD reduced order model are benchmarked against the fine resolution numerical simulation.
Both numerical accuracy and computational efficiency are used to assess the performance of the models.
}


\maketitle

\section{Introduction}
\label{sec:intro}
Proper orthogonal decomposition (POD) is one of the most successful reduced order modeling techniques in dynamical systems.
POD has been used to generate reduced order models (ROMs) for the simulation and control of many forced-dissipative nonlinear systems in science and engineering applications \cite{aubry1988dynamics,ito1998reduced,kunisch1999control,ly2001modeling,noack2003hierarchy,buffoni2006low,esfahanian2009equation,fang2009pod_om,lentine2010novel,stefanescu2013pod}.
POD extracts the most energetic modes in the system, which are expected to contain the dominant characteristics.
The globally supported POD modes are often constructed from high-fidelity numerical solutions (e.g., using finite difference/element/volume methods) and are problem dependent.
For many systems, it is possible to obtain a good approximation of their dynamics with few POD modes.
The systems built with these POD modes, called POD reduced order models (POD-ROMs) in what follows, are low dimensional and can provide an efficient framework for many applications.

The Galerkin POD-ROM (POD-ROM-G) is the simplest POD-ROM, which results from a Galerkin truncation followed by a projection of the truncated equation onto the space spanned by the POD modes.
The POD-ROM-G is an efficient tool for many applications of interest.
For fluid flows (see, e.g, \cite{noack2011reduced} and the exquisite survey in \cite{lassila2013model}), the POD-ROM-G works well for laminar fluid flows.
For turbulent flows, however, the POD-ROM-G yields inaccurate results.
As carefully explained in \cite{wang2012reduced}, for realistic turbulent flows, the high index POD modes that are not included in the POD-ROM-G do have a significant effect on the dynamics of the POD-ROM-G.
Several numerical stabilization strategies have been used to address this issue \cite{sirisup2004spectral,kalb2007intrinsic,bergmann2009improvement}.
Building on the analogy with large eddy simulation (LES) \cite{sagaut2006large,berselli2005mat} (see \cite{noack2008finite} for alternative approaches), several closure modeling strategies for POD-ROMs of turbulent flows have been proposed over the years \cite{rempfer1991proper,cazemier1997proper,cazemier1998proper,bergmann2009improvement,borggaard2011artificial,wang2011two,wang2012proper,ullmann2012pod,balajewicz2012stabilization,balajewicz2013low}, starting with the pioneering work in \cite{aubry1988dynamics}.
The main goal of this report is to propose and numerically investigate several closure models for POD-ROMs of fluid flows.
Three different classes of closure modeling strategies are considered.


The first strategy provides additional dissipation to the POD-ROM-G to account for the small scale dissipation effect of the discarded POD modes.
The main advantage of this strategy is that it has a negligibly small computational overhead.
Different viscosity kernels for POD-ROMs have been suggested in literature (see, e.g., \cite{bergmann2009improvement,sirisup2004spectral}).
In this study, these closure models together with three new closure models are presented in a unified framework and  their performance is investigated.

The second closure modeling strategy is also of eddy viscosity type.
The subgrid-scale operator used to account for the the effects of smaller scales is, however, different from that employed in the first closure modeling strategy.
This closure modeling approach is inspired from state-of-the-art LES models, such as the Smagorinksy model \cite{smagorinsky1963general} or its dynamic counterparts (see, e.g., \cite{germano1991dynamic}).
The Smagorinsky POD-ROM (POD-ROM-S) has been used in several studies \cite{noack2002low,ullmann2010pod,borggaard2011artificial,wang2011two,wang2012proper}.
To the best of our knowledge, the dynamic subgrid-scale POD-ROM was first used in \cite{wang2012proper}.
In this study, we investigate the POD-ROM-S together with one new closure model.
In general, these nonlinear closure models have a significant computational overhead (as explained in \cite{wang2011two}).
For the one-dimensional Burgers equation considered in this study, however, the computational overhead of these two closure models is negligible.

The third closure modeling strategy that we consider is based on the energy conservation concept.
This POD-ROM closure modeling approach was introduced by Cazemier \cite{cazemier1997proper} (see also \cite{cazemier1998proper}).
One of the main advantages of this closure model is that it does not require the specification of any free parameter, which is in stark contrast with the closure modeling strategies outlined above.
This closure model, however, has a higher computational overhead since it requires the computation of a penalty/drag term.
We note, however, that for the one-dimensional Burgers equation considered in this study this computational overhead is negligible.

Overall, there are 10 closure models, both new and current, in the three classes described above.
There are numerous other closure modeling strategies for POD-ROM of complex systems (see, e.g., \cite{noack2011reduced,wang2012reduced,lassila2013model}).
In general, when these closure models are introduced, they are deemed successful if they satisfy the following two criteria when compared with the fine resolution numerical simulation: (i) the new POD-ROM is relatively accurate; and (ii) the new POD-ROM has a significantly lower computational cost.
Given the number of competing closure modeling approaches, a natural {\it practical} question is which closure model should be used.
These intercomparison studies are scarce (see, e.g., \cite{wang2012proper} for an exception).
This study aims at answering the above practical question for the 10 closure models considered herein.

All the 10 POD-ROM closure models considered in this study are investigated in the numerical simulation of the Burgers equation.
We emphasize again that these closure models are developed for POD-ROMs of realistic turbulent flows.
In order to thoroughly assess their performance, however, as a first step, we consider the one-dimensional Burgers equation displaying challenging moving shock waves.
This simplified setting allows us to carefully assess the performance of the 10 closure models considered in this study and also carry out a parameter sensitivity study.
Of course, once we get a better understanding of the performance and limitations of these closure models, we will investigate them in realistic turbulent flow settings, such as those considered in \cite{wang2012proper}.
We also note that this progressive evaluation is common in the POD-ROM literature \cite{kunisch2001galerkin,borggaard2011artificial} or in the turbulence modeling literature \cite{kida1979asymptotic,love1980subgrid,gotoh1993statistics,bouchaud1995scaling,balkovsky1997intermittency,bec2007burgers,fauconnier2009family}.


The paper is organized as follows:
The mathematical model and the numerical methods for both spatial and temporal discretiations are summarized in Section \ref{sec:mat}.
The POD-ROM closure models are presented in Section \ref{sec:pod}.
These closure models are tested in Section \ref{sec:results} for two challenging examples displaying moving shock waves.
Finally, Section \ref{sec:cons} consists of some concluding remarks and future research directions.

\section{Mathematical models}
\label{sec:mat}
\subsection{Model equations}
All 10 POD-ROM closure models investigated in this report are tested on the Burgers equation:
\begin{equation}
\label{e:1}
    \frac{\partial u}{\partial t} + u \frac{\partial u}{\partial x} = \nu \frac{\partial^2 u}{\partial x^2} \, ,
\end{equation}
where $\nu$ is the viscosity parameter.
In this report, we consider two examples with different initial conditions:

\begin{eqnarray}
\mbox{Experiment 1:} & & \quad \quad    u(x,0) = \left\{
                                                \begin{array}{ll}
                                                  1, & \hbox{if} \quad  x \in [0,1/2] ; \\
                                                  0, & \hbox{if} \quad  x \in (1/2,1]
                                                \end{array}
                                              \right. \label{eq:a} \\
\mbox{Experiment 2:} & & \quad \quad    u(x,0) = u_0\exp{\bigg(\frac{-(x-x_c)^2}{\sigma}\bigg)\, ,} \quad \mbox{with} \quad x \in [0,1] \, ,\label{eq:b}
\end{eqnarray}
where $u_0=1$, $x_c = 0.3$, and $\sigma = 0.005$. The homogeneous Drichlet boundary conditions are applied for all the examples (i.e., $u(0,t) = u(1,t) = 0 \ \mbox{for} \ t>0$). Similar examples are considered in \cite{borggaard2011artificial}.

\subsection{Numerical methods}
\label{sec:numerical_methods}
We provide a brief description of the numerical methods employed in the investigation of the POD-ROM closure models.
As benchmark for all these closure models we use a direct numerical simulation (DNS) of the Burgers equation given in Eq.~(\ref{e:1}).
In order to minimize the spatial and temporal discretization errors in Eq.~(\ref{e:1}), a sixth-order compact difference and a third-order Runge-Kutta schemes are used for the spatial and temporal discretizations, respectively.
This will help to decouple the numerical effects from the POD modeling effects.
In the compact difference scheme, the first derivatives can be computed as follows \cite{lele1992compact}:
\begin{equation}
\alpha f'_{i-1} + f'_{i} +  \alpha f'_{i+1} = a\frac{f_{i+1}-f_{i-1}}{2h} + b\frac{f_{i+2}-f_{i-2}}{4h} \, ,
\label{eq:com1}
\end{equation}
which gives rise to an $\alpha$-family of tridiagonal schemes with $a=\frac{2}{3}(\alpha+2)$, and $b=\frac{1}{3}(4\alpha-1)$. The subscript $i$ represents the spatial index of an arbitrary grid point in the domain. Here, $\alpha=0$ leads to the explicit non-compact fourth-order scheme for the first derivative. A classical compact fourth-order scheme, which is also known as the Pad\'{e} scheme, is obtained by setting $\alpha=1/4$. The truncation error in the Eq.~(\ref{eq:com1}) is $\frac{4}{5!}(3\alpha-1)h^4 f^{(5)}$. A sixth-order compact scheme is obtained by choosing $\alpha=1/3$.

The second derivative compact centered scheme is given by
\begin{eqnarray}
\alpha f''_{i-1} + f''_{i} +  \alpha f''_{i+1} &=& a\frac{f_{i+1}-2f_{i} + f_{i-1}}{h^2} + b\frac{f_{i+2}-2f_{i} +f_{i-2}}{4h^2} \, ,
\label{eq:com2}
\end{eqnarray}
where $a=\frac{4}{3}(1-\alpha)$, and $b=\frac{1}{3}(10\alpha-1)$. For $\alpha=1/10$ the classical fourth-order Pad\'{e} scheme is obtained. The truncation error in the Eq.~(\ref{eq:com2}) is $-\frac{4}{6!}(11\alpha-2)h^4 f^{(6)}$. A sixth-order compact scheme is also obtained by choosing $\alpha=2/11$. The high-order sided derivative formulas are used for the Drichlet boundary conditions in order to complete the tridiagonal system of equations for both the first order derivative and the second order derivative formulas \cite{lele1992compact,carpenter1993stable,sari2009sixth}.

A system of semi-discrete ordinary differential equations (ODEs) is obtained after the spatial discretization of the Burgers equation. To implement the Runge-Kutta schemes, we cast the model equation in the following form
\begin{equation}
\frac{du}{dt} = \pounds (u) \, ,
\end{equation}
where $\pounds(u)$ is the discrete operator of spatial derivatives including nonlinear convective terms and linear diffusive terms. We assume that the numerical approximation for time level $\ell$ is known, and we seek the numerical approximation for time level $\ell+1$, after the time step $\Delta t$. The optimal third-order accurate total variation diminishing Runge-Kutta (TVDRK3) scheme \cite{gottlieb1998total} is then given as follows:
\begin{eqnarray}
u^{(1)} &=& u^{\ell} + \Delta t \pounds(u^{\ell}) \nonumber \\
u^{(2)} &=& \frac{3}{4}  u^{\ell} + \frac{1}{4} u^{(1)} + \frac{1}{4}\Delta t \pounds (u^{(1)}) \nonumber \\
u^{\ell+1} &=& \frac{1}{3}  u^{\ell} + \frac{2}{3} u^{(2)} + \frac{2}{3}\Delta t \pounds (u^{(2)}).
\label{eq:TVDRK}
\end{eqnarray}
We emphasize that we are not using any numerical stabilization to capture the shock in the DNS of the Burgers equation.
The dissipation mechanism in DNS is only obtained through the physical viscosity specified in the Burgers equation.
Therefore, in order to avoid numerical oscillations near the shock, a high resolution computation (DNS) is performed.

\section{Proper orthogonal decomposition (POD) reduced order models}
\label{sec:pod}
\subsection{Computing POD basis functions}
A POD can be constructed from the field variable $u$ at different times (snapshots). In this paper, these snapshots are obtained by solving Eq.~(\ref{e:1}) using the high-order compact difference scheme outlined in Section~\ref{sec:numerical_methods}.
In the time marching process, the $i$th record of the field is denoted by $u^{i}(x)$ for $i=1,2,...,N$, where $N$ is the number of the snapshots used to construct the POD basis.
First, the flow field data are decomposed into the mean part and the fluctuating part:
\begin{equation}
\label{e:9}
    u(x,t)=\bar{u}(x) + \hat{u}(x,t), \quad \bar{u}(x)=\frac{1}{N}\sum_{i=1}^{N}u(x,t_i) \, ,
\end{equation}
where $\bar{u}$ is the mean part (a function of space) and $\hat{u}$ is the fluctuating part (a function of space and time).
In order to obtain the POD bases functions, we first build a correlation matrix from the snapshots as follows:
\begin{equation}
\label{e:2}
    C_{ij}=\int_{\Omega} \hat{u}^{i}(x)\hat{u}^{j}(x)dx \, ,
\end{equation}
where $\Omega$ is the entire spatial domain and $i$ and $j$ refer to the $i$th and $j$th snapshots. The correlation matrix $C$ is a non-negative symmetric $N\times N$ matrix.
Defining the inner product of any two fields $f$ and $g$ as
\begin{equation}
\label{e:3}
    (f,g)=\int_{\Omega}f(x)g(x)dx \, ,
\end{equation}
Eq.~(\ref{e:2}) can also be written as $C_{ij}=(\hat{u}^{i},\hat{u}^{j})$. In this study, we use the well-known trapezoidal integration rule for computing the inner products numerically.  Solving the eigenvalue problem for this $C$ matrix provides the optimal POD basis functions. This has been shown in detail in the POD literature (see, e.g., \cite{sirovich1987turbulence,holmes1998turbulence,ravindran2000reduced}). The eigenvalue problem can be written in the following form:
\begin{equation}
\label{e:4}
    CW=W\Lambda \, ,
\end{equation}
where $\Lambda=\mbox{diag}[\lambda_1, \lambda_2, ..., \lambda_N ]$,
$W=[w^{1},  w^{2}, ..., w^{N}]$, $\lambda_i$ is the $i$th eigenvalue and $w^{i}$ is the corresponding $i$th eigenvector.
The eigenvalues are stored in descending order, $\lambda_1 \geq \lambda_2 \geq ... \geq \lambda_N$.
The POD basis functions can be written as
\begin{equation}
\label{e:5}
    \phi_1(x) = \sum_{i=1}^{N}w^{1}_{i}\hat{u}^{i}(x), \quad \phi_2(x) = \sum_{i=1}^{N}w^{2}_{i}\hat{u}^{i}(x), \quad ..., \quad \phi_N(x) = \sum_{i=1}^{N}w^{N}_{i}\hat{u}^{i}(x) \, ,
\end{equation}
where $w^{j}_{i}$ is the $i$th component of eigenvector $w^{j}$. The eigenvectors must be normalized in such a way that the basis functions satisfy the following orthogonality condition:
\begin{equation}
\label{e:6}
    (\phi_k,\phi_l)=\left\{
                      \begin{array}{ll}
                        1, & k=l \\
                        0, & k\neq l \, .
                      \end{array}
                    \right.
\end{equation}
It can be shown that, for Eq.~(\ref{e:6}) to be true, the eigenvector $\boldsymbol w^{j}$ must satisfy the following equation \cite{esfahanian2009equation}:
\begin{equation}
\label{e:7}
\sum_{i=1}^{N}w^{j}_{i}w^{j}_{i}=\frac{1}{\lambda_j}.
\end{equation}
In practice, most of the subroutines for solving the eigensystem given in Eq.~(\ref{e:4}) return the eigenvector matrix $W$ having all the eigenvectors normalized to unity. In that case, the orthogonal POD bases can be obtained as
\begin{equation}
\label{e:8}
    \phi_j(x) = \frac{1}{\sqrt{\lambda_j}}\sum_{i=1}^{N}w^{j}_{i}\hat{u}^{i}(x)
\end{equation}
where $\phi_j(x)$ is the $j$th POD basis function.

\subsection{Galerkin projection for reduced order models}
These POD basis functions account for the essential dynamics of the system.
To build a ROM, we truncate the system by considering the first $R$ POD basis functions with $R\ll N$.
These first $R$ POD modes correspond to the $R$ largest eigenvalues, $\lambda_1$, $\lambda_2$, ..., $\lambda_R$.
Using these first $R$ POD basis functions the field variables can be approximated as follows:
\begin{equation}
\label{e:11}
    \hat{u}(x,t)=\sum_{k=1}^{R}a_k(t)\phi_{k}(x) \, ,
\end{equation}
where $a_k$ are the time dependent coefficients, and $\phi_{k}$ are the space dependent modes.
To derive the POD-ROM, we first rewrite the Burgers equation (i.e., Eq.~(\ref{e:1})) as
\begin{equation}
\label{e:12}
    \frac{\partial u}{\partial t} = \nu L[u] + N[u;u] \, ,
\end{equation}
where for $f$ and $g$ arbitrary functions $L[f]=\frac{\partial^2 f}{\partial x^2}$ is the linear operator and $N[f;g]= - f \frac{\partial g}{\partial x}$ is the nonlinear operator.
By applying the Galerkin projection to the system (i.e., multiplying Eq.~(\ref{e:1}) with the basis function and integrating over the domain), we obtain the Galerkin POD-ROM, denoted POD-ROM-G:
\begin{equation}
\label{e:13}
    \left(\frac{\partial u}{\partial t},\phi_k\right) = (\nu L[u],\phi_k) + (N[u;u],\phi_k), \quad \mbox{for} \quad k=1,2, ..., R \, .
\end{equation}
Substituting Eq.~(\ref{e:9}) into Eq.~(\ref{e:8}), and simplifying the resulting equation by using the orthogonality condition given in Eq.~(\ref{e:6}), the POD-ROM-G in Eq.~(\ref{e:13}) can be written as follows:
\begin{equation}
\label{e:14}
    \frac{d a_{k}}{dt} = b_{k}^{1} + b_{k}^{2} + \sum_{i=1}^{R}(L_{ik}^{1} +L_{ik}^{2})a_{i} + \sum_{i=1}^{R}\sum_{j=1}^{R} N_{ijk}a_{i}a_{j}, \quad \mbox{for} \quad k=1,2, ..., R \, ,
\end{equation}
where
\begin{eqnarray}
b_{k}^{1}  &=&(\nu L[\bar{u}],\phi_{k}) \label{e:14a} \\
b_{k}^{2}  &=&(N[\bar{u};\bar{u}],\phi_{k}) \label{e:14b}  \\
L_{ik}^{1} &=&(\nu L[\phi_{i}],\phi_{k}) \label{e:14c}  \\
L_{ik}^{2} &=&(N[\bar{u};\phi_{i}] + N[\phi_{i} ; \bar{u}],\phi_{k}) \label{e:14d}  \\
N_{ijk}    &=&(N[\phi_{i};\phi_{j}],\phi_{k}) \label{e:14e} .
\end{eqnarray}
The POD-ROM-G given by Eq.~(\ref{e:14}) consists of $R$ coupled ordinary differential equations and can be solved by a standard numerical method (such as the third-order Runge-Kutta scheme that was used in this study). We emphasize that the number of degrees of freedom of the system has been substantially decreased.
The vectors, matrices and tensors in Eqs.~(\ref{e:14a})-(\ref{e:14e}) are generally precomputed, which results in a dynamical system that can be solved very efficiently.
To complete the dynamical system given by Eq.~(\ref{e:14}), the initial condition is given by using the following projection:
\begin{equation}
a_{k}(t=0)= (u(x,t=0)-\bar{u}(x),\phi_{k}) \, ,
\end{equation}
where $u(x,t=0)$ is the physical initial condition of the problem.

\subsection{Closure models}
\label{sec:closure}

In this study, we consider three classes of POD-ROM closure models.
To the authors' best knowledge, some of these closure models are new.

To unify the notation, we rewrite the POD-ROM given by Eq.~(\ref{e:14}) in the following form:
\begin{equation}
\label{e:15}
    \frac{d a_{k}}{dt} = b_{k}^{1} + b_{k}^{2} + \tilde{b}_{k}^{3} + \sum_{i=1}^{R}(L_{ik}^{1} +L_{ik}^{2} +\tilde{L}_{ik}^{3})a_{i} + \sum_{i=1}^{R}\sum_{j=1}^{R} N_{ijk}a_{i}a_{j} , \quad \mbox{for} \quad k=1,2, ..., R \, ,
\end{equation}
where $\tilde{b}_{k}^{3}$ and $\tilde{L}_{ik}^{3}$ are the constraint and linear coefficient terms related to the closure models.
Eq.~(\ref{e:15}), with various choices for $\tilde{b}_{k}^{3}$ and $\tilde{L}_{ik}^{3}$, represents the POD-ROM considered in this manuscript.
We emphasize that the coefficients $\tilde{b}_{k}^{3}=\tilde{L}_{ik}^{3}=0$ for the POD-ROM-G.

\subsubsection{Constant eddy viscosity models}
\label{sec:viscosity_kernel}
The viscosity kernel closure models account for the effect of the truncated POD modes by using an eddy viscosity ansatz.
In these closure models, the coefficient of the diffusion operator is constant in space and time, but can be POD mode dependent.
To simplify the notation, the coefficient of the diffusion operator is written as the product of two factors that are constant in space and time: $\nu_{e}$, which represents the eddy viscosity amplitude and does not depend on the POD mode; and $\psi_{k}$, which represents the viscosity coefficient and can depend on the POD mode.
With this notation, the POD-ROM coefficients can be written as follows:
\begin{eqnarray}
\tilde{b}_{k}^{3}  &=&(\nu_{e} \psi_{k} L[\bar{u}],\phi_{k}) \label{e:18a} \\
\tilde{L}_{ik}^{3} &=&(\nu_{e} \psi_{k} L[\phi_{i}],\phi_{k}) \, . \label{e:18b}
\end{eqnarray}
By adjusting the eddy viscosity amplitude $\nu_{e}$ in the POD-ROM, a better accuracy can be obtained.

The first and simplest model in this category is the Heisenberg POD-ROM (POD-ROM-H).
This closure model, which was introduced in \cite{aubry1988dynamics}, uses the following viscosity kernel:
\begin{equation}
\label{e:19}
\psi_{k}^{POD-ROM-H}=1 \, .
\end{equation}
We note that, since in LES of turbulent flows this closure model is called the \emph{mixing length} model, the POD-ROM-H is also called the mixing length POD-ROM in the literature \cite{wang2012proper}.

An improvement of the POD-ROM-H was proposed by Rempfer \cite{rempfer1991proper}.
In this closure model (denoted POD-ROM-R), a linear viscosity kernel is considered:
\begin{equation}
\label{e:20}
\psi_{k}^{POD-ROM-R}=\frac{k}{R} \, .
\end{equation}
We emphasize that the viscosity kernel in Eq.~(\ref{e:20}) is POD mode dependent.

In this study, we propose two new POD-ROM closure models, which are variants of the POD-ROM-R.
The first closure model (denoted POD-ROM-RQ) uses a quadratic viscosity kernel:
\begin{equation}
\label{e:21}
\psi_{k}^{POD-ROM-RQ}=\left(\frac{k}{R}\right)^2.
\end{equation}
The second closure model (denoted POD-ROM-RS) uses a square-root viscosity kernel:
\begin{equation}
\label{e:22}
\psi_{k}^{POD-ROM-RS}=\left(\frac{k}{R}\right)^{1/2}.
\end{equation}

The structural differences among POD-ROM-R, POD-ROM-RQ, and POD-ROM-RS are obvious, since we have normalized the viscosity kernels $\psi_{k}$ between 0 to 1:
For the POD modes with the highest energy content, the POD-ROM-RQ dissipates most energy, followed by the POD-ROM-R and POD-ROM-RS (in this order).
For the POD modes with the lowest energy content, the POD-ROM-RS dissipates most energy, followed by the POD-ROM-R and POD-ROM-RQ (in this order).
We also note that the POD-ROM-RS is limiting the amount of eddy viscosity introduced by the closure model.
In this sense, the POD-ROM-RS is similar in spirit with the concept of ``clipping" used in LES models of eddy viscosity type \cite{sagaut2006large,borggaard2009bounded}.

The spectral vanishing viscosity POD-ROMs, which were introduced in \cite{karamanos2000spectral,sirisup2004spectral}, are also viscosity kernel closure models.
The spectral vanishing viscosity concept was introduced by Tadmor \cite{tadmor1989convergence} using the inviscid Burgers equation.
The extension of this closure model to the POD setting yields the following POD-ROM (denoted POD-ROM-T):
\begin{equation}
\label{e:23}
\psi_{k}^{POD-ROM-T}=\left\{
                   \begin{array}{ll}
                     0, & k \leq M \\
                     1, & k > M \, ,
                   \end{array}
                 \right.
\end{equation}
where $M\leq R$ is another free parameter in this model.
The POD-ROM-T is similar to the POD-ROM-R (and its variants) in that it adds a POD mode dependent amount of artificial viscosity.
We note that, to the best of our knowledge, the POD-ROM-T has not been used in the POD setting before.


Sirisup and Karniadakis \cite{sirisup2004spectral} extended the spectral vanishing viscosity concept put forth in \cite{tadmor1989convergence} and \cite{maday1993legendre} to the POD setting.
The resulting model, which we call the Maday-Karniadakis POD-ROM (denoted POD-ROM-MK) uses the following viscosity kernel function:
\begin{equation}
\label{e:24}
\psi_{k}^{POD-ROM-MK}=\left\{
                   \begin{array}{ll}
                     e^{-(k-R)^2/(k-M)^2}, & k \leq M \\
                     0, & k > M \, .
                   \end{array}
                 \right.
\end{equation}

The last closure model in this class is an extension to the POD setting of the model introduced by Chollet and Lesieur \cite{lesieur1996new,chollet1985two} and employed in a spectral vanishing viscosity framework in \cite{karamanos2000spectral}.
The resulting POD-ROM, which we call the Chollet-Lesieur POD-ROM (denoted by POD-ROM-CL) employs the viscosity kernel function
\begin{equation}
\label{e:25}
\psi_{k}^{POD-ROM-CL}=\kappa_{0}^{-3/2}[\kappa_{1}+\kappa_{2}e^{-\kappa_{3}/(k/R)}] \, ,
\end{equation}
where the coefficients have the following values \cite{karamanos2000spectral}: $\kappa_{0}=1.1135$, $\kappa_{1}=0.441$, $\kappa_{2}=15.2$, and $\kappa_{3}=3.03$.
We note that, to the best of our knowledge, the POD-ROM-CL is new.

\subsubsection{Smagorinsky type models}
\label{sec:smagorinsky}

This section outlines two closure models that are significantly different from the viscosity kernel models presented in Section~\ref{sec:viscosity_kernel}:
The closure models considered in this section introduce an artificial viscosity that is variable in both space and time, whereas the models in Section~\ref{sec:viscosity_kernel} use an artificial viscosity that is constant in space and time.

The Smagorinsky model (and its improvements) is by far the most popular closure model in LES of turbulent flows.
It is thus natural that this model has been extended to the POD setting \cite{noack2002low,ullmann2010pod,borggaard2011artificial,wang2011two,wang2012proper}.
To present the resulting POD-ROM, we consider the following nonlinear operator:
\begin{equation}
\label{e:27}
    S[f,g]=\bigg|\frac{\partial f}{\partial x}\bigg| \frac{\partial^2 g}{\partial x^2} \, .
\end{equation}
The Smagorinsky POD-ROM (denoted POD-ROM-S) uses the following coefficients in the generic POD-ROM given in Eq.~(\ref{e:15}):
\begin{eqnarray}
\tilde{b}_{k}^{3}  &=&(\nu_{e} S[\bar{u};\bar{u}],\phi_{k}) \label{e:30} \\
\tilde{L}_{ik}^{3} &=&(\nu_{e} S[\bar{u};\phi_{i}] + \nu_{e} S[\phi_{i} ; \bar{u}],\phi_{k}) \, , \label{e:31}
\end{eqnarray}
where $\nu_{e}$ is the constant eddy viscosity amplitude.


By incorporating Rempfer's idea of using a POD mode dependent eddy viscosity coefficient, we propose the Smagorinsky-Rempfer POD-ROM (denoted POD-ROM-SR):
\begin{eqnarray}
\tilde{b}_{k}^{3}  &=&(\nu_{e} \psi_{k}^{POD-ROM-R} S[\bar{u};\bar{u}],\phi_{k}) \label{e:33} \\
\tilde{L}_{ik}^{3} &=&(\nu_{e} \psi_{k}^{POD-ROM-R} S[\bar{u};\phi_{i}] + \nu_{e} \psi_{k}^{POD-ROM-R} S[\phi_{i} ; \bar{u}],\phi_{k}) \, . \label{e:34}
\end{eqnarray}
We note that, to the best of our knowledge, the POD-ROM-SR is new.

We emphasize that, for general three-dimensional flows, the POD-ROM-S and POD-ROM-SR are significantly more expensive than the closure models in Section~\ref{sec:viscosity_kernel}.
The reason is that the nonlinear Smagorinsky term in the POD-ROM requires the evaluation of its associated tensor at each time step.
(See, e.g., \cite{wang2011two} for efficient algorithms for this type of POD-ROM closure models.)
In the one-dimensional setting of the Burgers equation that we consider in this report, the nonlinear Smagorinsky term can be precomputed.
Thus, the resulting POD-ROMs are practically as efficient as those in Section~\ref{sec:viscosity_kernel}.


\subsubsection{Cazemier's penalty model}
A closure model that is different from those in Sections~\ref{sec:viscosity_kernel} and \ref{sec:smagorinsky} was proposed in \cite{cazemier1997proper,cazemier1998proper}.
This POD-ROM, which we call the Cazemier POD-ROM (denoted POD-ROM-C), is based on the concept of energy conservation and adds the following penalty term to the generic POD-ROM  given by Eq.~(\ref{e:14}):
\begin{equation}
\label{e:40}
    \frac{d a_{k}}{dt} = b_{k}^{1} + b_{k}^{2} + \sum_{i=1}^{R}(L_{ik}^{1} +L_{ik}^{2})a_{i} + \sum_{i=1}^{R}\sum_{j=1}^{R} N_{ijk}a_{i}a_{j} + H_{k}a_{k}, \quad \mbox{for} \quad k=1,2, ..., R \, .
\end{equation}
The linear damping coefficient is given by
\begin{eqnarray}
H_{k}=-\frac{1}{N \lambda_{k}} \sum_{i=1}^{R} \sum_{j=1}^{R} N_{ijk}\langle a_i a_j a_k \rangle - (L_{ik}^{1} +L_{ik}^{2}) \, ,
\end{eqnarray}
where $a_{k}^{n}$ are computed as
\begin{eqnarray}
\label{e:41}
a_{k}^{n} = (u(x,t_n)-\bar{u}(x),\phi_{k})
\end{eqnarray}
and $\langle a_i a_j a_k \rangle$ can be precomputed from the snapshots using the following ensemble average:
\begin{eqnarray}
\langle a_i a_j a_k \rangle = \frac{1}{N} \sum_{n=1}^{N} a_{i}^{n} a_{j}^{n} a_{k}^{n} \, .
\end{eqnarray}
One of the main advantages of the POD-ROM-C is that it does not require any free parameter.
A potential drawback, however, is that the POD-ROM-C has a higher computational overhead.

\begin{figure}[!ht]
\centering
\includegraphics[width=0.6\textwidth]{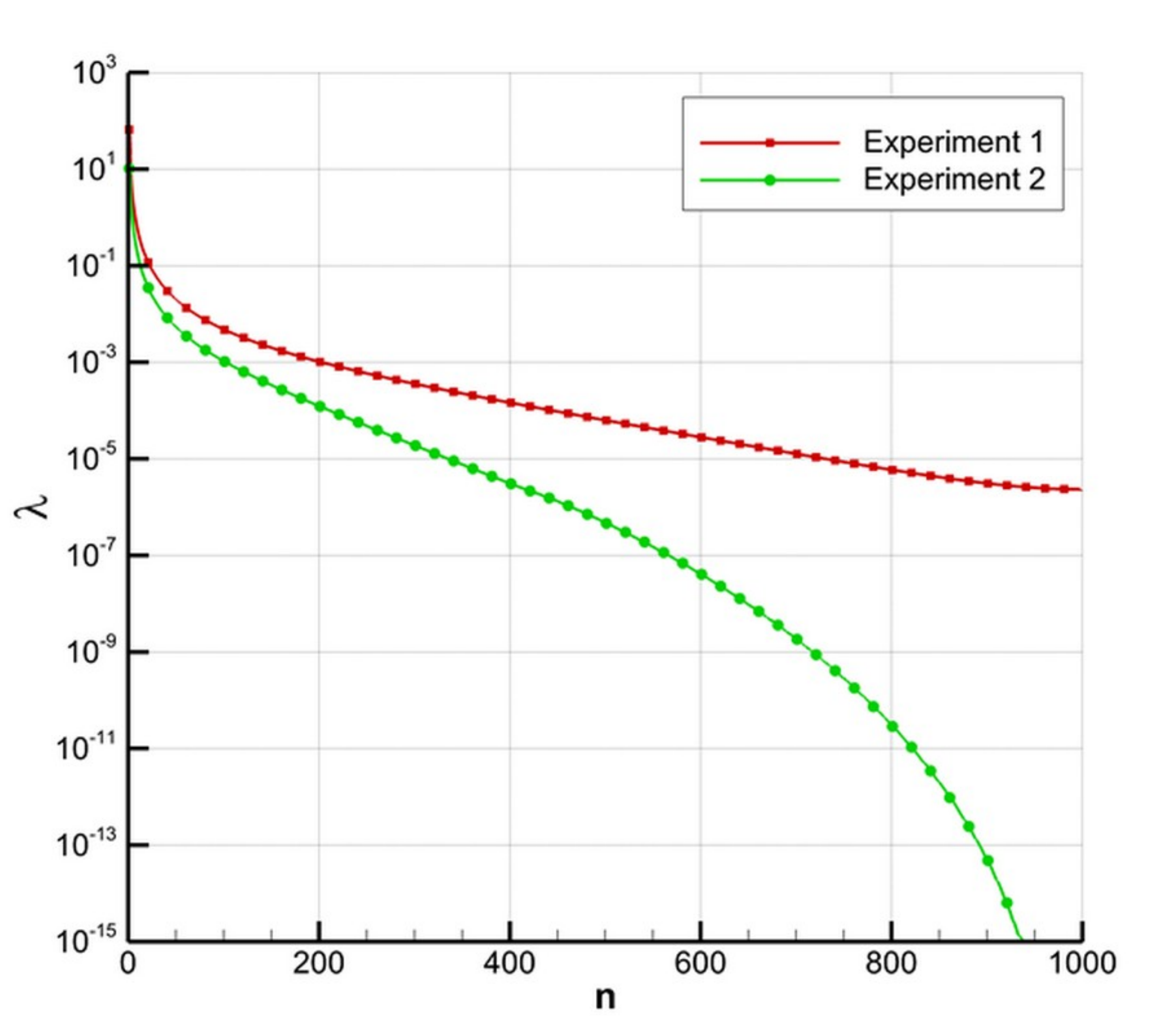}
\caption{Eigenvalues of the correlation matrix $C$ for 1000 snapshots.}
\label{fig:1}
\end{figure}

\begin{figure}[!ht]
\centering
\mbox{
\subfigure[Experiment 1]{\includegraphics[width=0.5\textwidth]{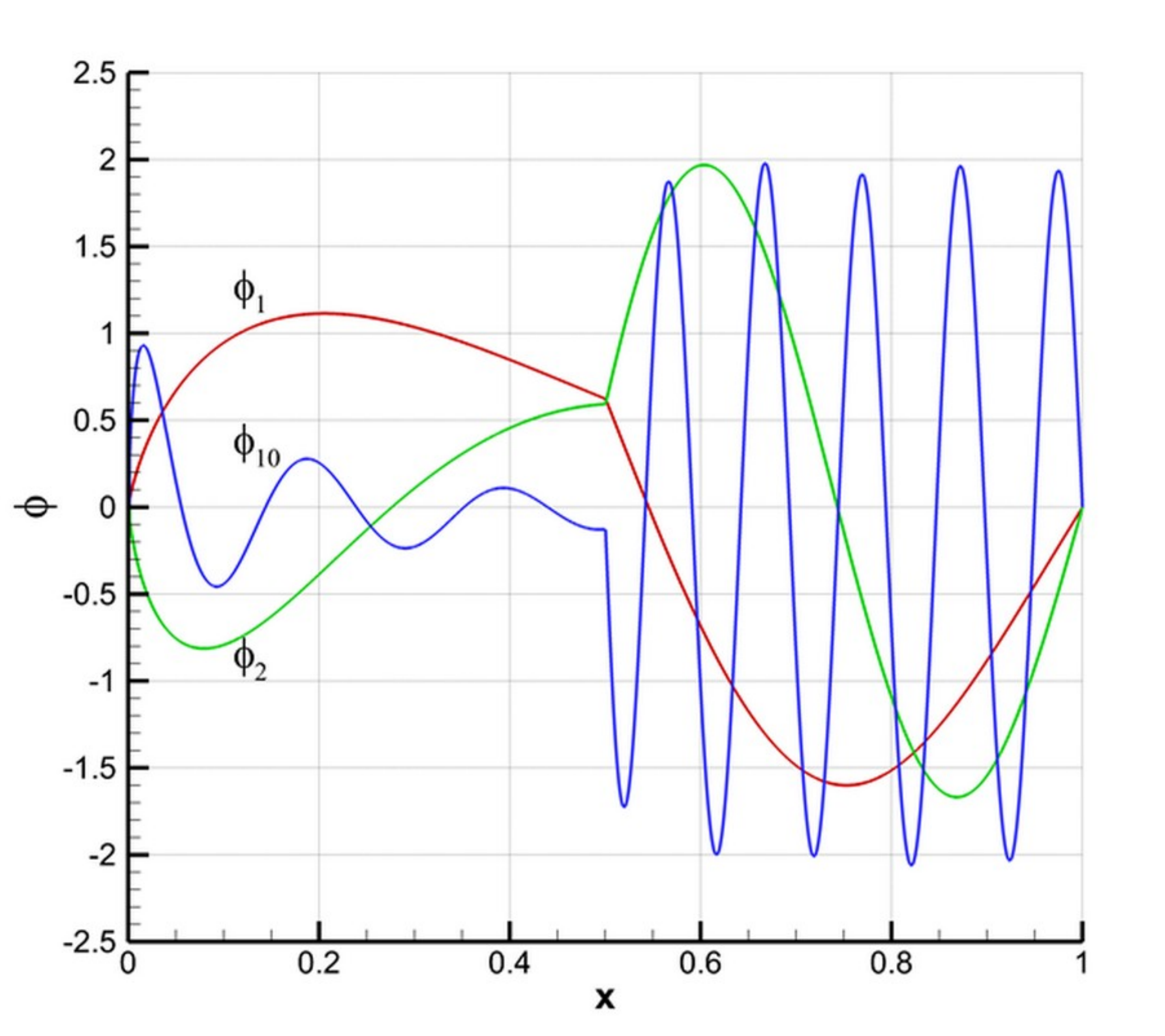}}
\subfigure[Experiment 2]{\includegraphics[width=0.5\textwidth]{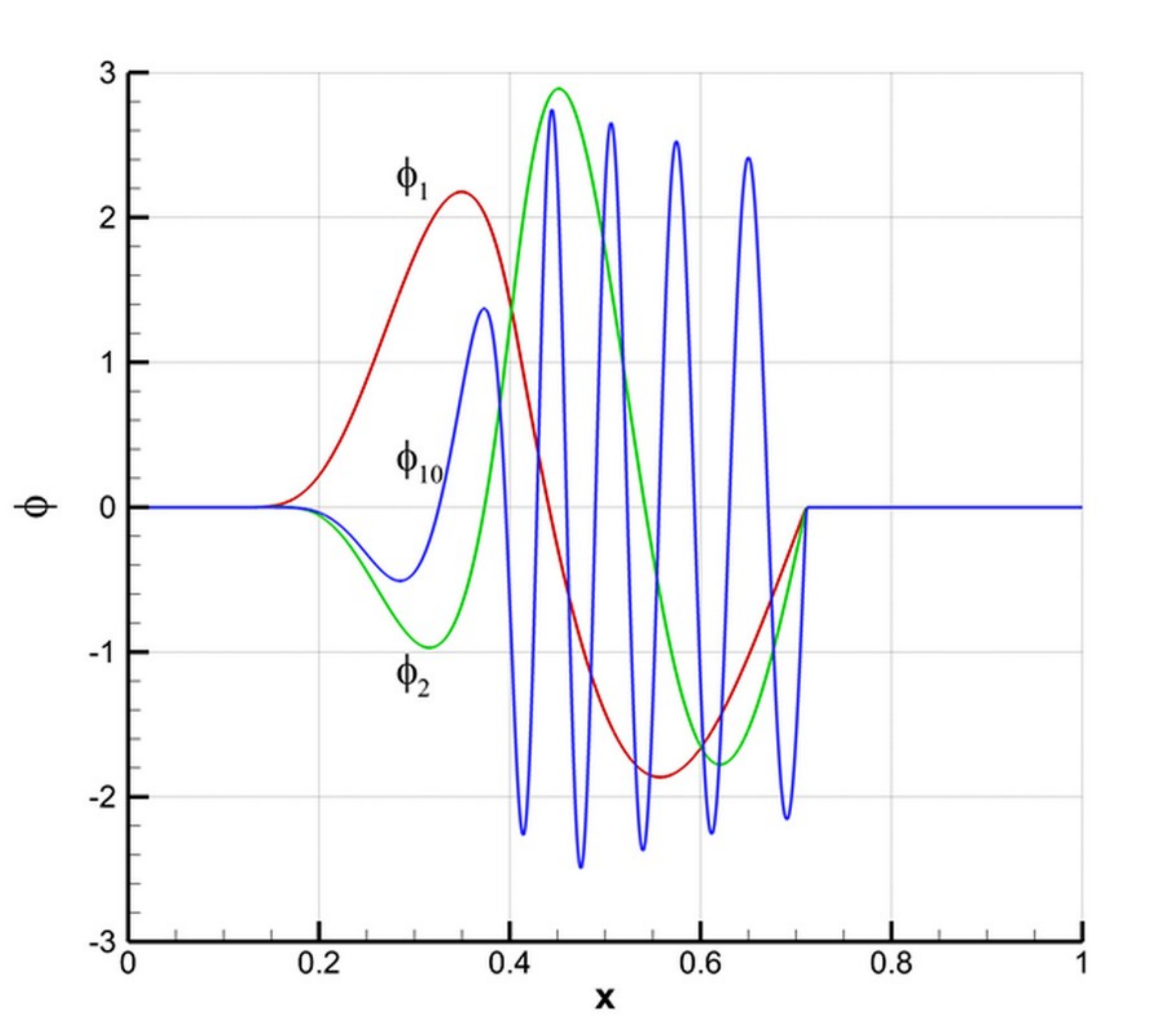}}
}
\caption{Illustrative examples of POD basis functions.}
\label{fig:2}
\end{figure}

\section{Results}
\label{sec:results}

In this section, we present a numerical investigation of the POD-ROM closure models outlined in Section~\ref{sec:pod}.
Both the numerical accuracy and the computational efficiency of the POD-ROMs are considered.
Results for the POD-ROM-G (i.e., the POD-ROM without any closure model) and for the DNS (the benchmark) are also included for comparison purposes.
All the models are tested on the Burgers equation with two different initial conditions: Eq.~(\ref{eq:a}) and Eq.~(\ref{eq:b}), which correspond to Experiment 1 and Experiment 2, respectively.
Both settings yield shock wave phenomena that are challenging to capture with a standard POD-ROM-G.

A sixth-order compact difference scheme is used for the spatial discretization of the models and a third-order Runge-Kutta scheme is employed for the temporal discretization.
The computational domain, $[0,1]$, is uniformly discretized by using 8192 grid points (which yields a mesh-size $h=1/8192$).
The time step is $\Delta t = 5\times 10^{-5}$.
The viscosity parameter in the Burgers equation is $\nu = 10^{-4}$ for both experiments.

To build the POD basis, we collect 1000 snapshots in the time interval $[0,1]$ taken at equidistant time instances.
The correlation matrix $C$ is constructed using these 1000 snapshots.
The eigenvalues of the correlation matrix are shown in Fig.~\ref{fig:1} for both experiments.
Fig.~\ref{fig:2} shows some illustrative POD basis functions (i.e., $\phi_{1}(x)$, $\phi_{2}(x)$, and $\phi_{10}(x)$) for both experiments.

\begin{figure}[!t]
\centering
\mbox{
\subfigure[DNS]{\includegraphics[width=0.35\textwidth]{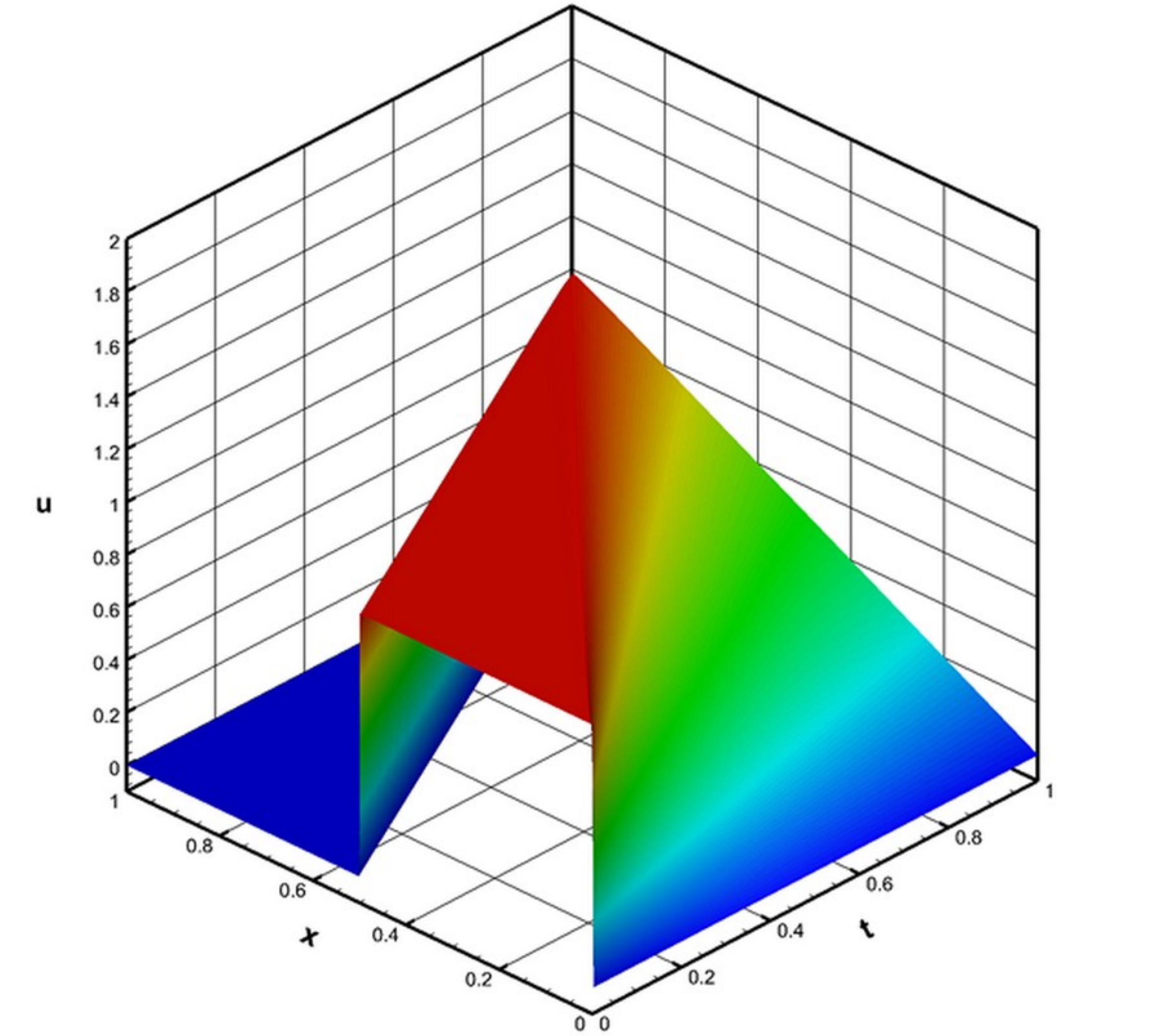}}
\subfigure[POD-ROM-G (5 modes)]{\includegraphics[width=0.35\textwidth]{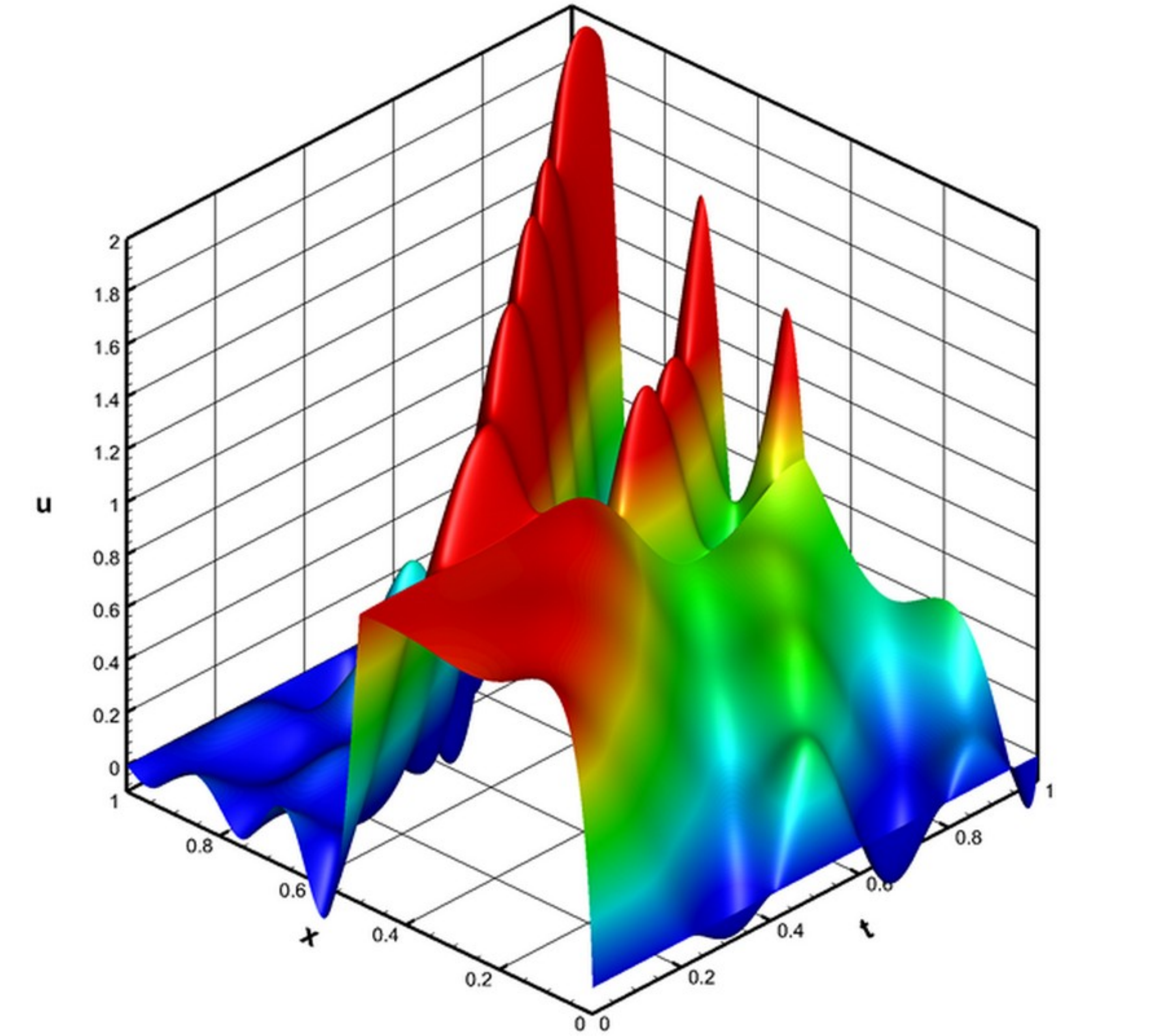}}
\subfigure[POD-ROM-G (10 modes)]{\includegraphics[width=0.35\textwidth]{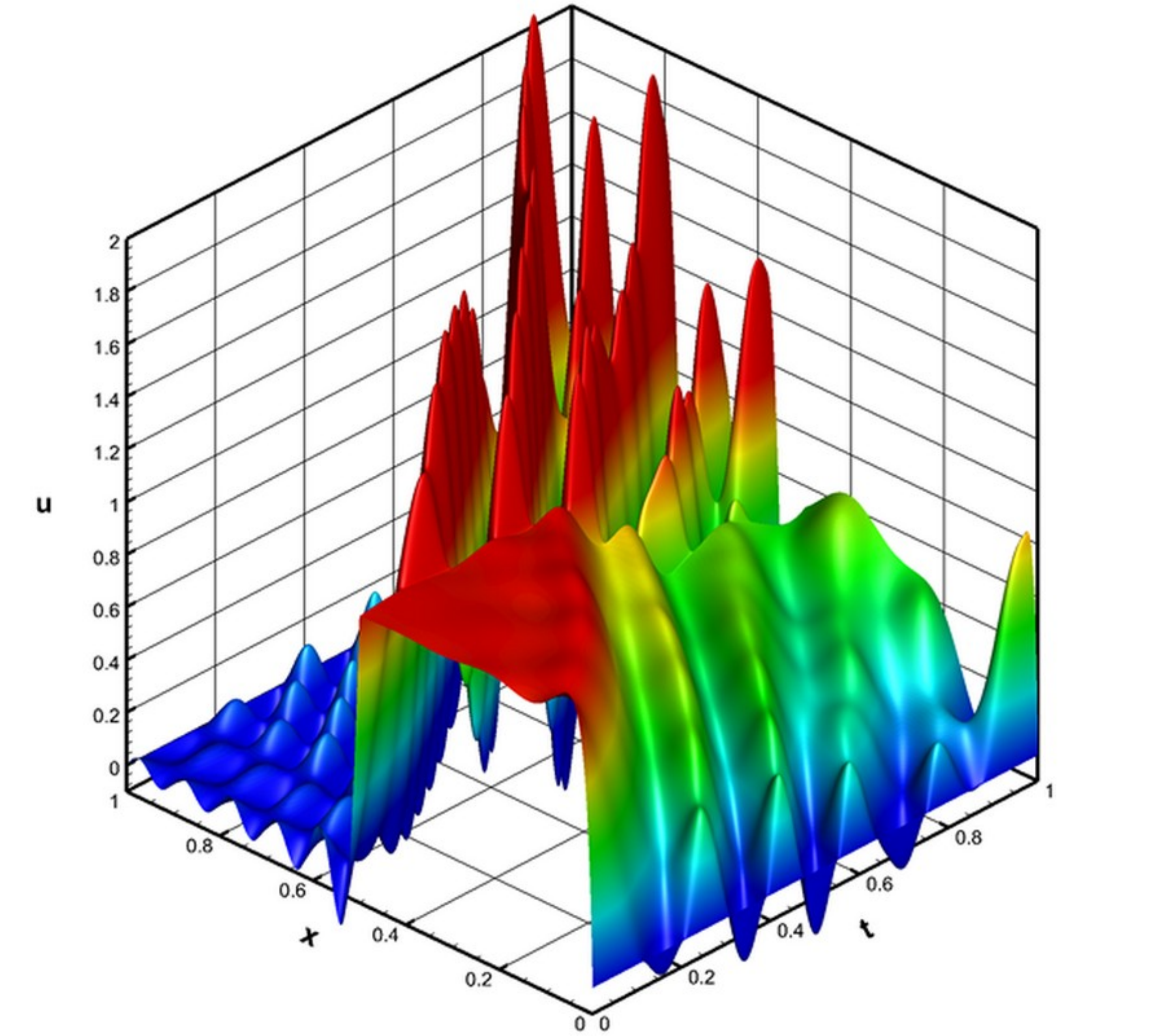}}
}\\
\mbox{
\subfigure[POD-ROM-G (20 modes)]{\includegraphics[width=0.35\textwidth]{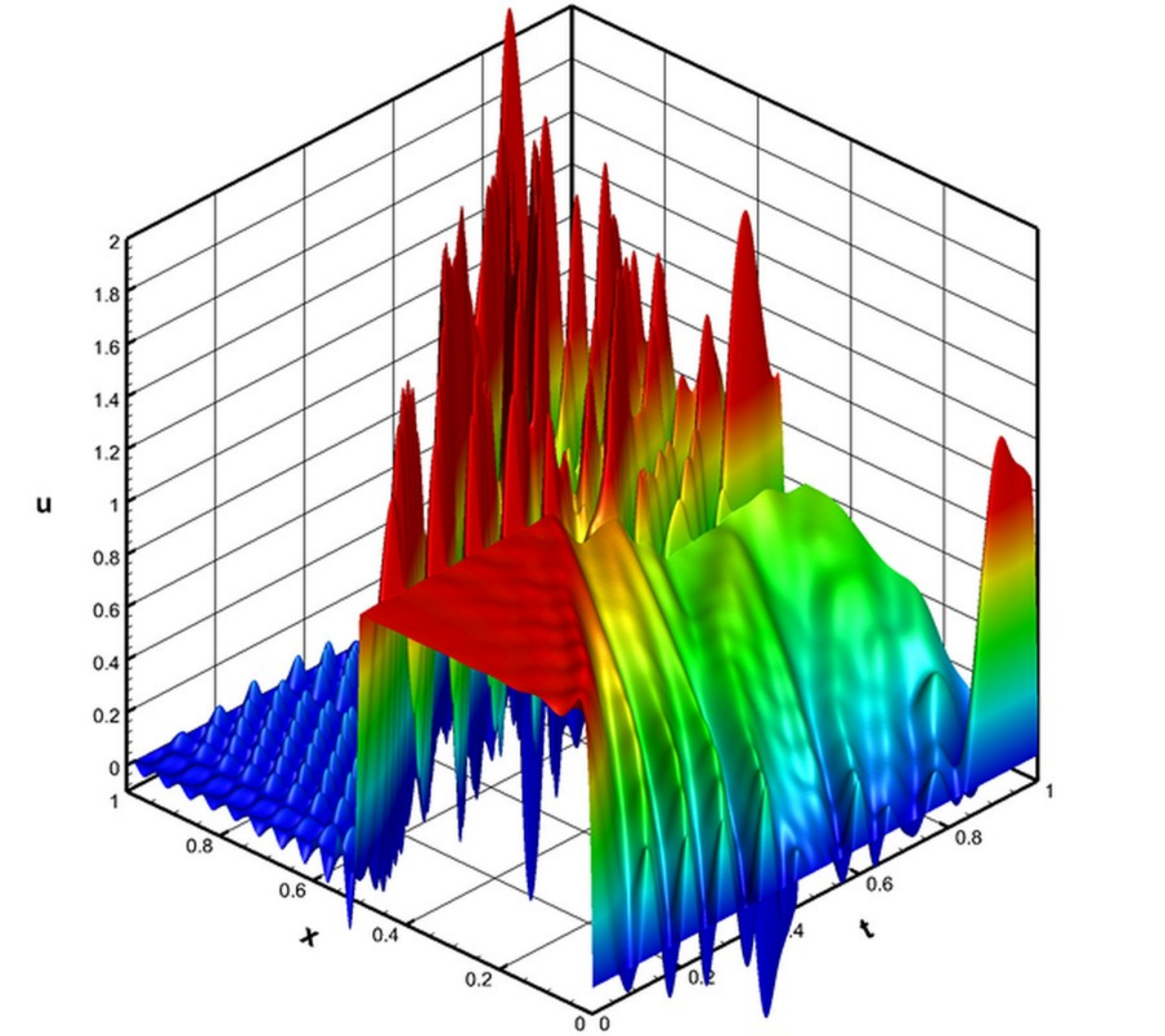}}
\subfigure[POD-ROM-G (30 modes)]{\includegraphics[width=0.35\textwidth]{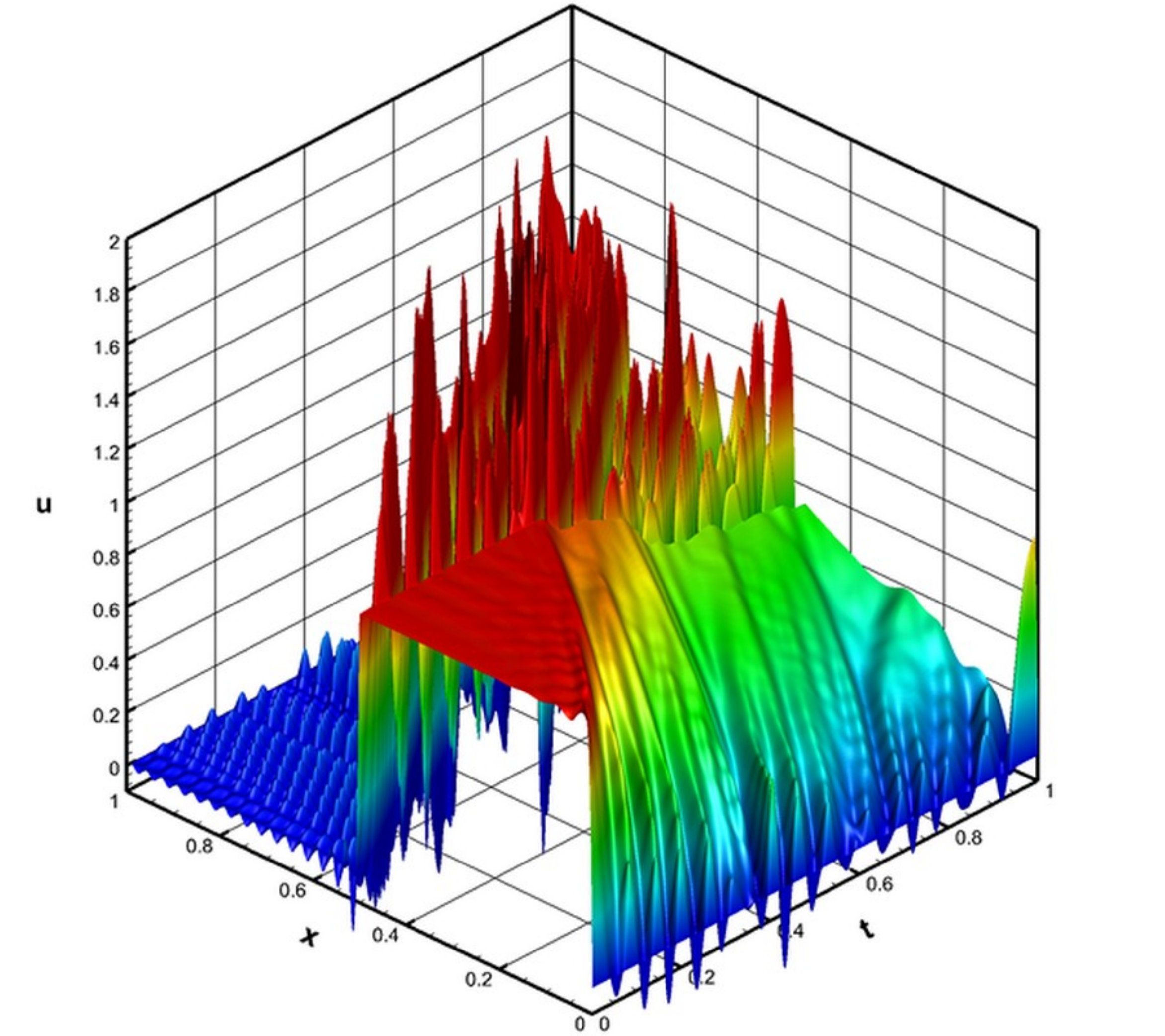}}
\subfigure[POD-ROM-G (40 modes)]{\includegraphics[width=0.35\textwidth]{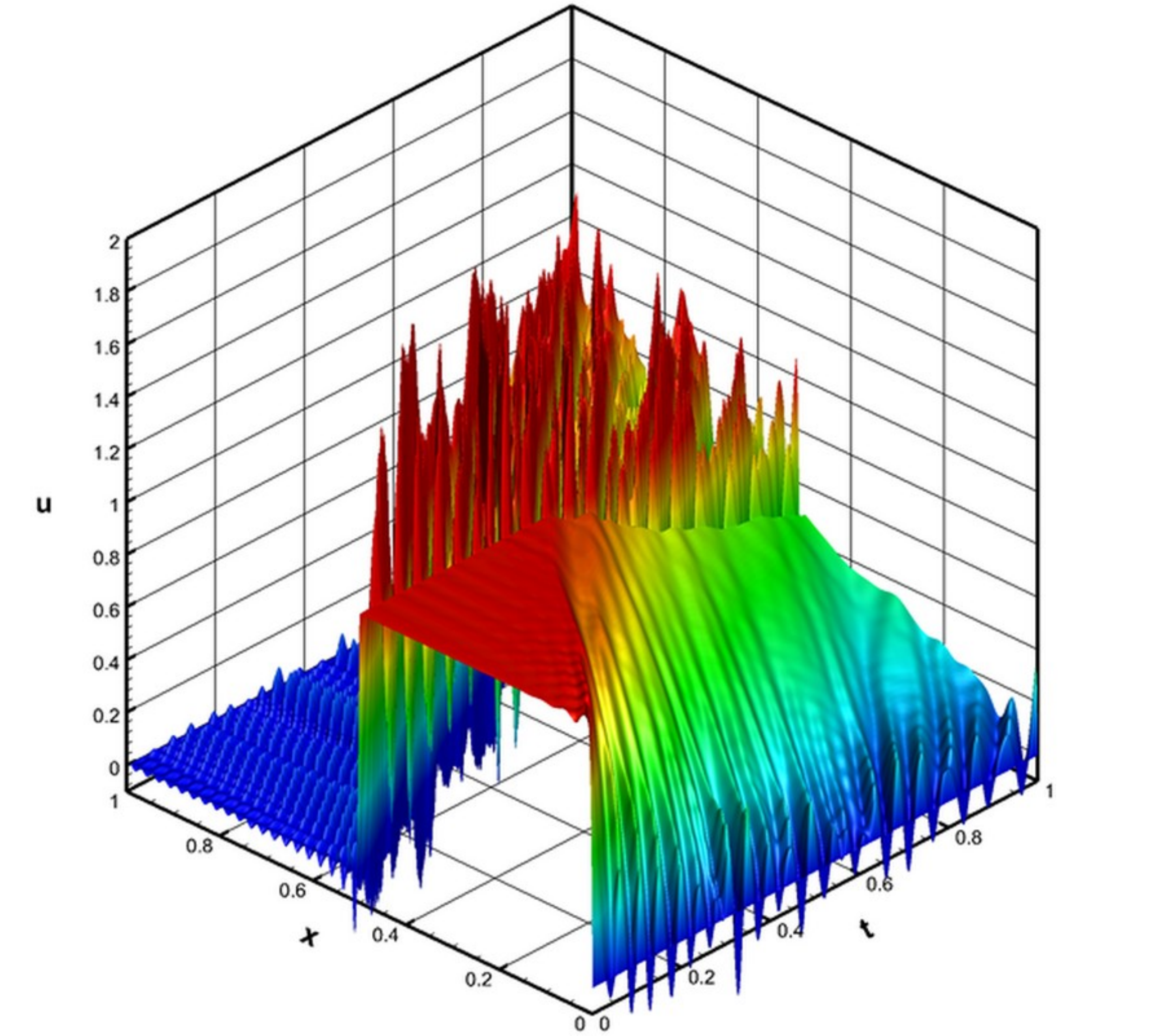}}
}\\
\mbox{
\subfigure[POD-ROM-G (80 modes)]{\includegraphics[width=0.35\textwidth]{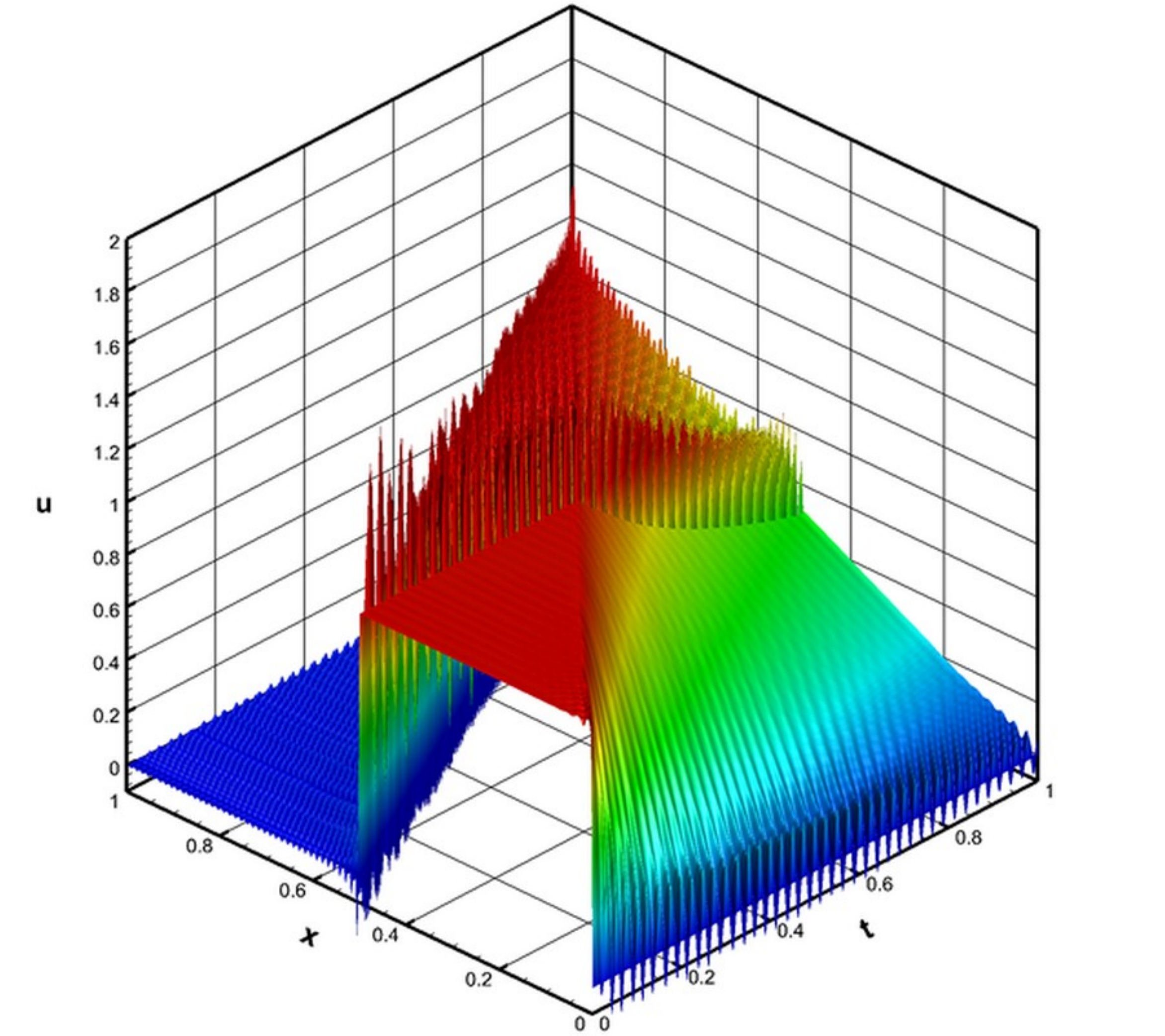}}
\subfigure[POD-ROM-G (160 modes)]{\includegraphics[width=0.35\textwidth]{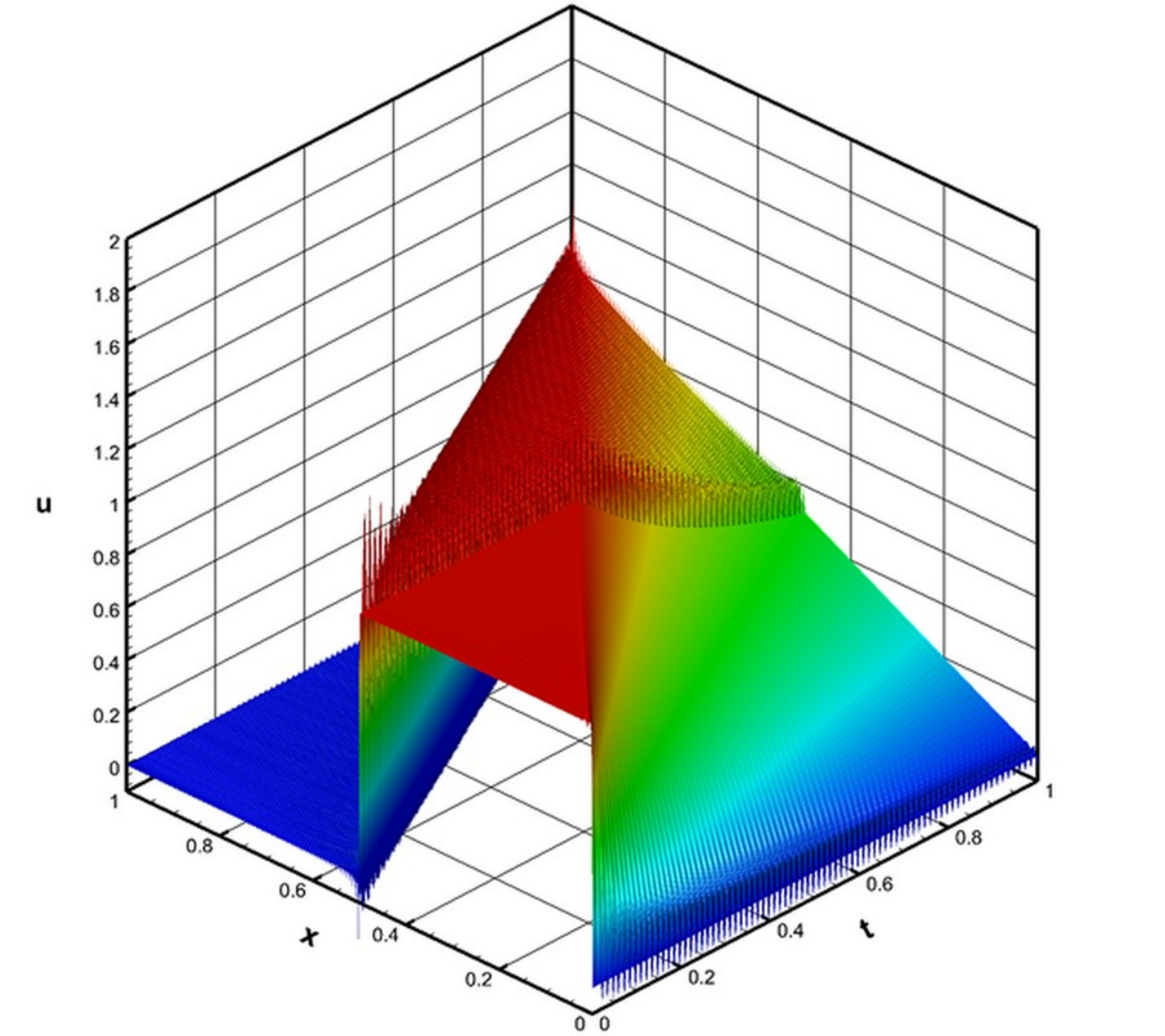}}
\subfigure[POD-ROM-G (320 modes)]{\includegraphics[width=0.35\textwidth]{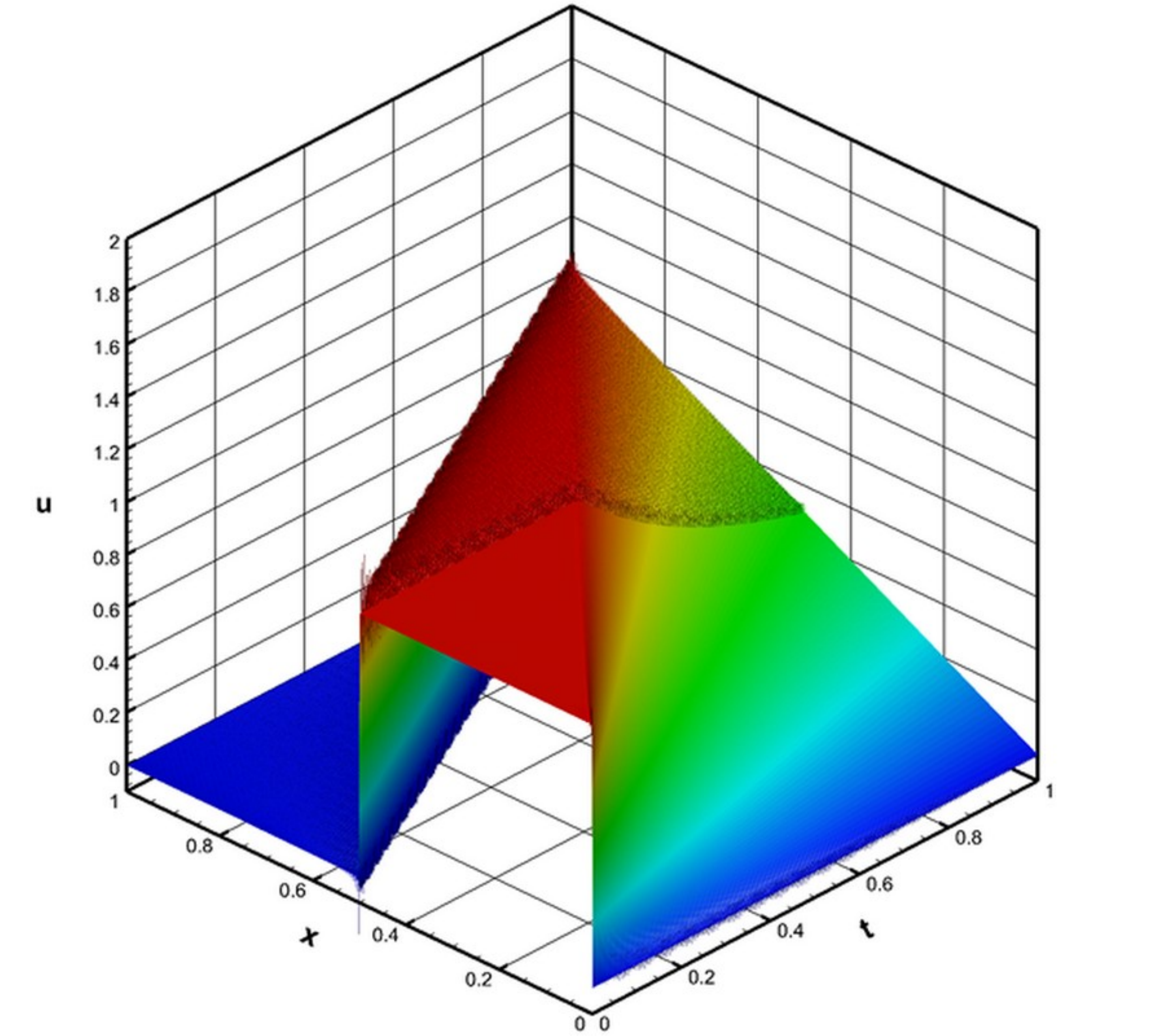}}
}
\caption{Experiment 1: POD-ROM-G with various numbers of POD modes.}
\label{fig:3}
\end{figure}

\begin{figure}[!t]
\centering
\mbox{
\subfigure[DNS]{\includegraphics[width=0.35\textwidth]{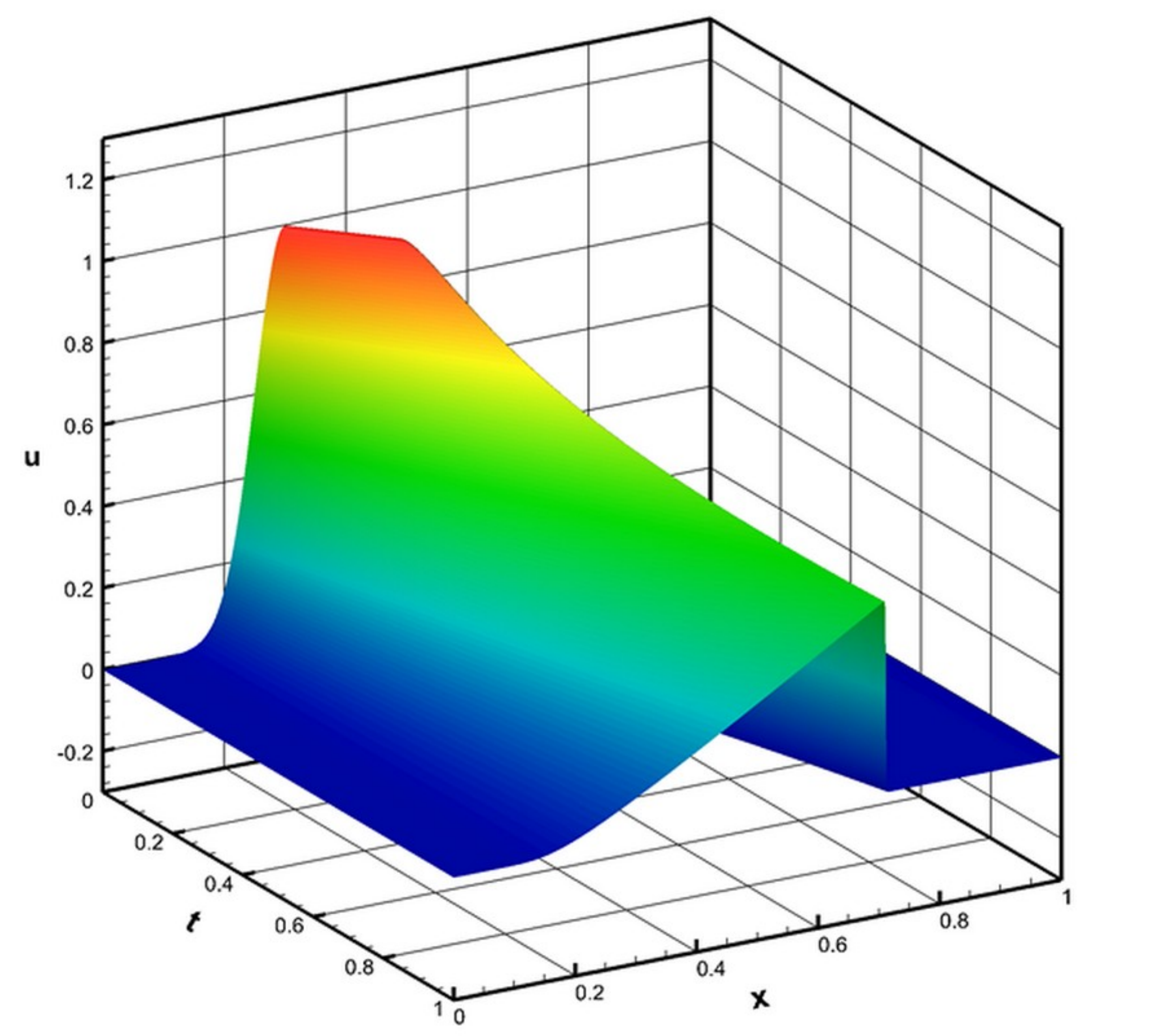}}
\subfigure[POD-ROM-G (5 modes)]{\includegraphics[width=0.35\textwidth]{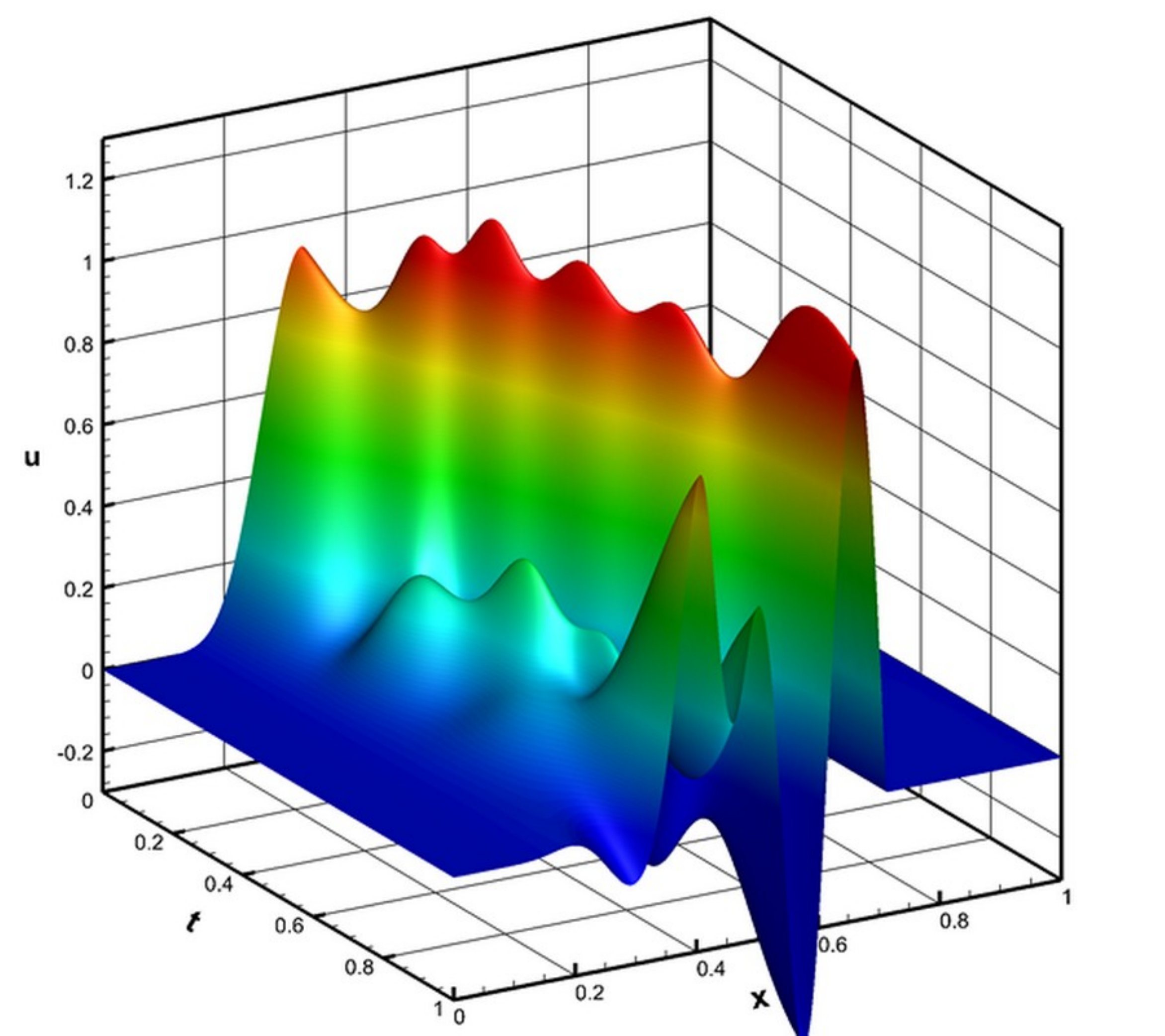}}
\subfigure[POD-ROM-G (10 modes)]{\includegraphics[width=0.35\textwidth]{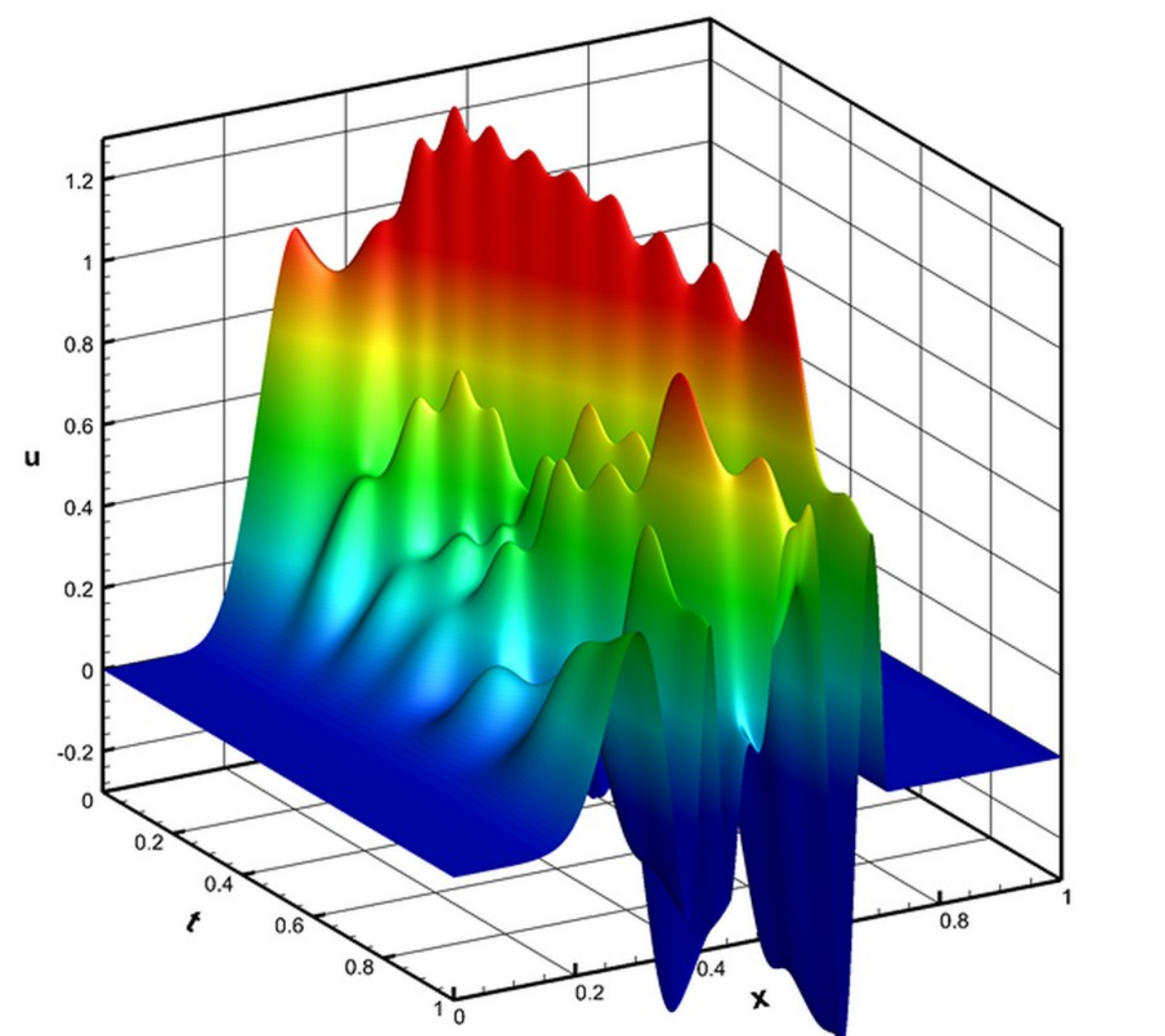}}
}\\
\mbox{
\subfigure[POD-ROM-G (20 modes)]{\includegraphics[width=0.35\textwidth]{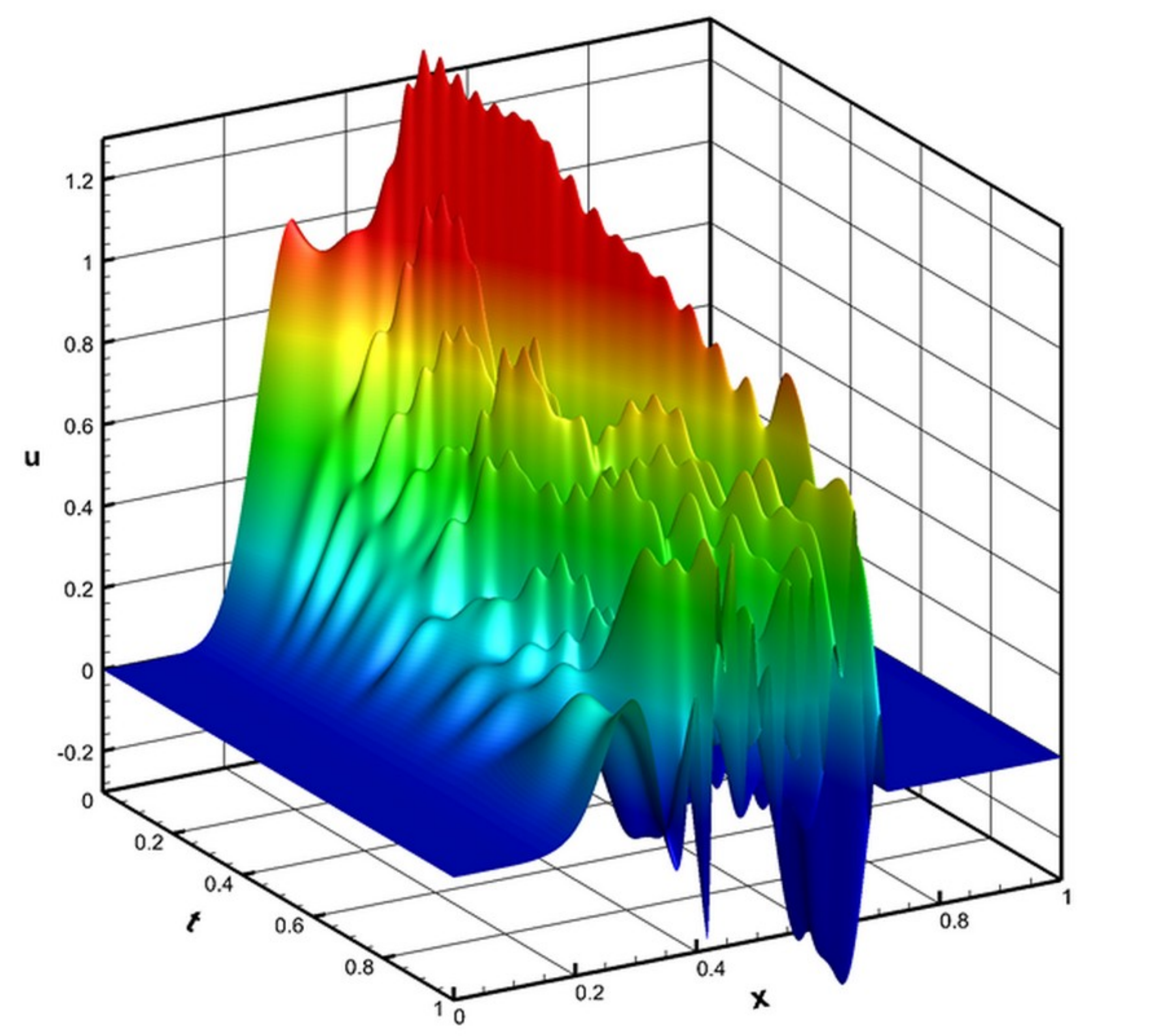}}
\subfigure[POD-ROM-G (30 modes)]{\includegraphics[width=0.35\textwidth]{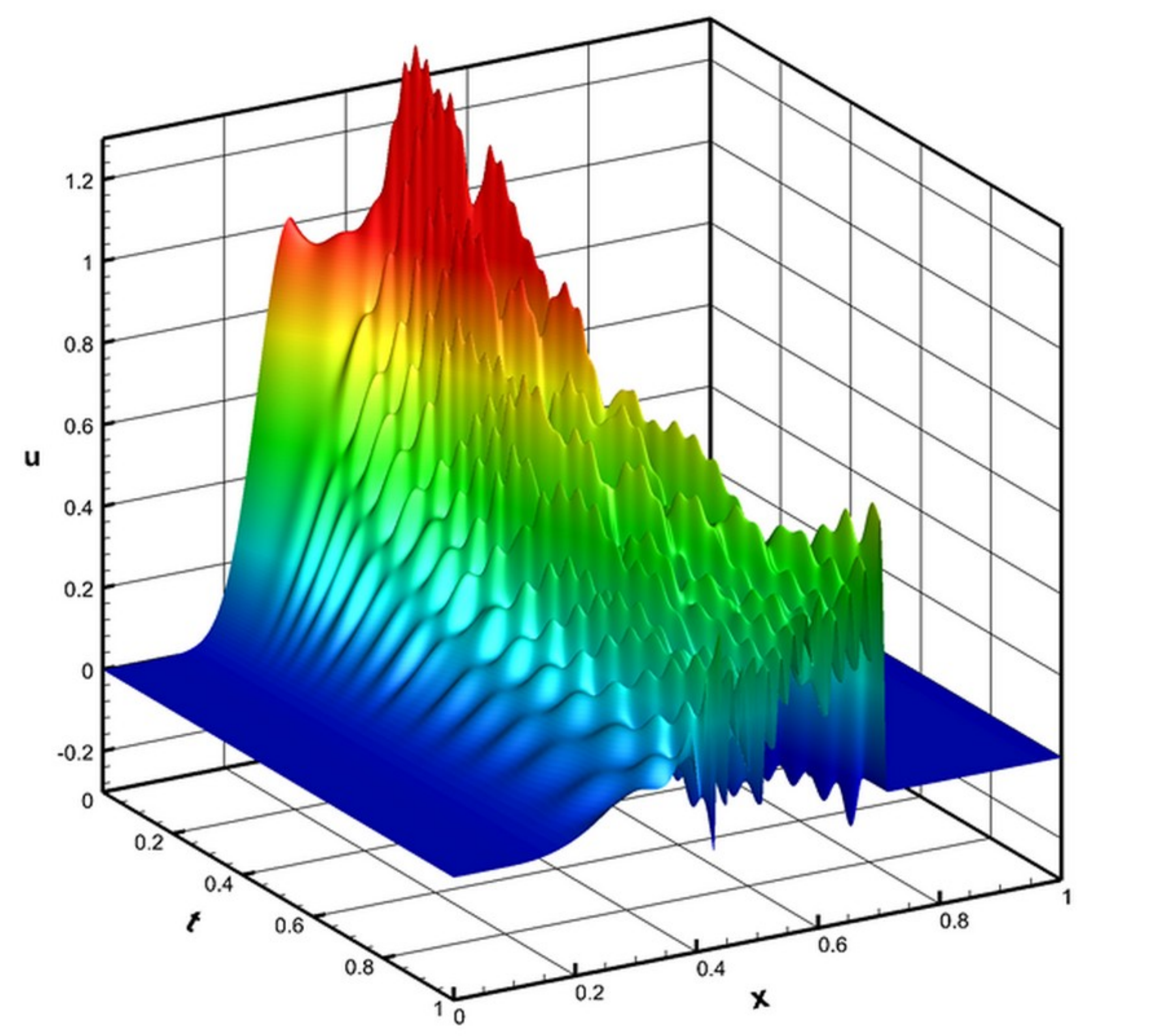}}
\subfigure[POD-ROM-G (40 modes)]{\includegraphics[width=0.35\textwidth]{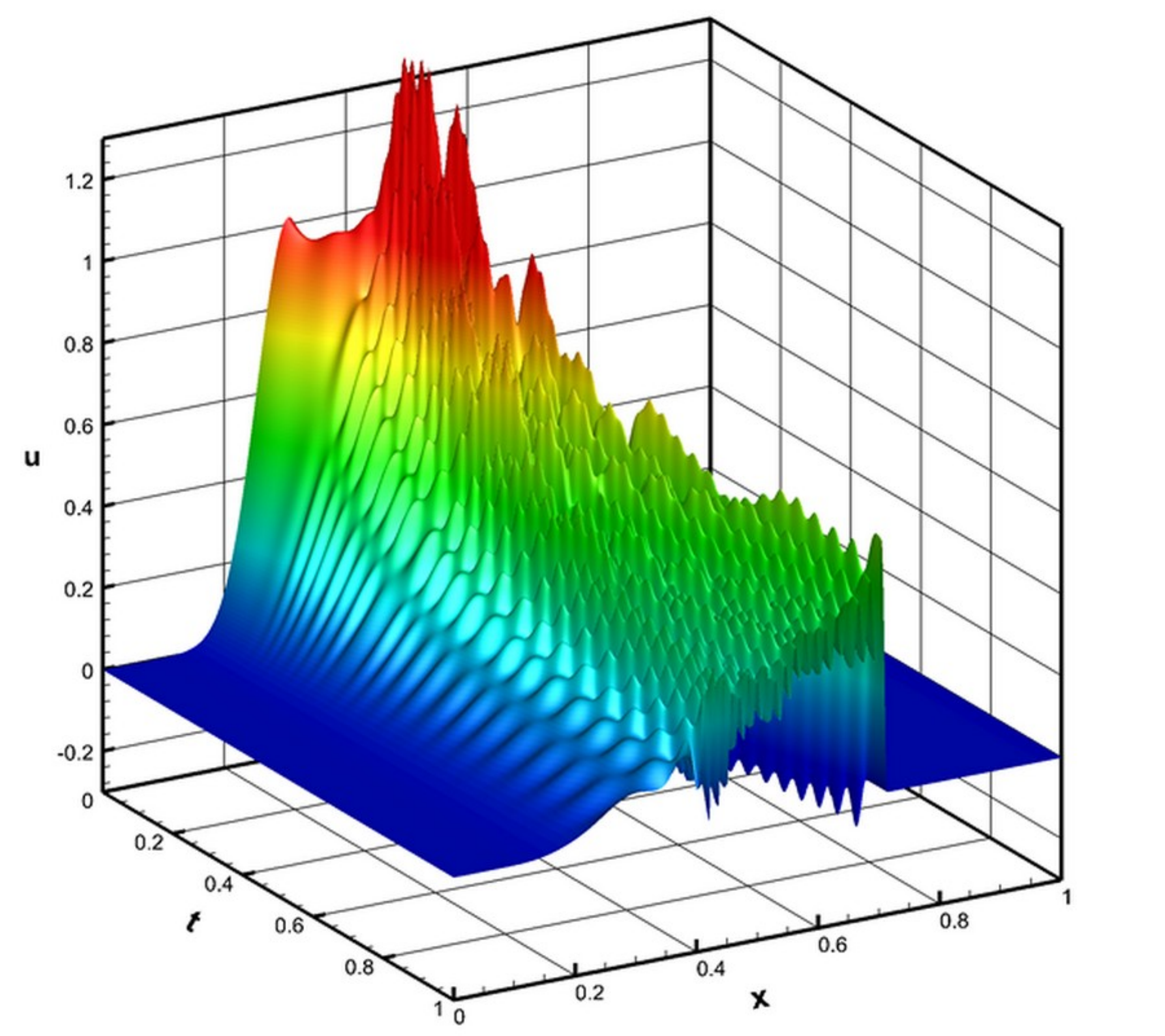}}
}\\
\mbox{
\subfigure[POD-ROM-G (80 modes)]{\includegraphics[width=0.35\textwidth]{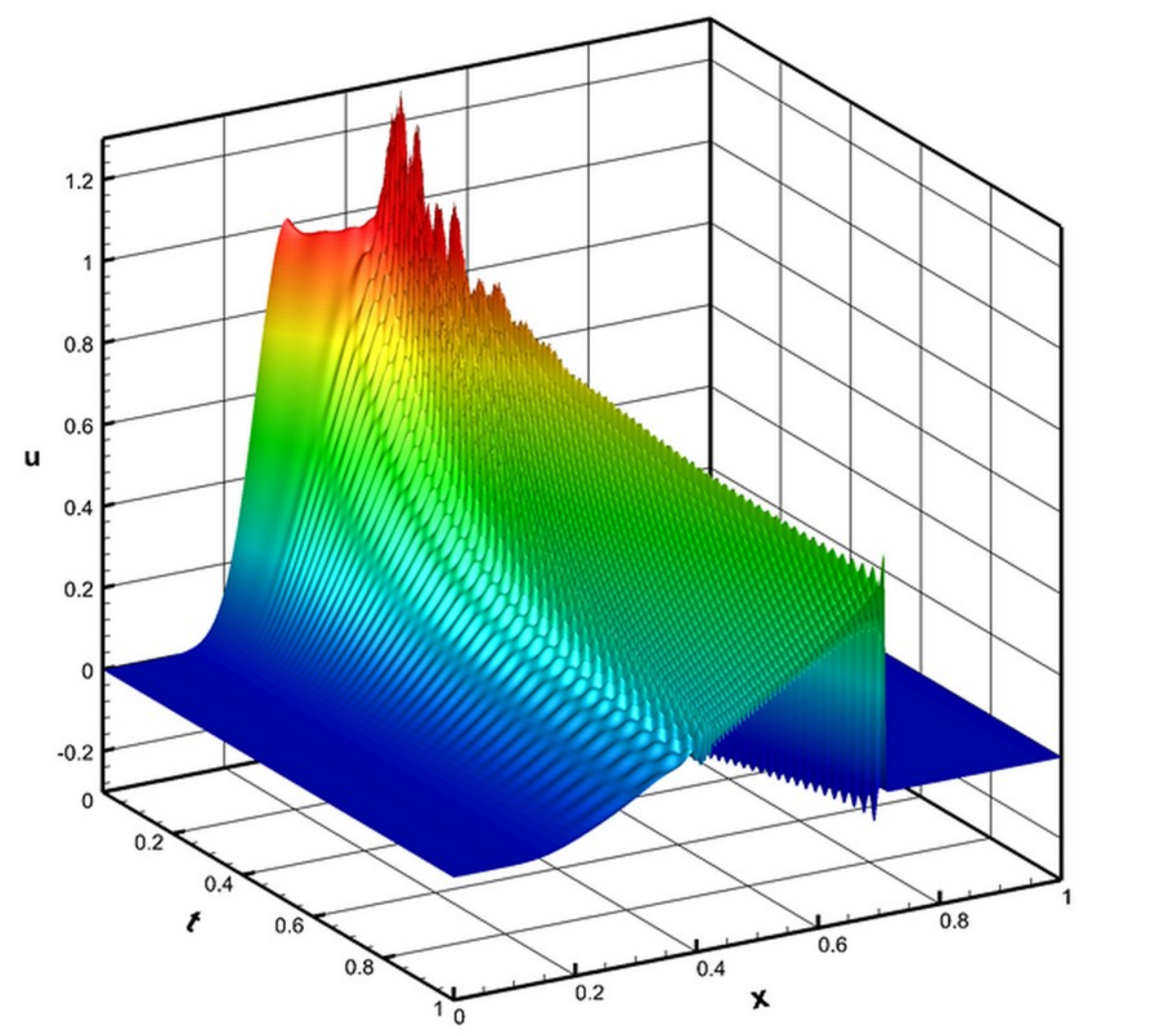}}
\subfigure[POD-ROM-G (160 modes)]{\includegraphics[width=0.35\textwidth]{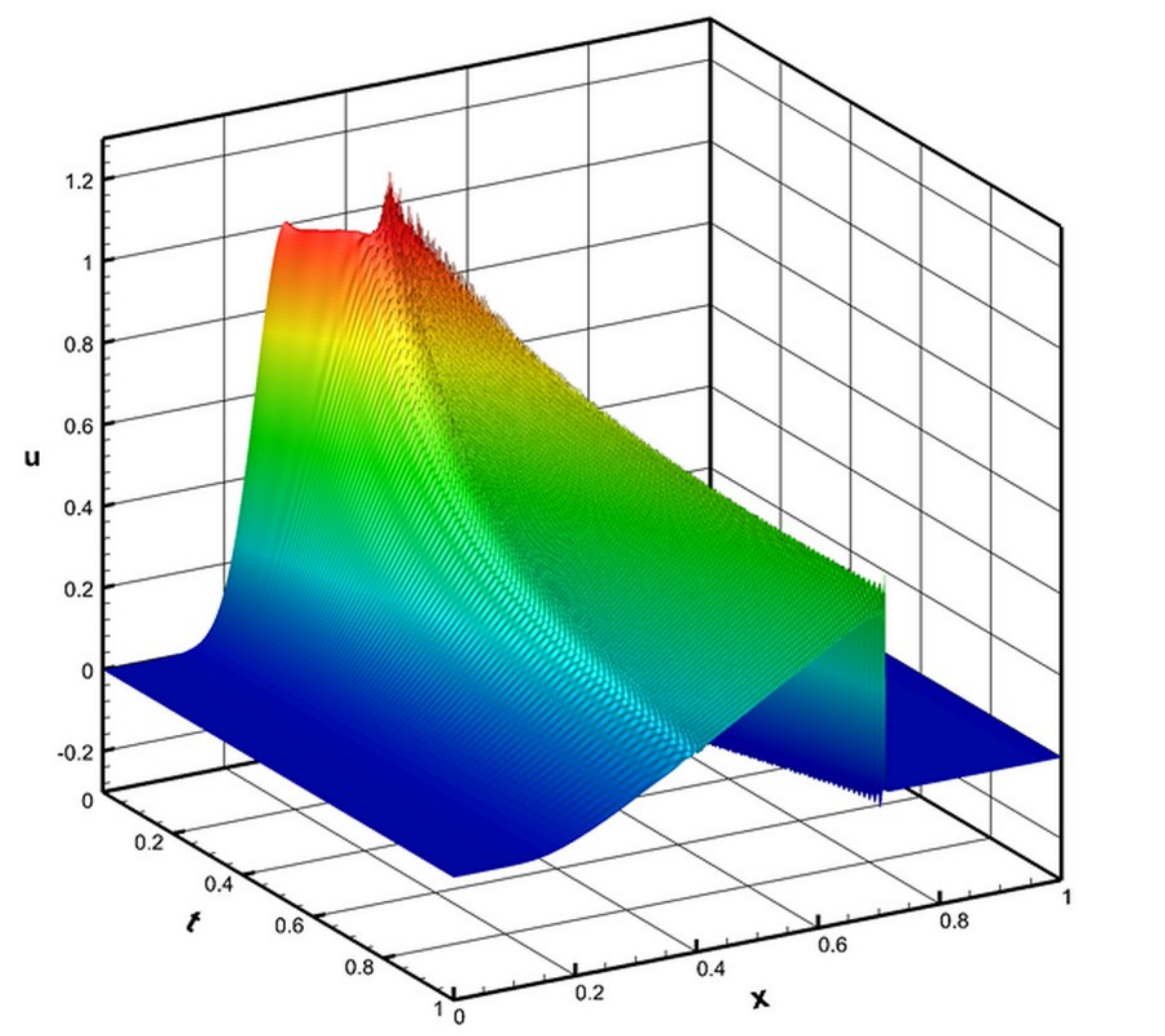}}
\subfigure[POD-ROM-G (320 modes)]{\includegraphics[width=0.35\textwidth]{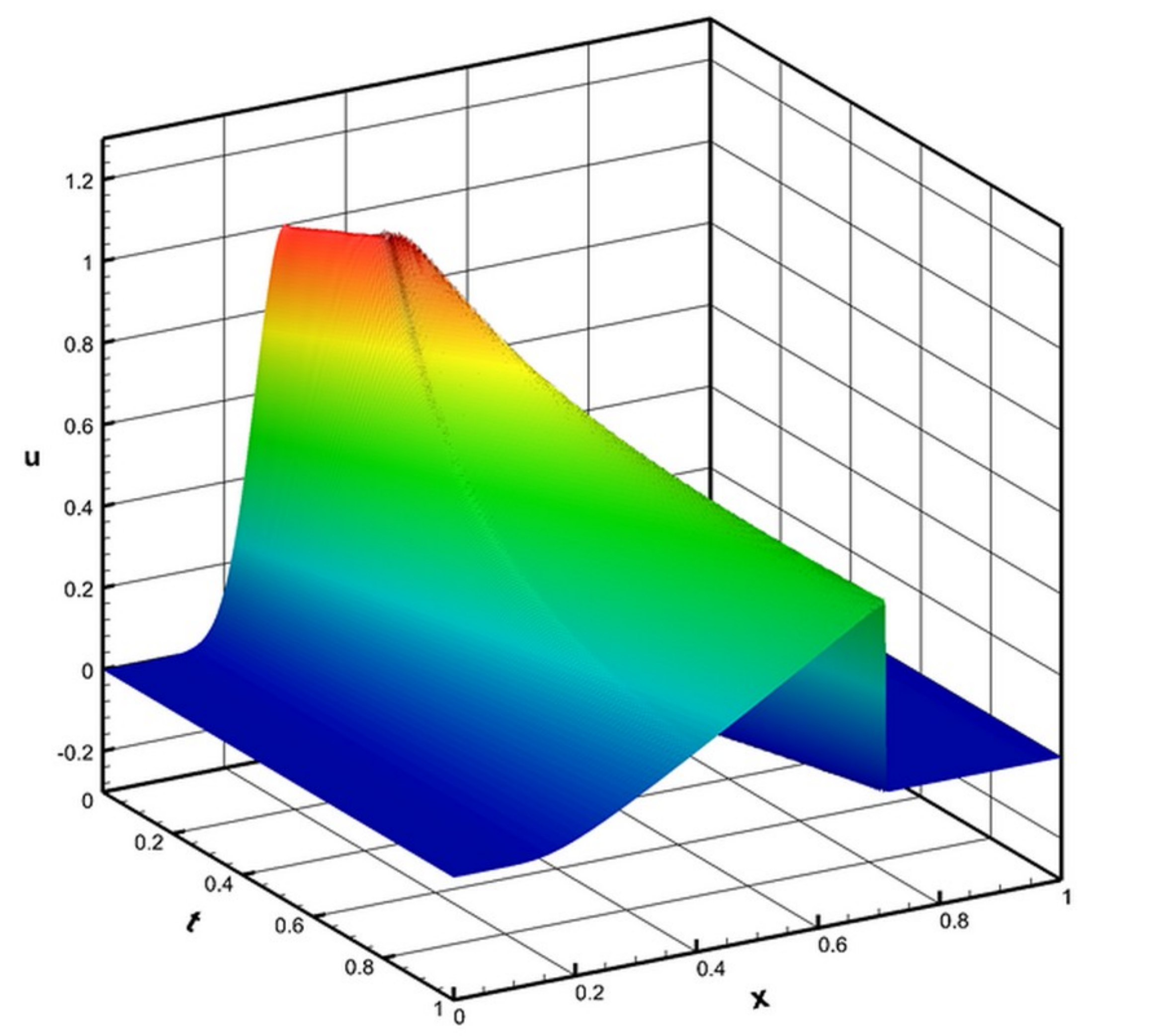}}
}
\caption{Experiment 2: POD-ROM-G with various numbers of POD modes.}
\label{fig:4}
\end{figure}

First, we present the results for the standard POD-ROM-G given by Eq.~\ref{e:14}.
The percentage of the captured energy in the POD-ROM-G is shown in Table~\ref{tab:1} for both experiments and for various numbers of POD modes.
Table~\ref{tab:1} also shows the CPU times for the POD-ROM-G.
The computational cost of the DNS computations (with a resolution of 8192 grid points) are 131.6 s and 95.3 s for Experiment 1 and Experiment 2, respectively.
As expected, the POD-ROM-G is not practical in terms of computational cost when the number of POD basis functions, $R$, is large.
Therefore, practical POD-ROMs are designed for $R\ll N$.
Fig.~\ref{fig:3} and Fig.~\ref{fig:4} show the results for the POD-ROM approximation for various values of $R$ for Experiment 1 and Experiment 2, respectively.
The reference DNS solutions are also included in both figures for comparison purposes.
When few POD modes are used (i.e., $R$ is small), the POD-ROM-G performs poorly, displaying significant numerical oscillations for both experiments.
These figures clearly show that the POD-ROM-G modeling error becomes smaller and smaller by increasing the number of POD modes, $R$.
As pointed out above, however, for practical purposes, an efficient POD-ROM should be designed by using a small number of POD modes.
Next, we present results for the POD-ROM closure models discussed in Section \ref{sec:closure}.

\begin{table}[!t]
\centering
\caption{
The computational efficiency and the percentage of captured energy for the POD-ROM-G for various number of POD modes.
The computational cost of the DNS are 131.6 s and 95.3 s for Experiment 1 and Experiment 2, respectively. \vspace*{0.1cm}
}
\label{tab:1}
\begin{tabular}{clrlr}
\hline\noalign{\smallskip}
\multirow{2}{*}{Number of modes}  &
\multicolumn{2}{l}{\underline{Experiment 1  \quad \quad \quad \quad \quad}} &
\multicolumn{2}{l}{\underline{Experiment 2  \quad \quad \quad \quad \quad}} \\ \noalign{\smallskip}
 (R) & $\frac{\sum_{i=1}^{R} \lambda_i}{\sum_{i=1}^{N}\lambda_i}\times 100$  & CPU time (s) & $\frac{\sum_{i=1}^{R}}{\sum_{i=1}^{N}}\times 100$ & CPU time (s) \\
\noalign{\smallskip}\hline\noalign{\smallskip}
  5     & 91.250726  & 0.1239     & 86.541659 & 0.1429 \\
  10    & 95.615358  & 0.7109     & 93.611926 & 0.6638 \\
  20    & 97.867613  & 4.5643     & 97.170311 & 4.6672 \\
  30    & 98.629576  & 15.0837    & 98.317899 & 15.1217 \\
  40    & 99.011706  & 40.2829    & 98.871930 & 35.0466 \\
  80    & 99.581931  & 293.9203   & 99.641204 & 301.1842 \\
  160   & 99.854665  & 2328.0041  & 99.933295 & 2299.7983 \\
  320   & 99.967961  & 18217.3255 & 99.996588 & 19875.0094 \\
\noalign{\smallskip}\hline
\end{tabular}
\end{table}

We emphasize that the closure models presented in Section \ref{sec:closure} have a free modeling parameter $\nu_e$.
Exceptions are the POD-ROM-G, which has no closure term at all, and the POD-ROM-C.
First, for all the closure models, we perform a sensitivity study on the modeling parameter $\nu_e$. The discrete root mean squared (RMS) errors (with respect to the DNS results) are computed at the final time $t=1$.

\begin{figure}[!t]
\centering
\mbox{
\subfigure[Experiment 1]{\includegraphics[width=0.5\textwidth]{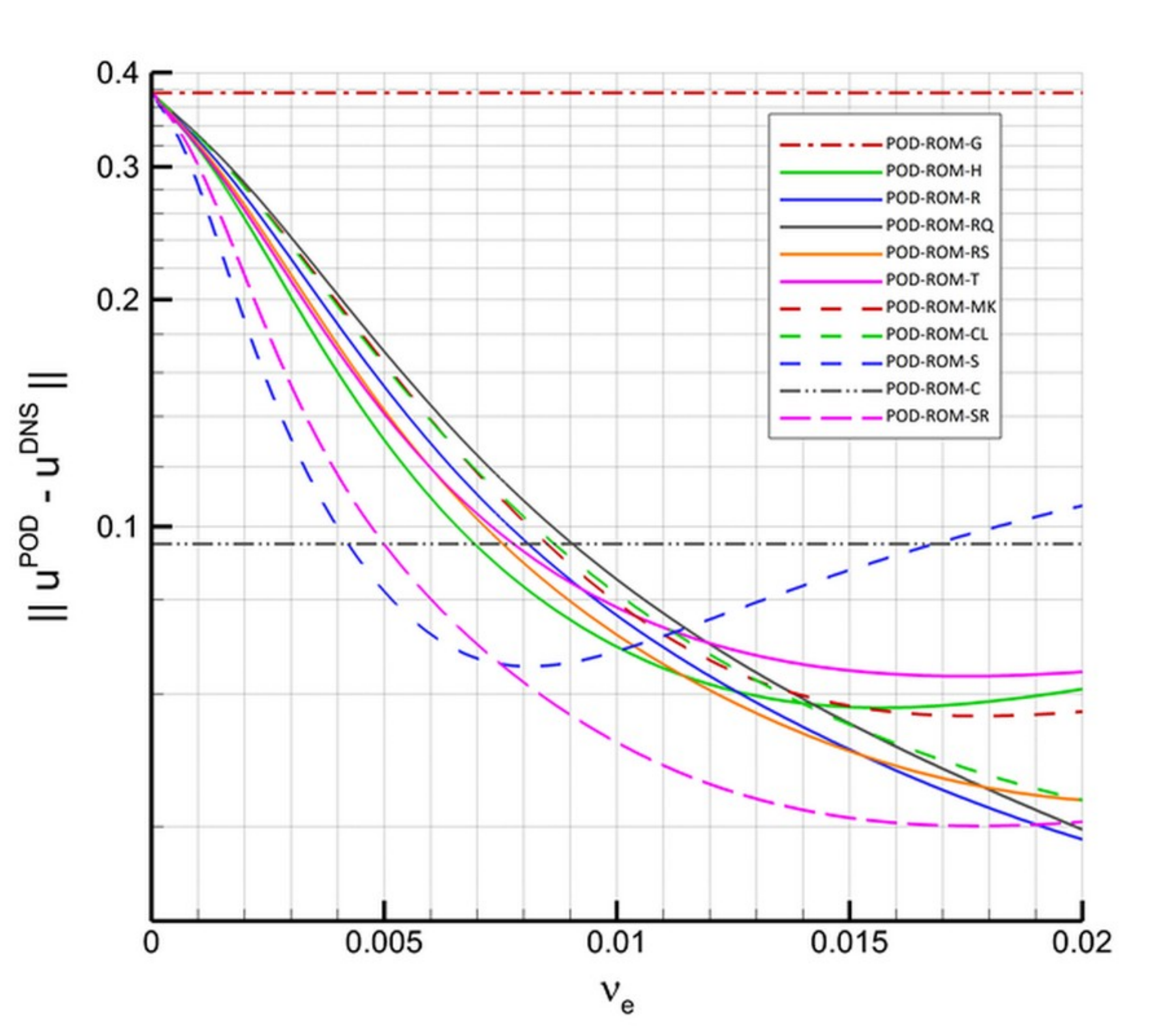}}
\subfigure[Experiment 2]{\includegraphics[width=0.5\textwidth]{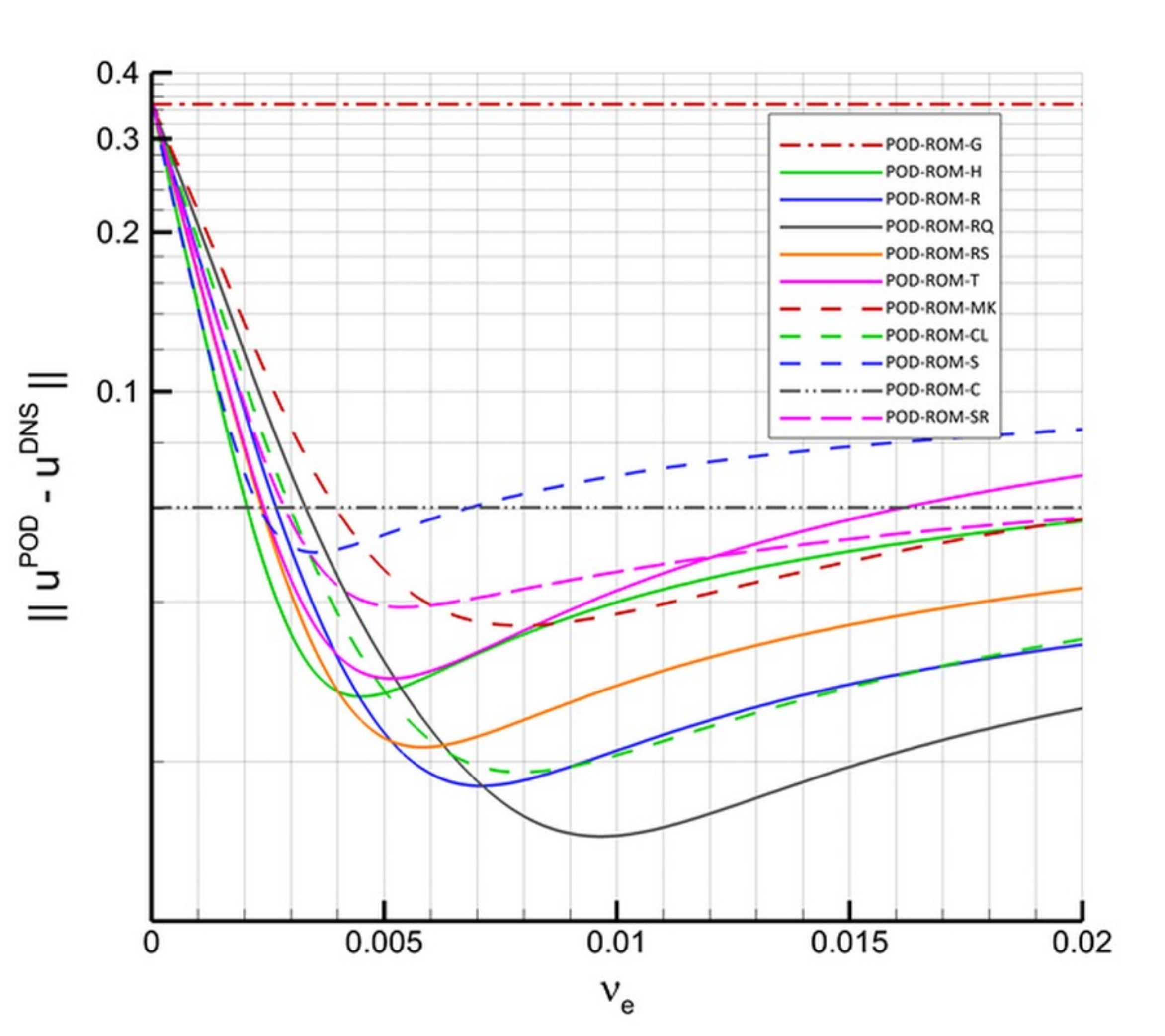}}
}
\caption{Sensitivity analysis of the free parameter in POD-ROM closure models using $R=5$ modes. Results obtained using the POD-ROM-G and POD-ROM-C models are also included for comparison purposes.}
\label{fig:5}
\end{figure}

\begin{figure}[!t]
\centering
\mbox{
\subfigure[Experiment 1]{\includegraphics[width=0.5\textwidth]{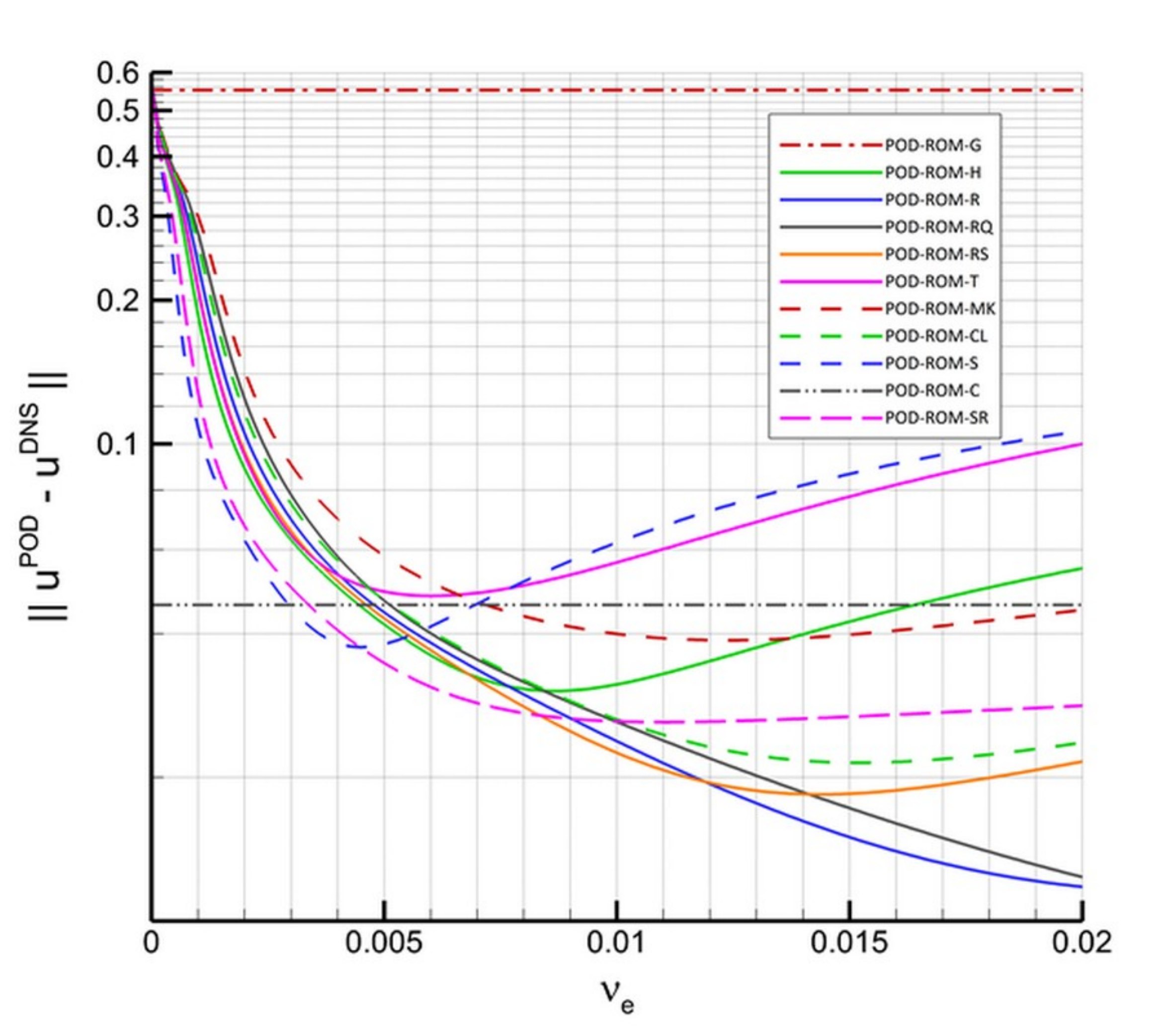}}
\subfigure[Experiment 2]{\includegraphics[width=0.5\textwidth]{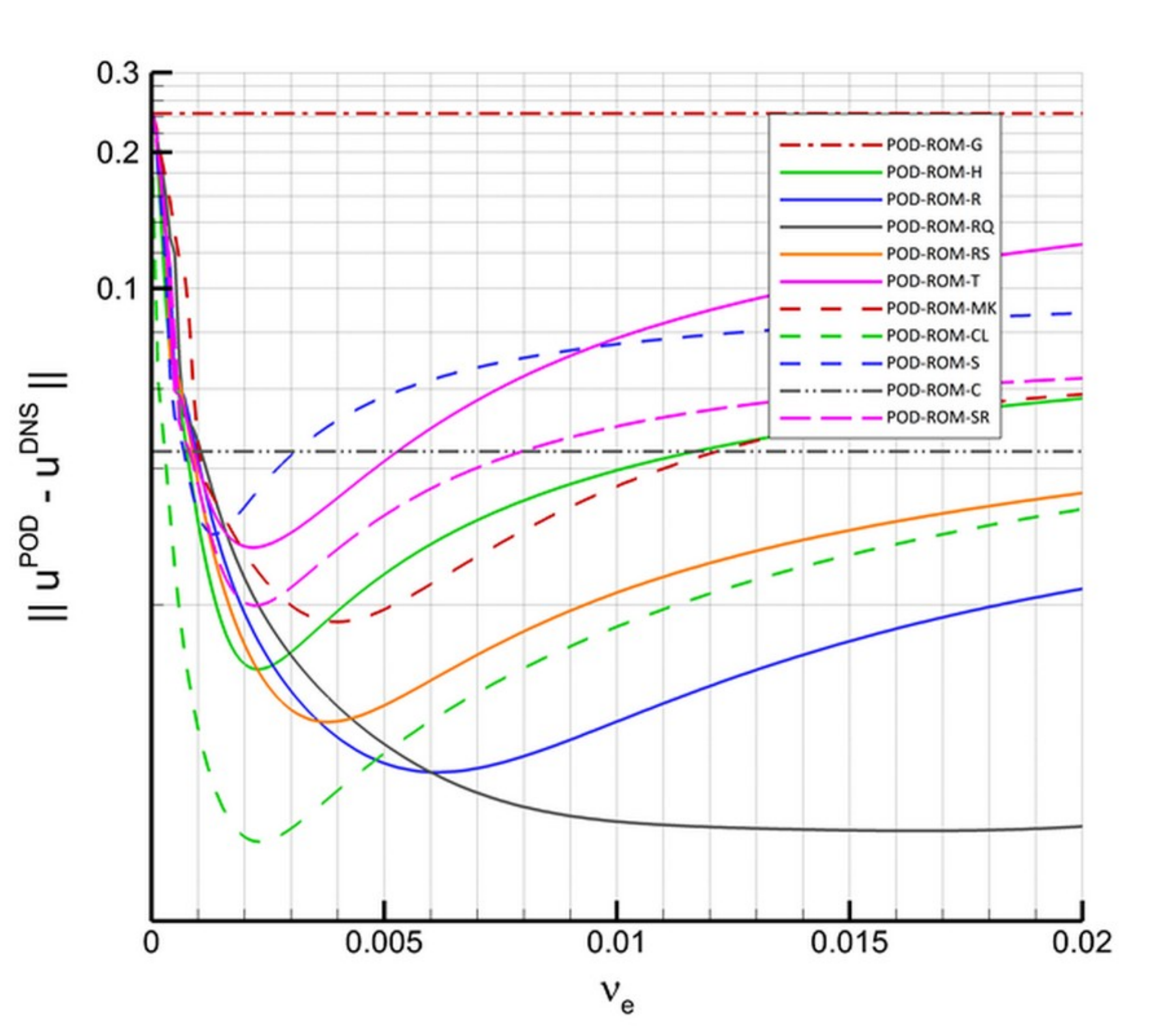}}
}
\caption{Sensitivity analysis of the free parameter in POD-ROM closure models using $R=10$ modes. Results obtained using the POD-ROM-G and POD-ROM-C models are also included for comparison purposes.}
\label{fig:6}
\end{figure}

\begin{figure}[!t]
\centering
\mbox{
\subfigure[Experiment 1]{\includegraphics[width=0.5\textwidth]{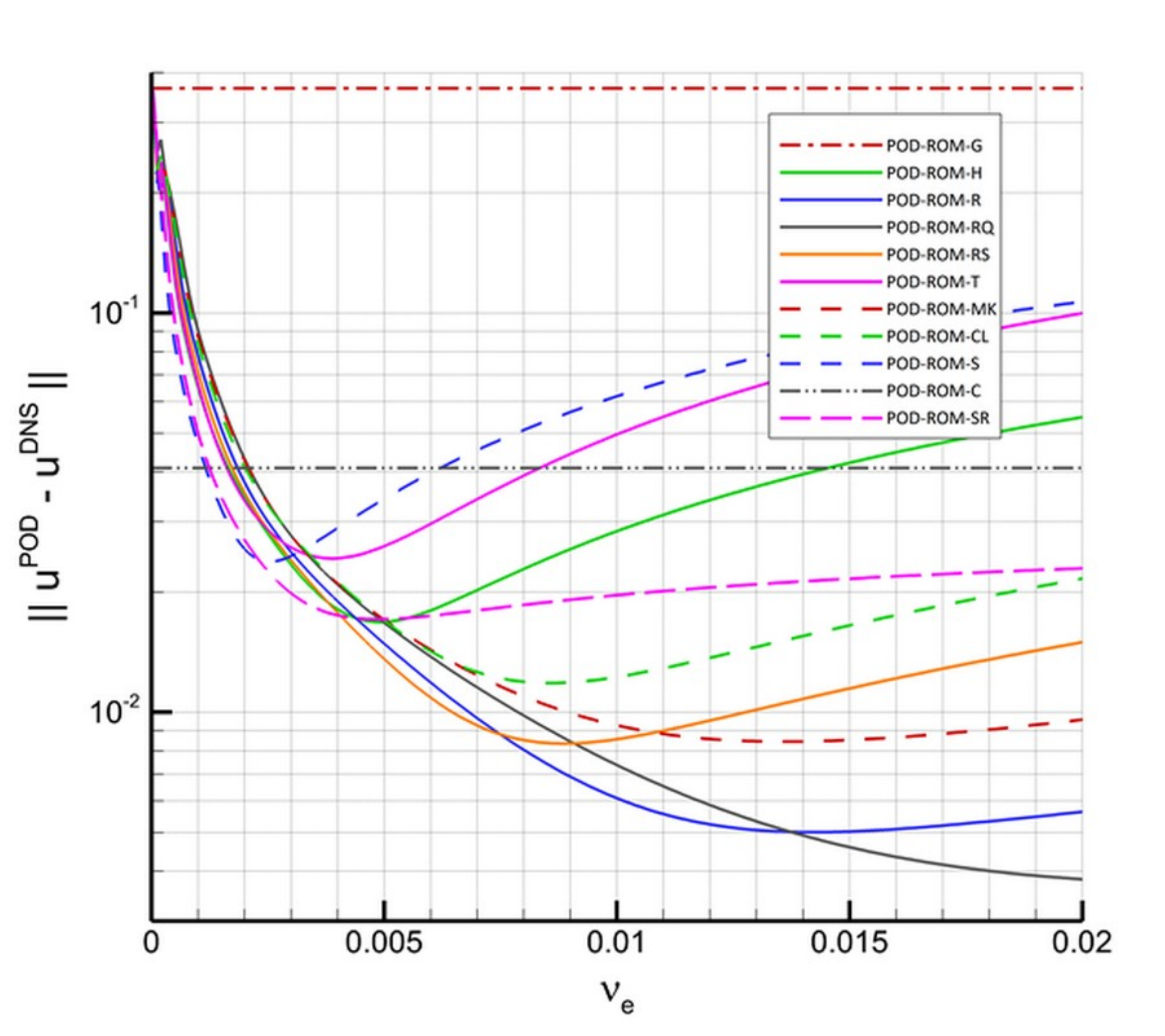}}
\subfigure[Experiment 2]{\includegraphics[width=0.5\textwidth]{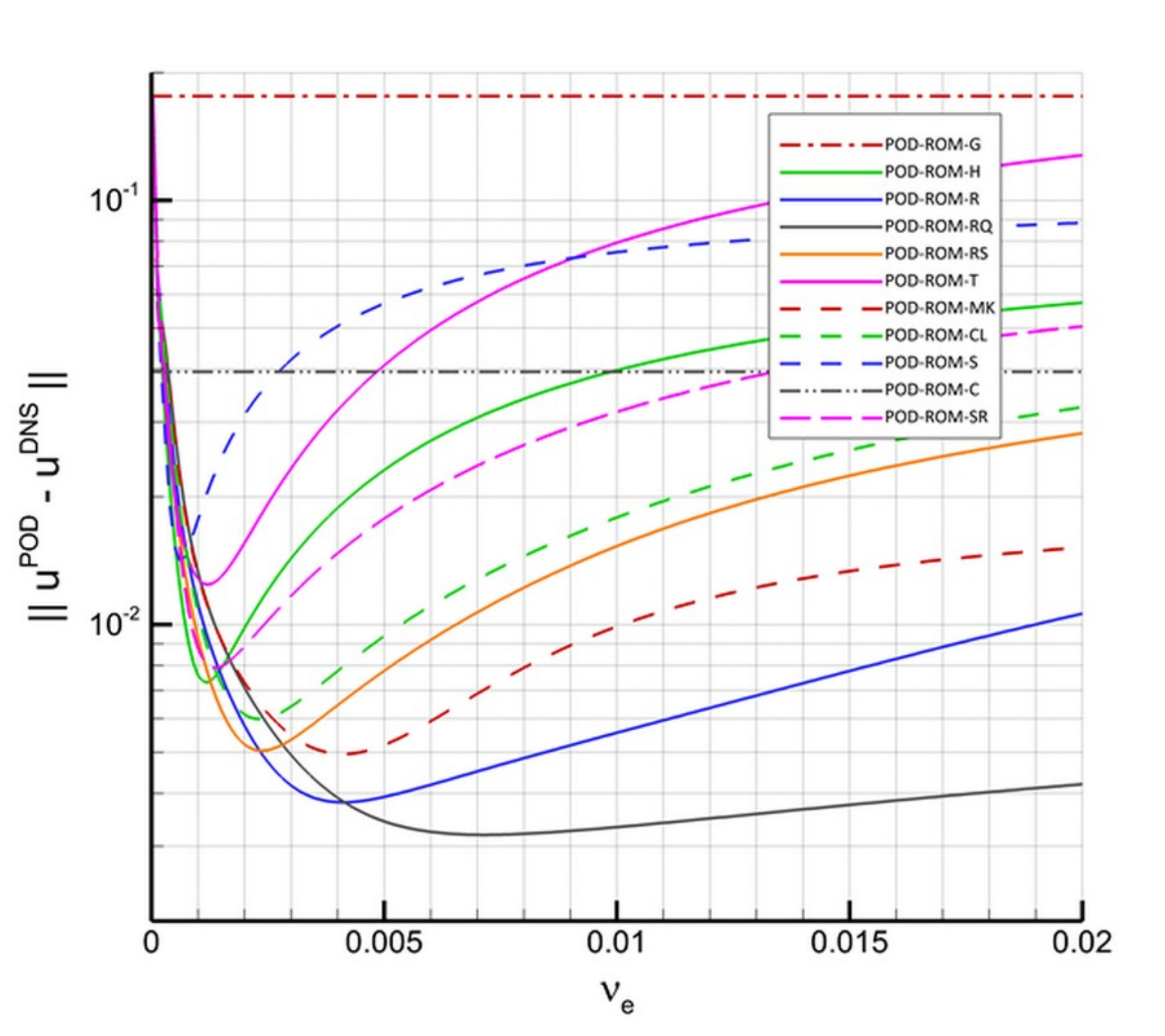}}
}
\caption{Sensitivity analysis of the free parameter in POD-ROM closure models using $R=20$ modes. Results obtained using the POD-ROM-G and POD-ROM-C models are also included for comparison purposes.}
\label{fig:7}
\end{figure}

Fig.~\ref{fig:5} shows the RMS errors with respect to the modeling parameter $\nu_e$ for all the closure models with $R=5$ for both experiments.
The two straight lines in the Fig.~\ref{fig:5} show the RMS errors of the POD-ROM-G and the POD-ROM-C, which have no dependence on $\nu_e$.
As expected, all the closure models perform better that the standard POD-ROM-G.
For appropriate values of $\nu_e$, each POD-ROM performs better than the POD-ROM-C.
Fig.~\ref{fig:6} and Fig.~\ref{fig:7} display the same type of data for $R=10$ and $R=20$, respectively.
It can be concluded from these figures that POD-ROM-R and POD-ROM-RQ provide the most accurate results among the closure models utilized in this report.
When the optimal $\nu_e$ value is considered, the POD-ROM-R is slightly more accurate than the POD-ROM-RQ in Experiment 1, and significantly less accurate in Experiment 2.

\begin{figure}
\centering
\mbox{
\subfigure[POD-ROM-H]{\includegraphics[width=0.35\textwidth]{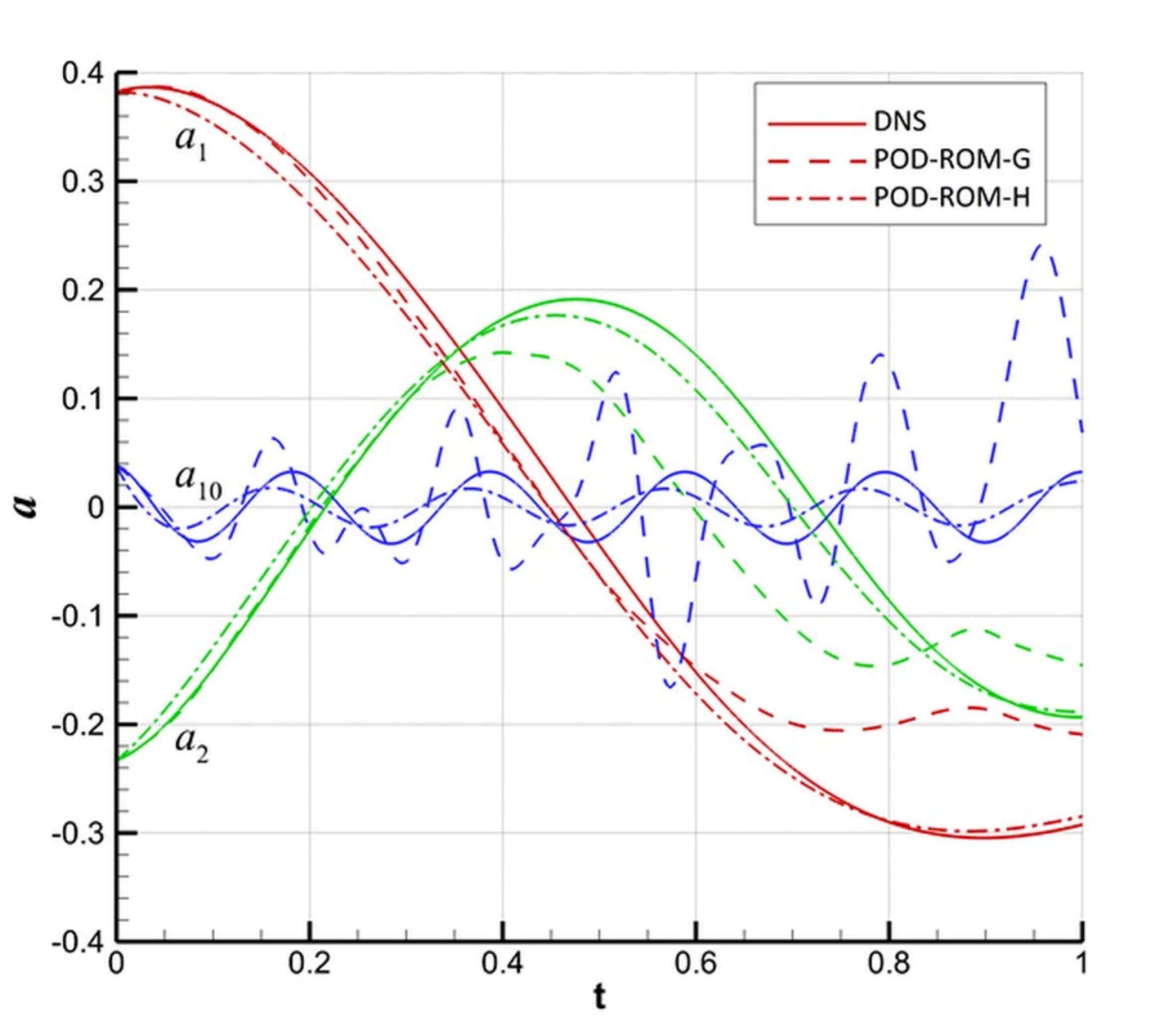}}
\subfigure[POD-ROM-R]{\includegraphics[width=0.35\textwidth]{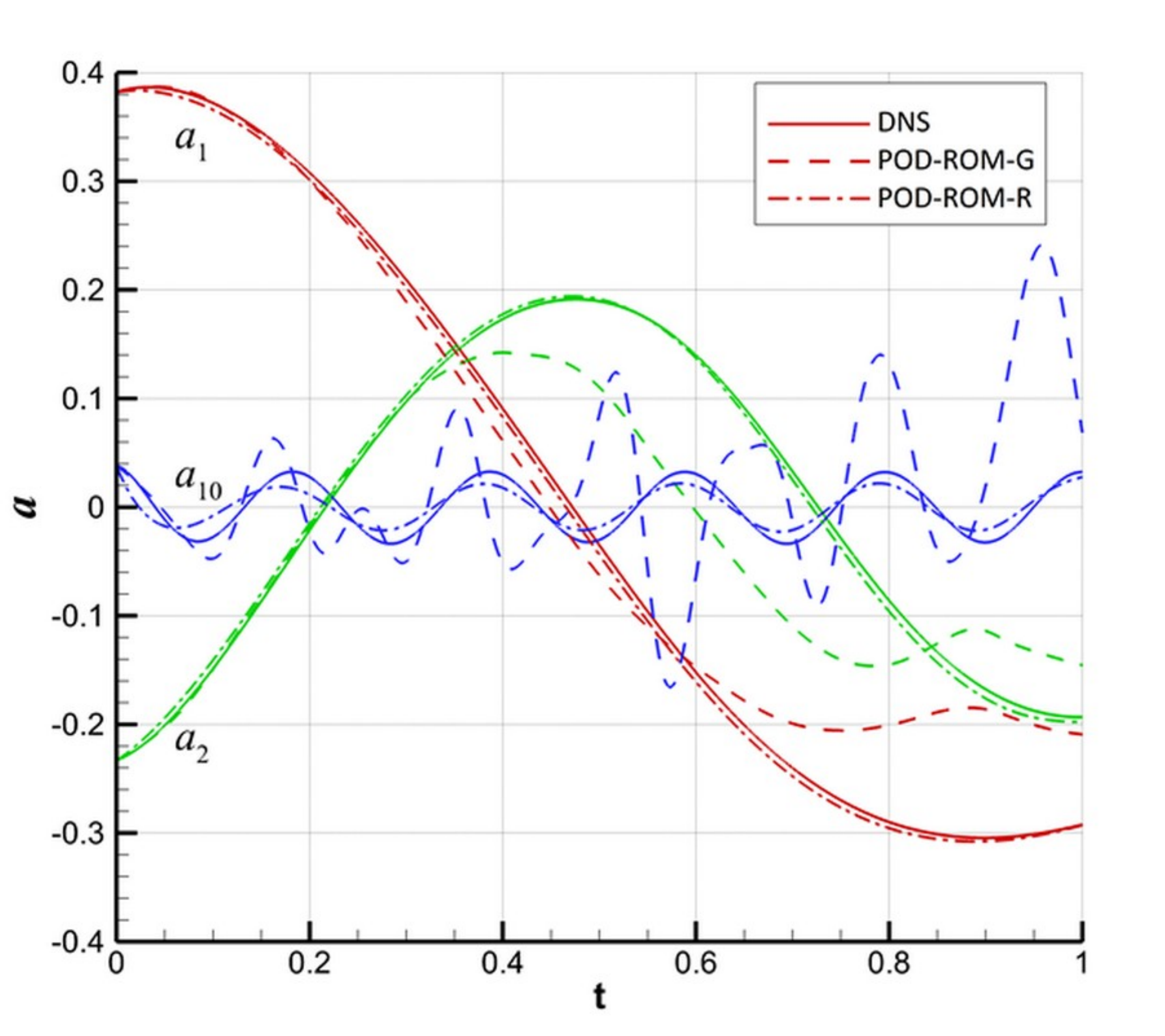}}
\subfigure[POD-ROM-RQ]{\includegraphics[width=0.35\textwidth]{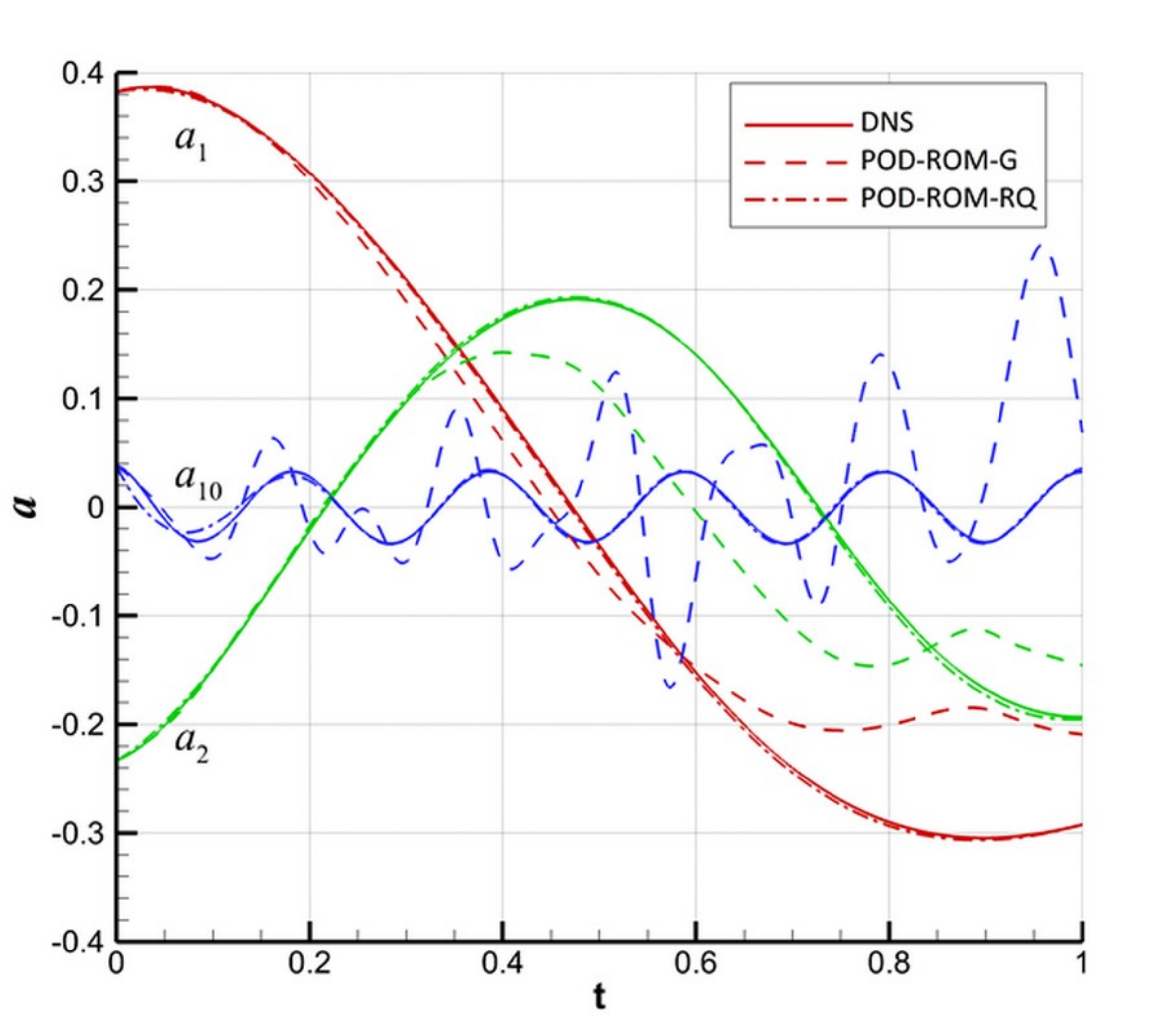}}
}\\
\mbox{
\subfigure[POD-ROM-SR ]{\includegraphics[width=0.35\textwidth]{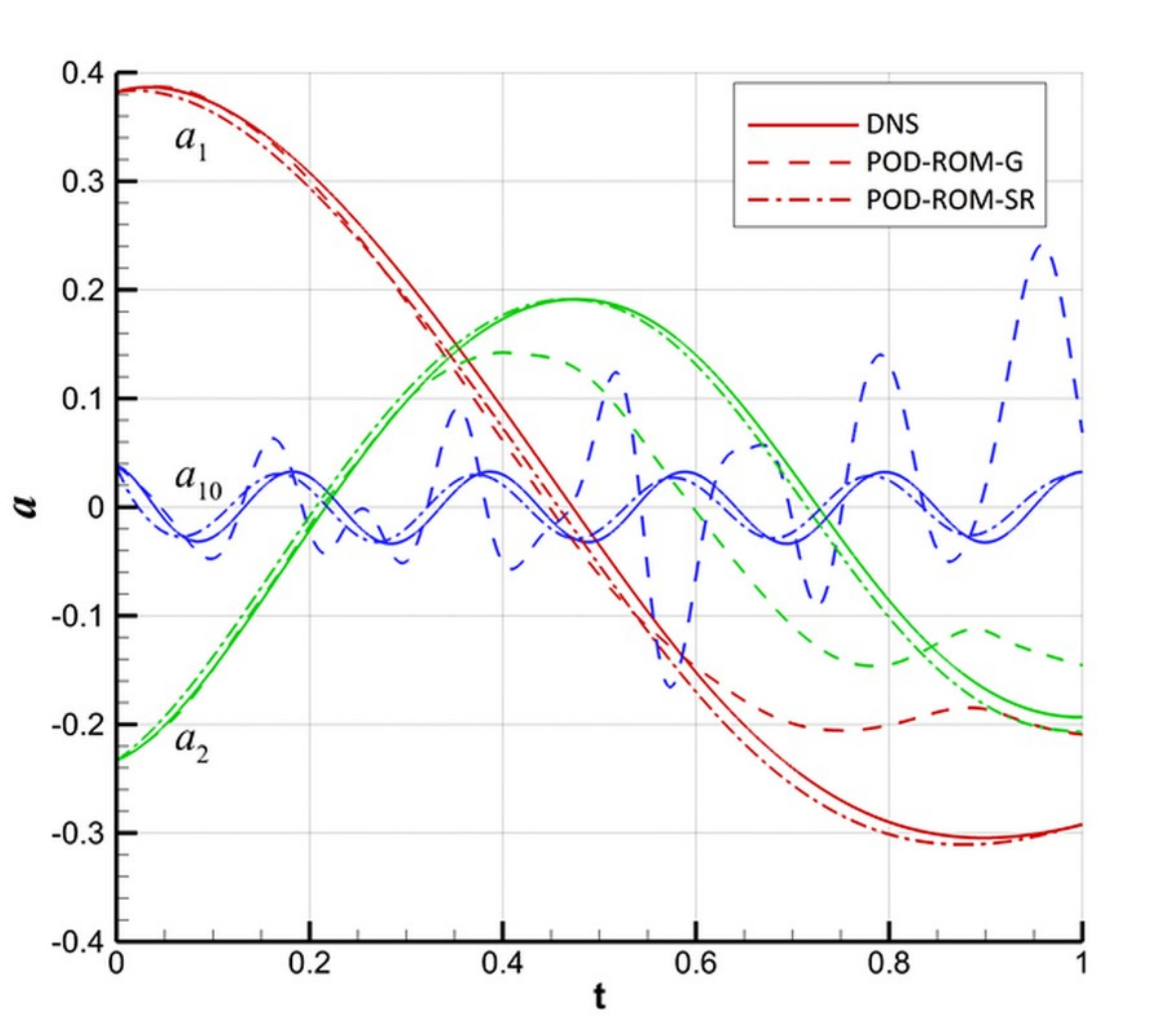}}
\subfigure[POD-ROM-T]{\includegraphics[width=0.35\textwidth]{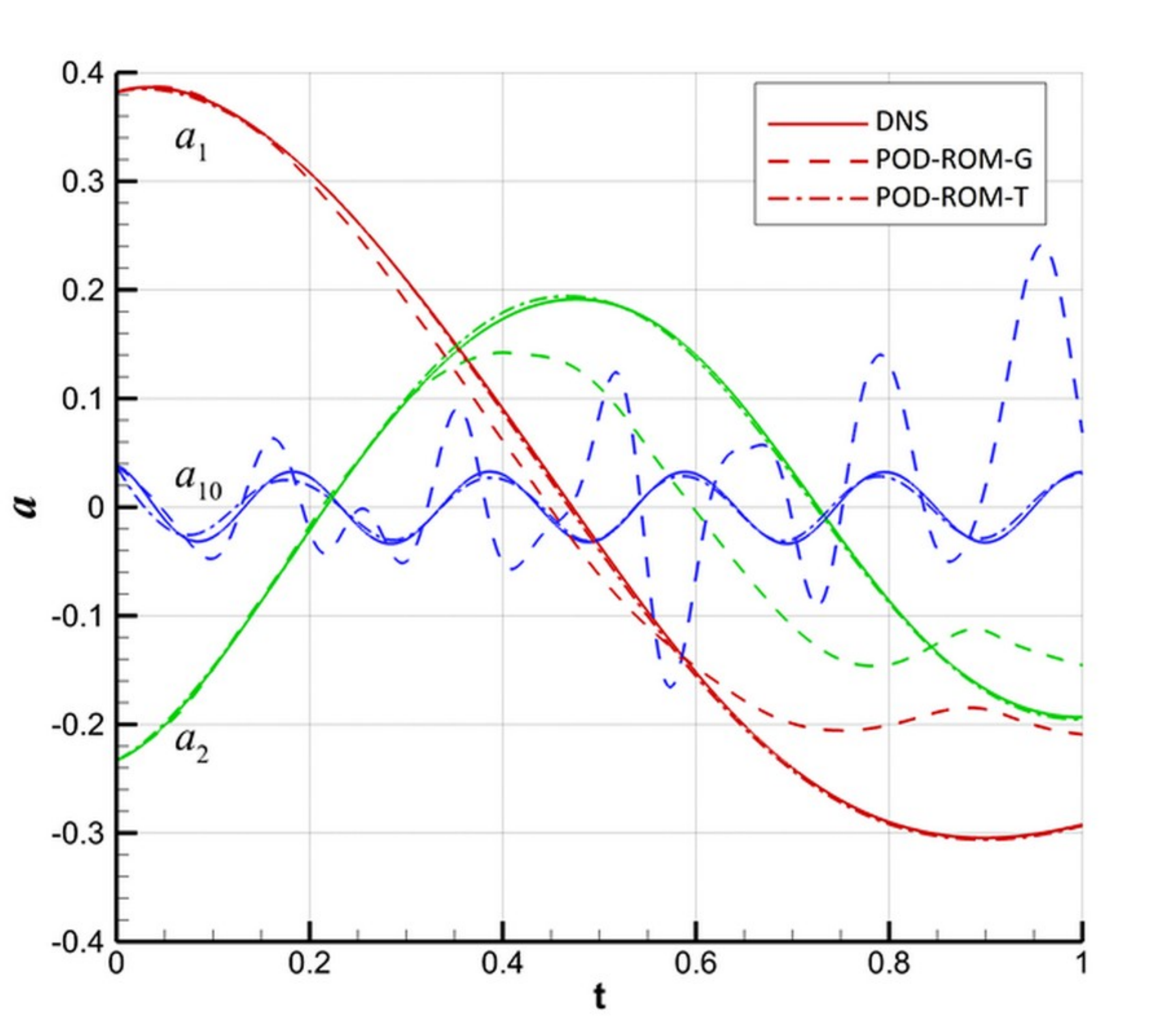}}
\subfigure[POD-ROM-MK]{\includegraphics[width=0.35\textwidth]{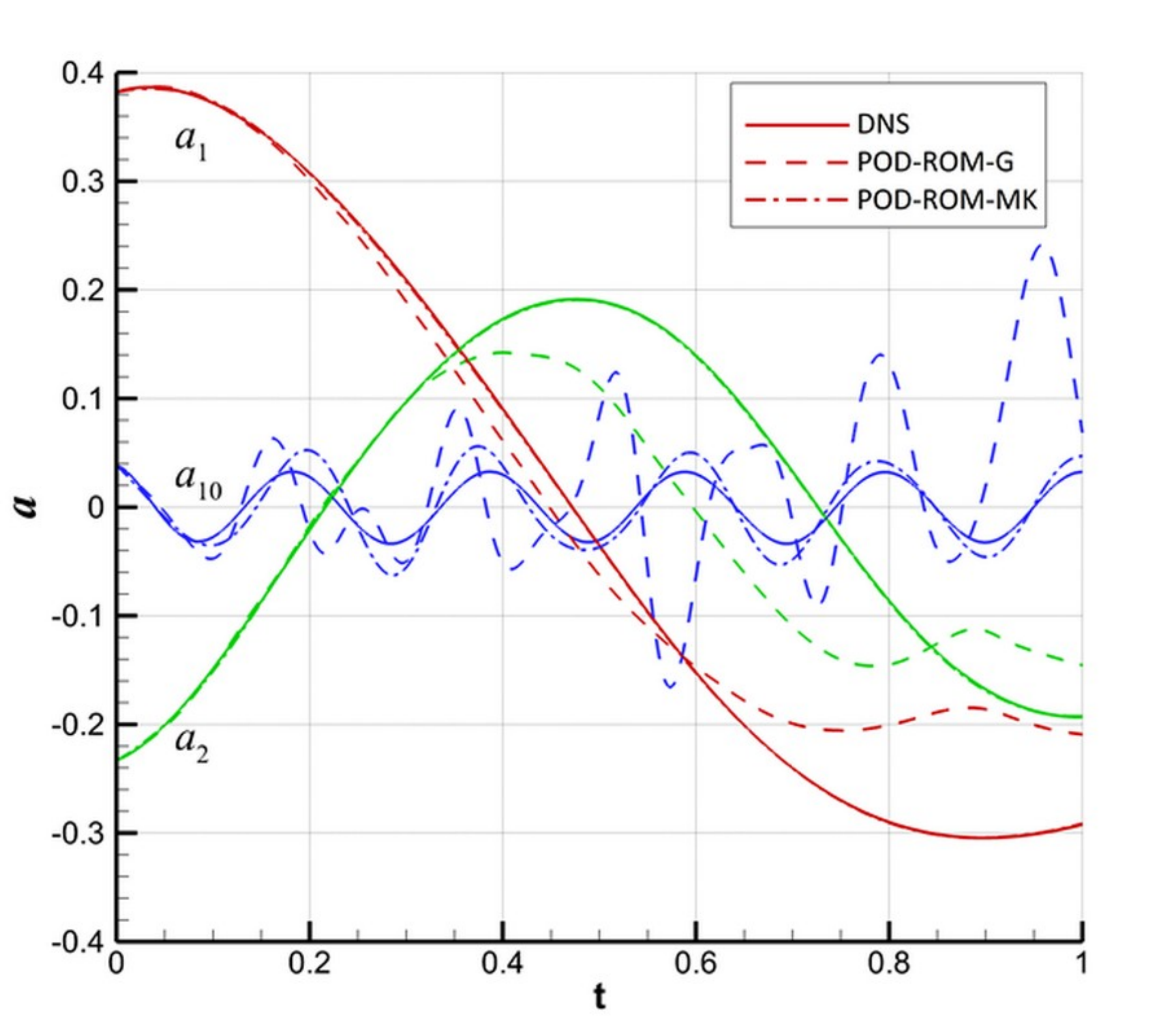}}
}\\
\mbox{
\subfigure[POD-ROM-CL]{\includegraphics[width=0.35\textwidth]{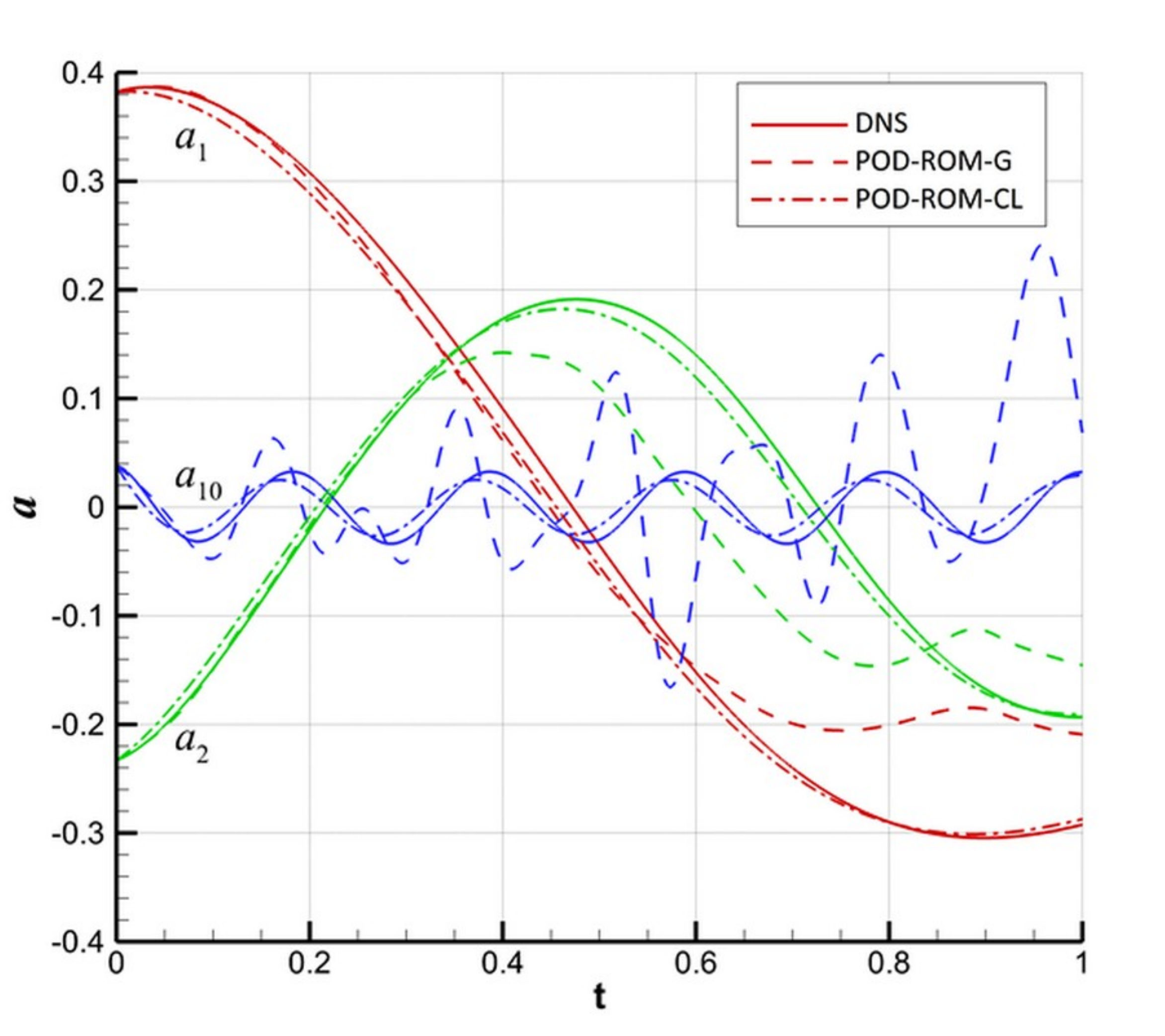}}
\subfigure[POD-ROM-S]{\includegraphics[width=0.35\textwidth]{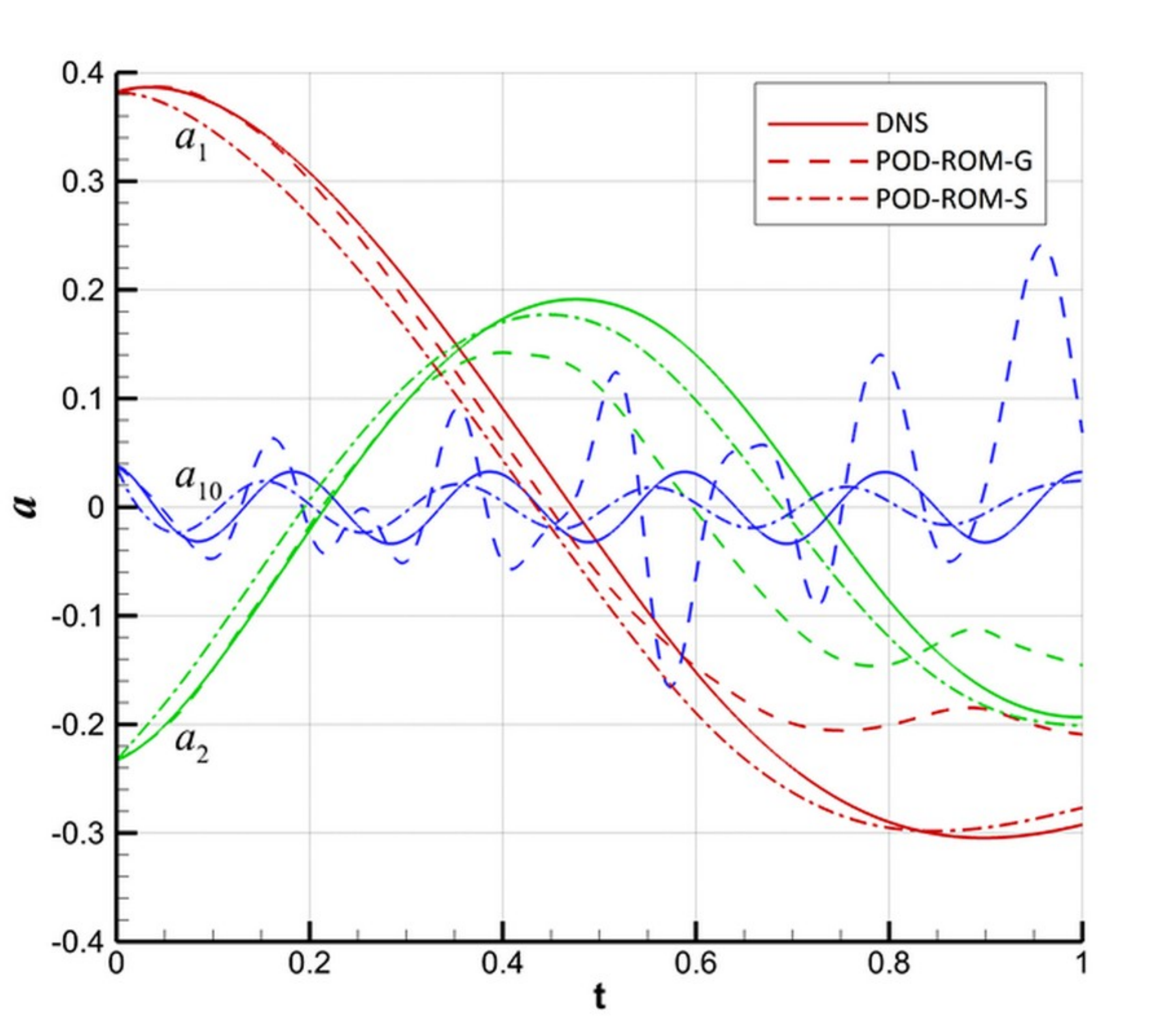}}
\subfigure[POD-ROM-C]{\includegraphics[width=0.35\textwidth]{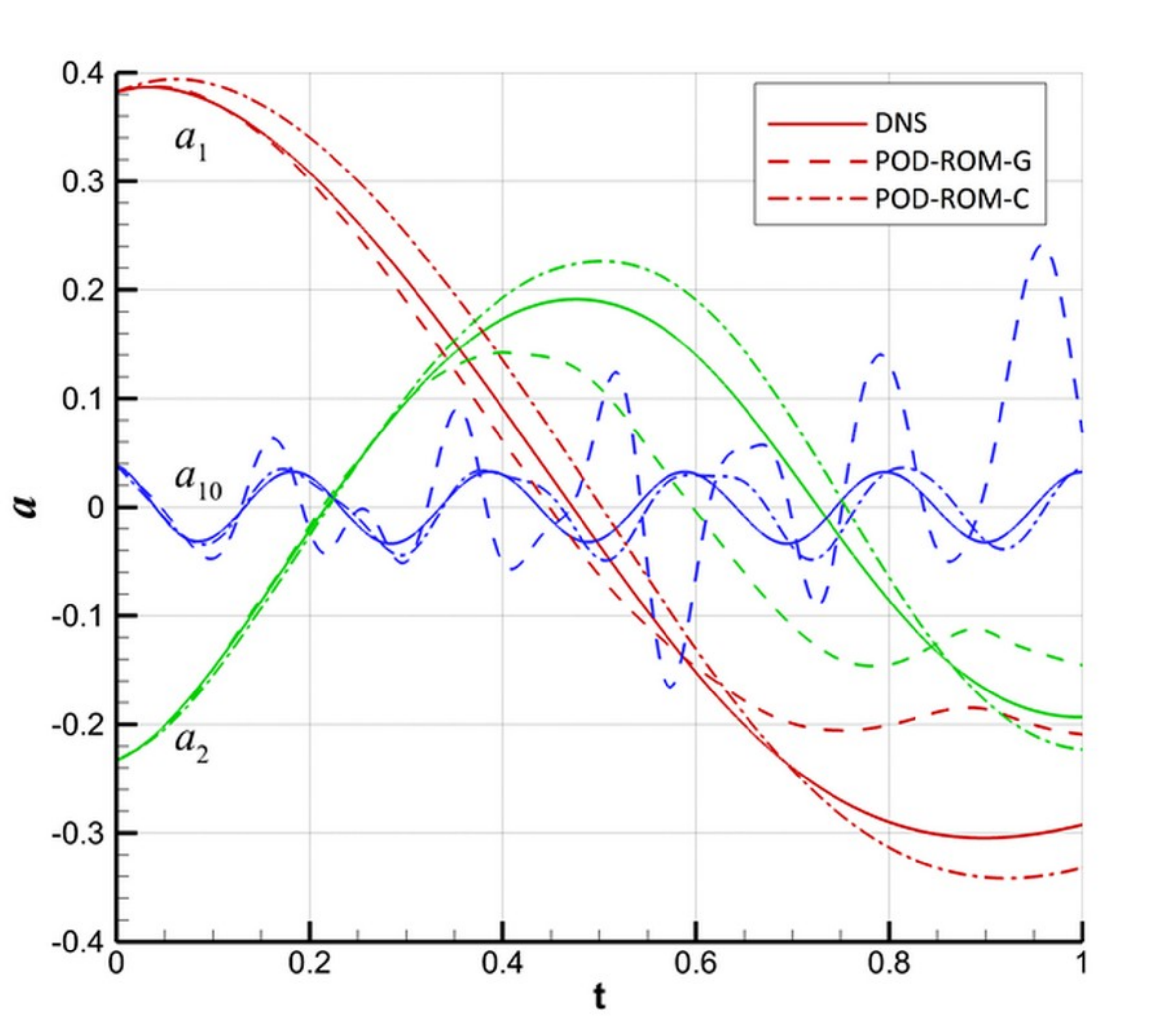}}
}
\caption{
Experiment 1:
Time evolutions of $a_1, a_2$ and $a_{10}$ coefficients of the POD-ROMs.
DNS and POD-ROM-G results are also included for comparison purposes.
}
\label{fig:8}
\end{figure}

\begin{figure}
\centering
\mbox{
\subfigure[POD-ROM-H]{\includegraphics[width=0.35\textwidth]{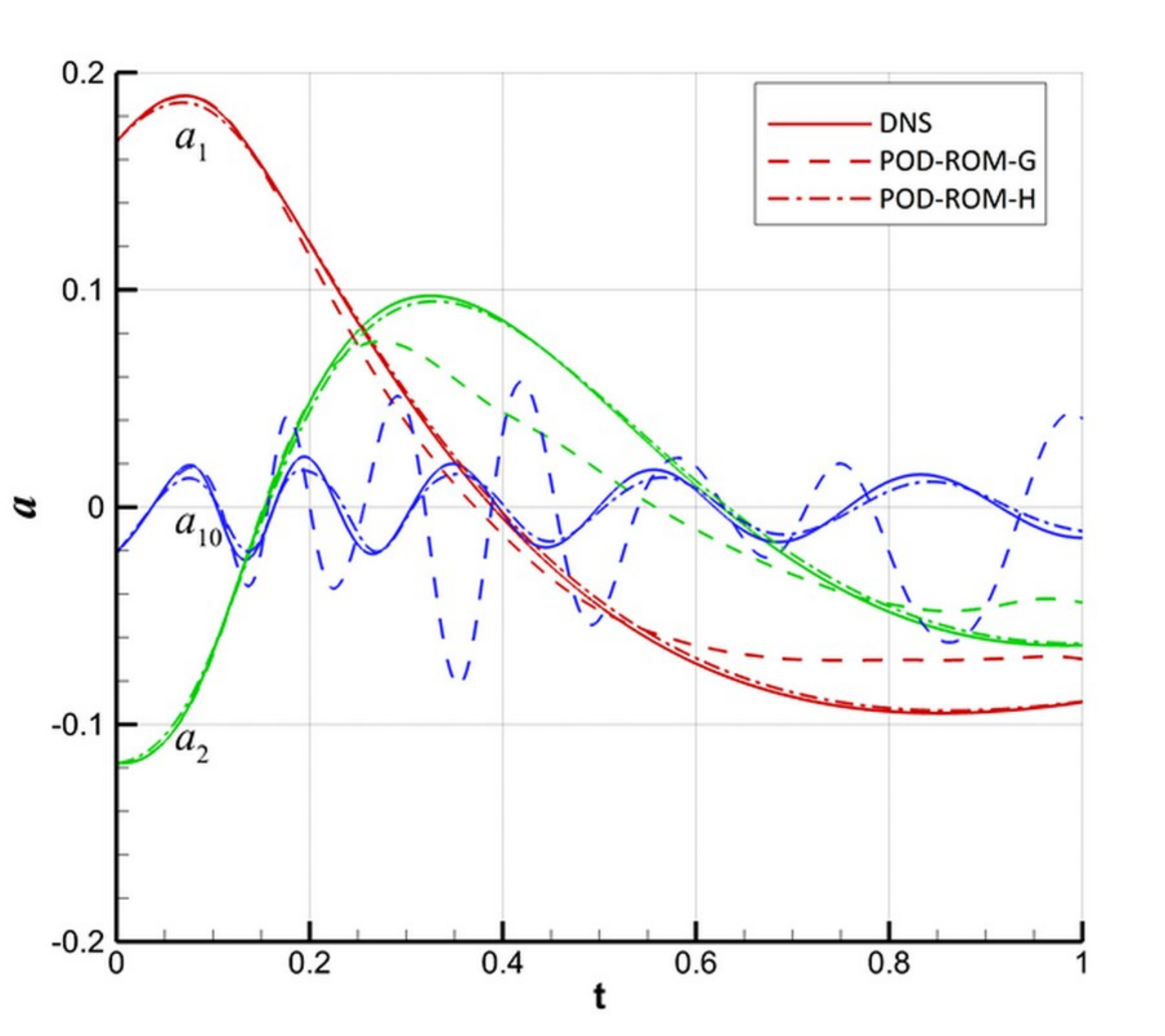}}
\subfigure[POD-ROM-R]{\includegraphics[width=0.35\textwidth]{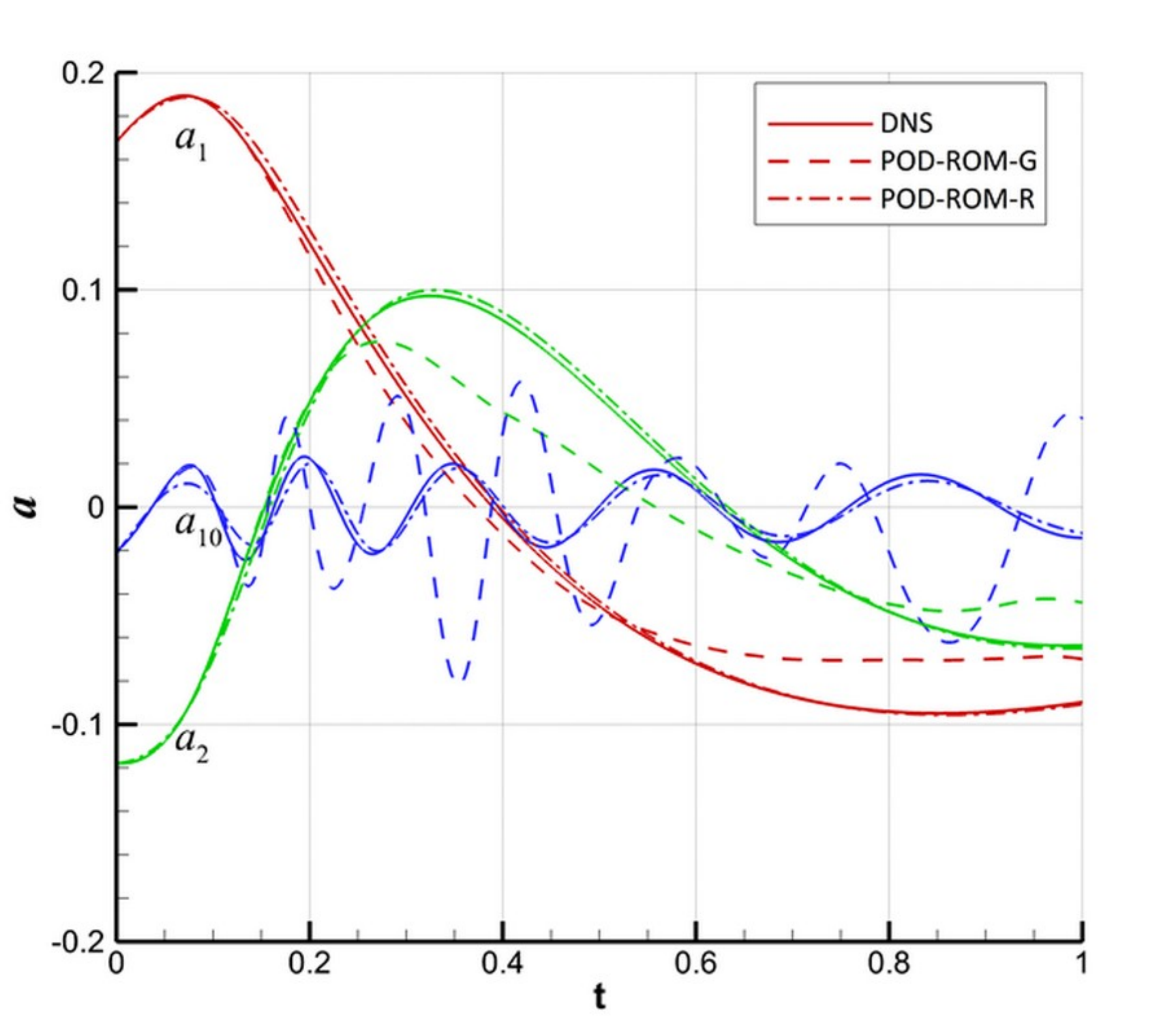}}
\subfigure[POD-ROM-RQ]{\includegraphics[width=0.35\textwidth]{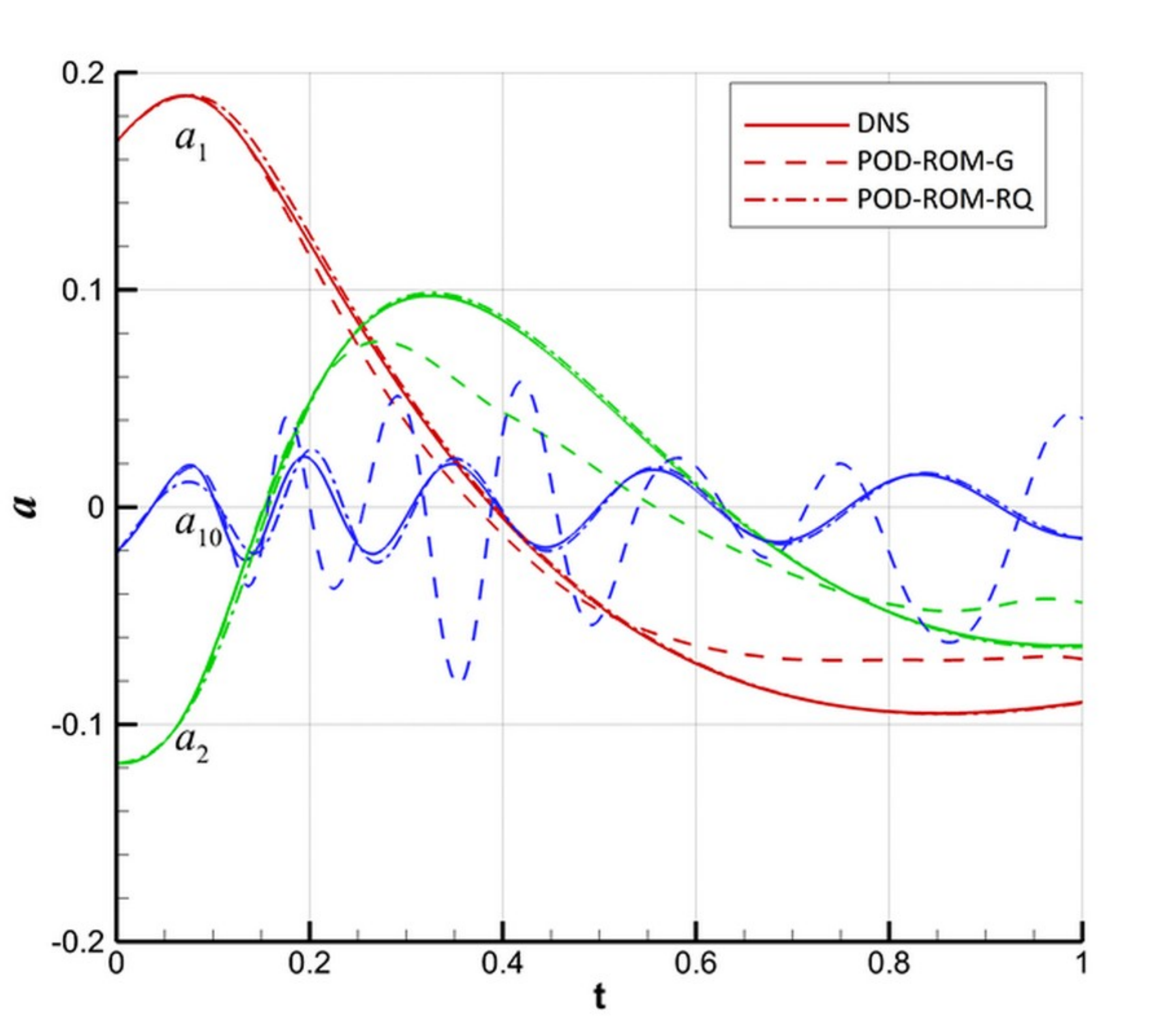}}
}\\
\mbox{
\subfigure[POD-ROM-SR ]{\includegraphics[width=0.35\textwidth]{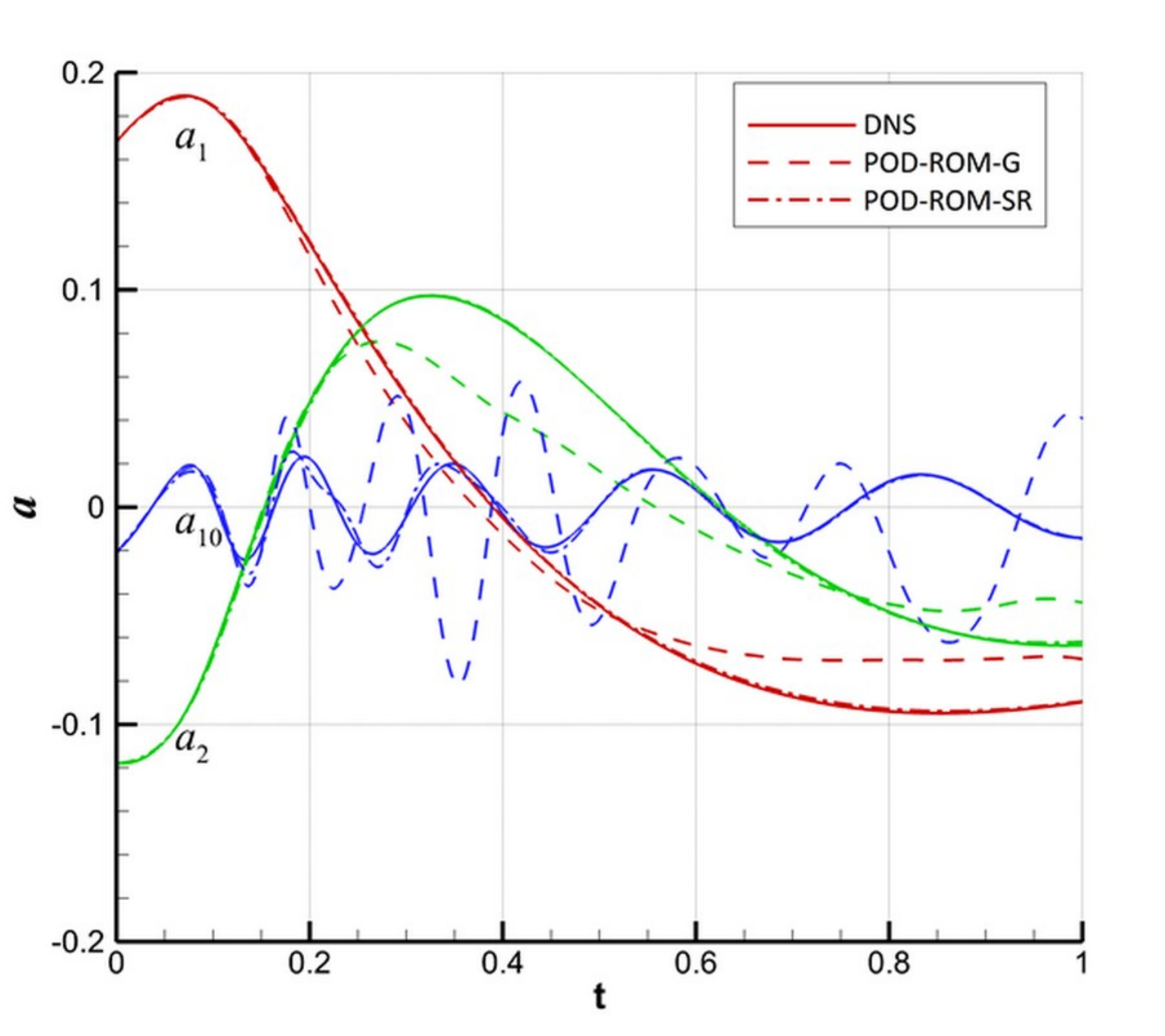}}
\subfigure[POD-ROM-T]{\includegraphics[width=0.35\textwidth]{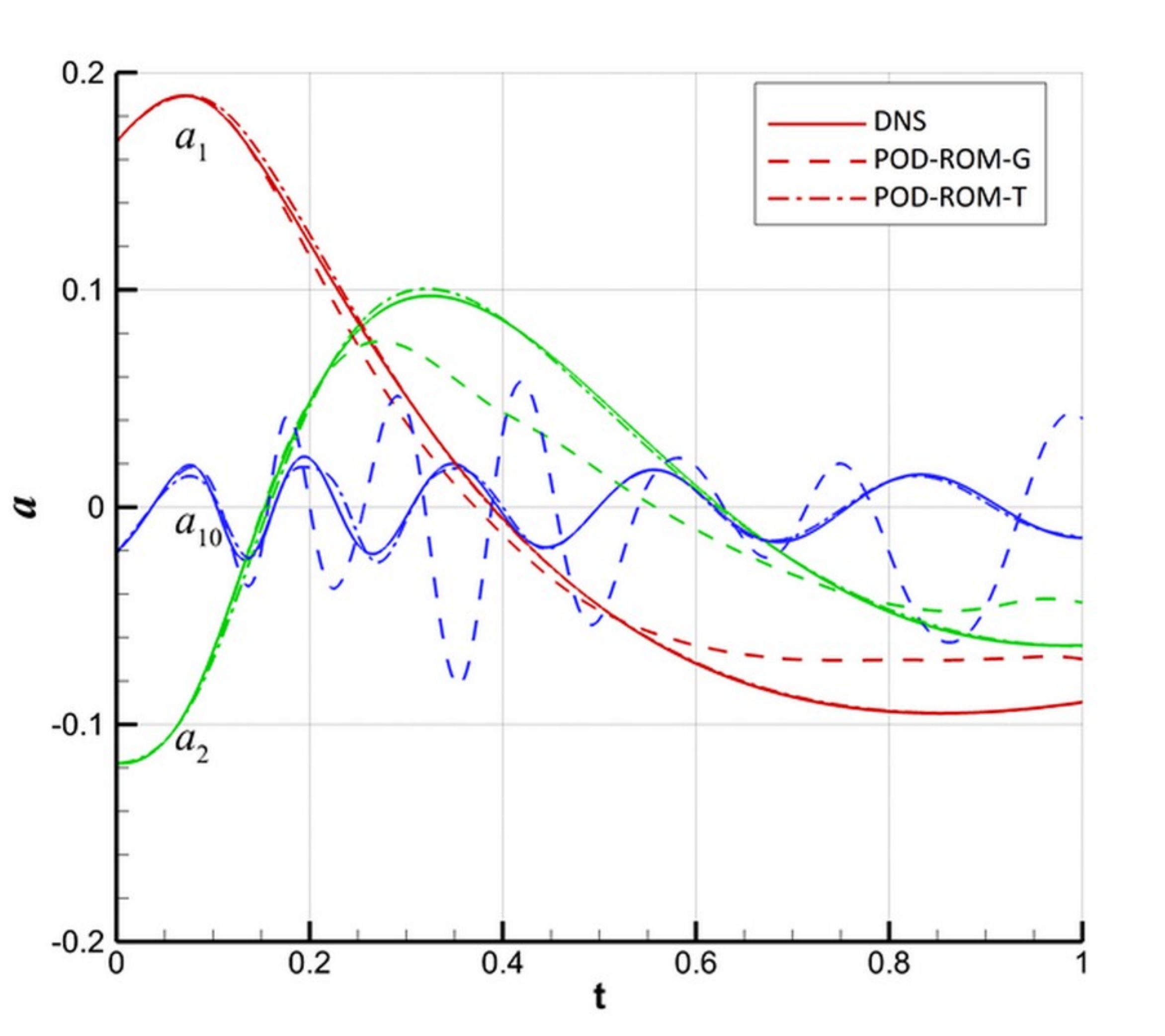}}
\subfigure[POD-ROM-MK]{\includegraphics[width=0.35\textwidth]{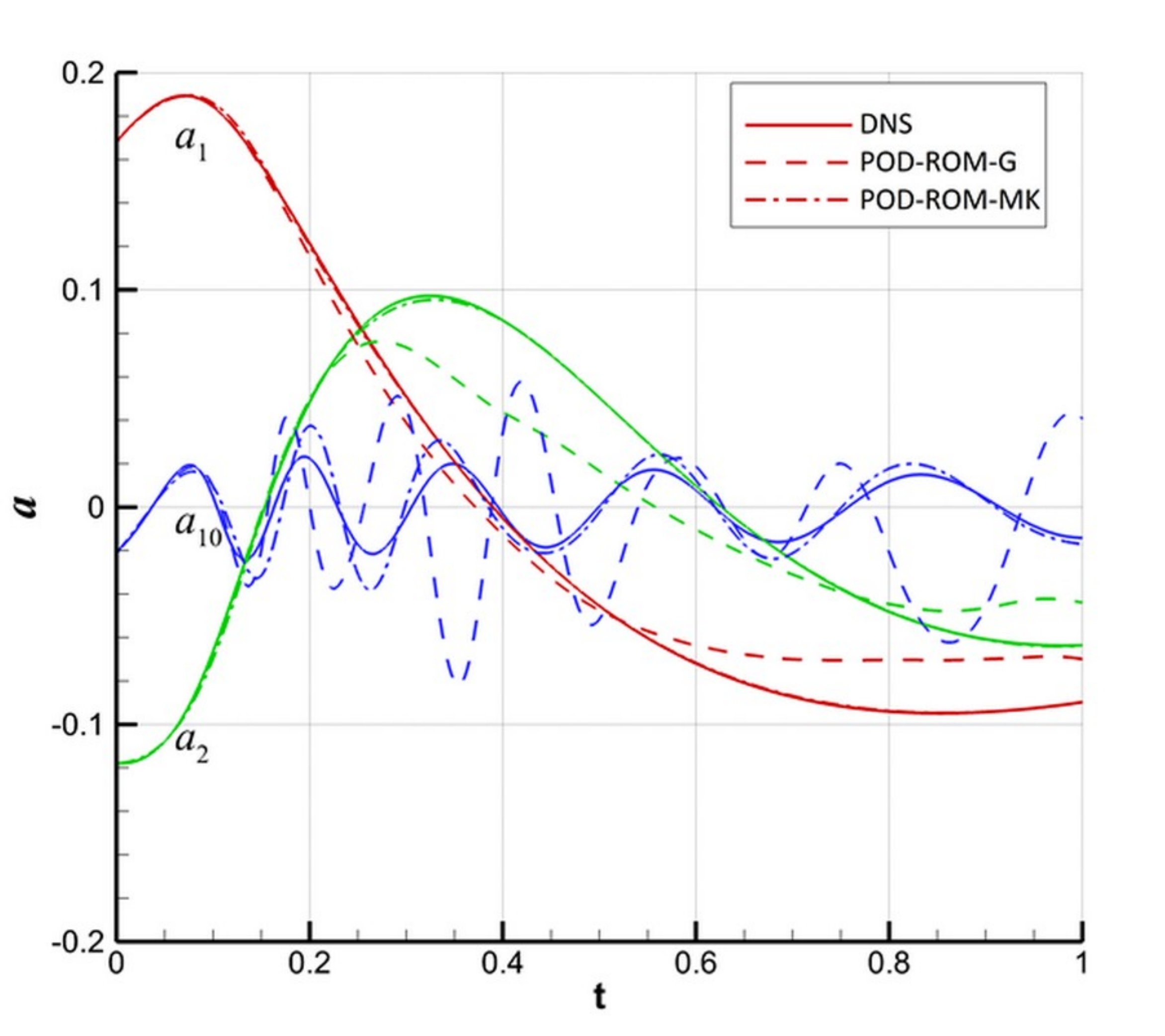}}
}\\
\mbox{
\subfigure[POD-ROM-CL]{\includegraphics[width=0.35\textwidth]{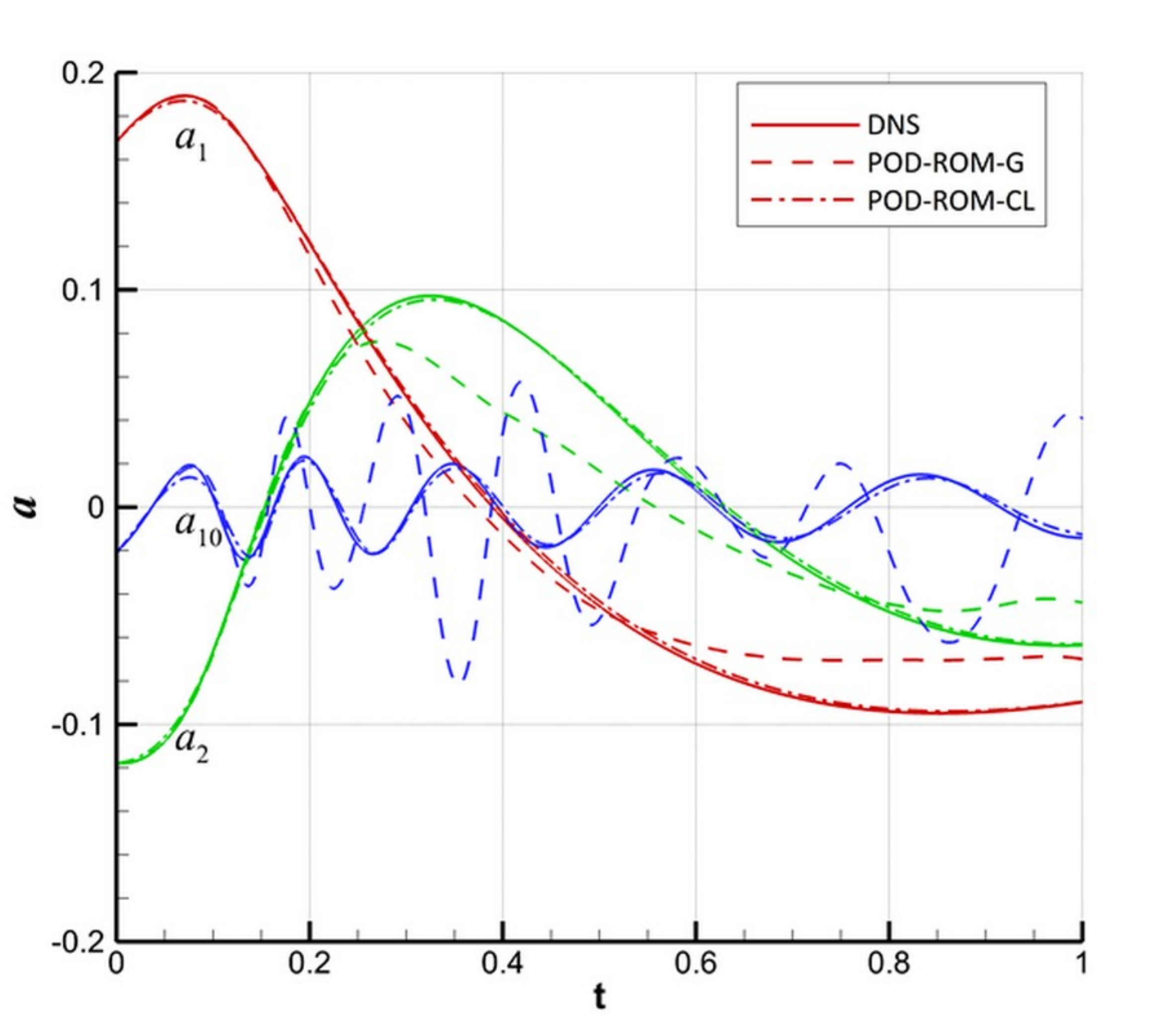}}
\subfigure[POD-ROM-S]{\includegraphics[width=0.35\textwidth]{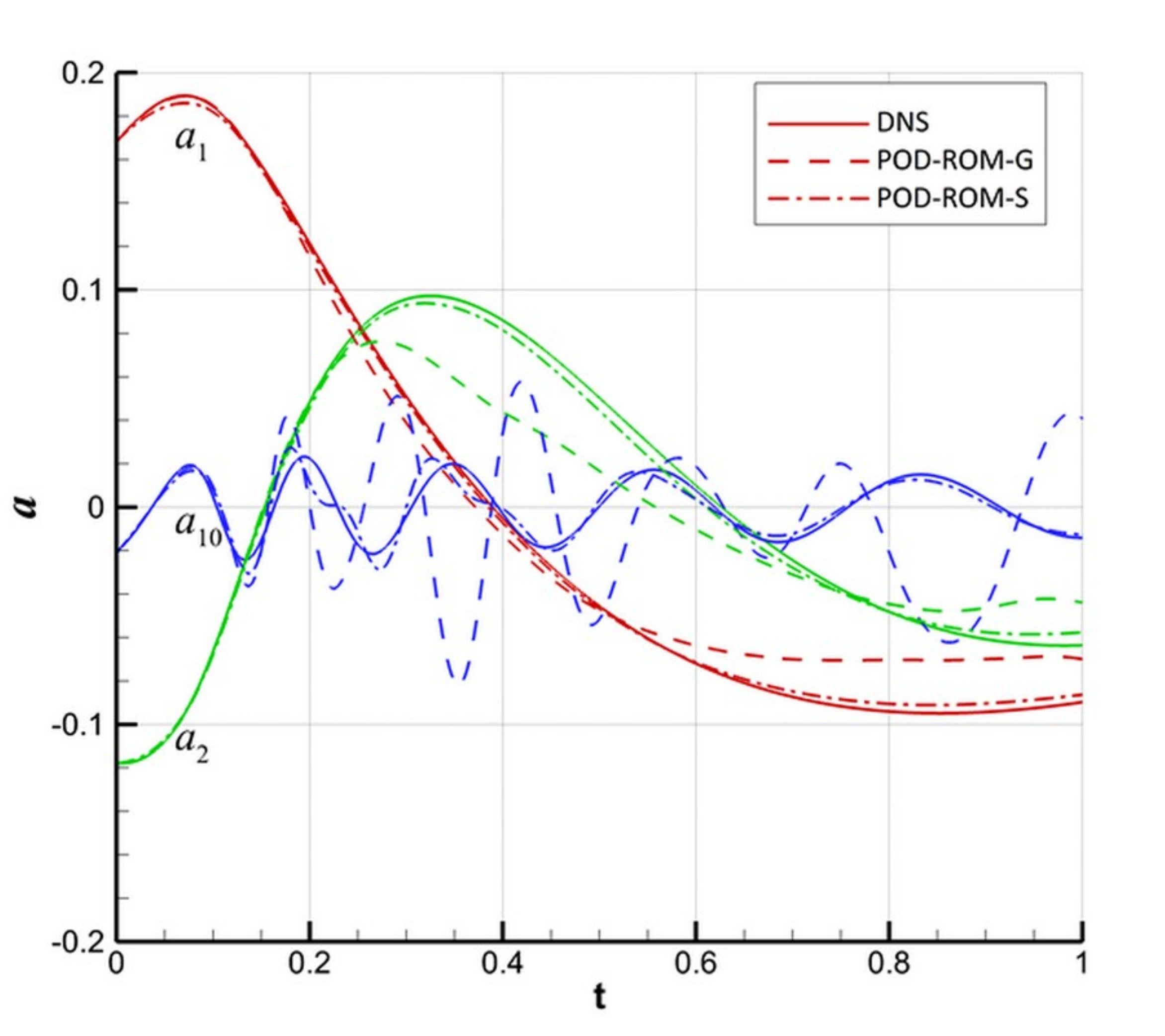}}
\subfigure[POD-ROM-C]{\includegraphics[width=0.35\textwidth]{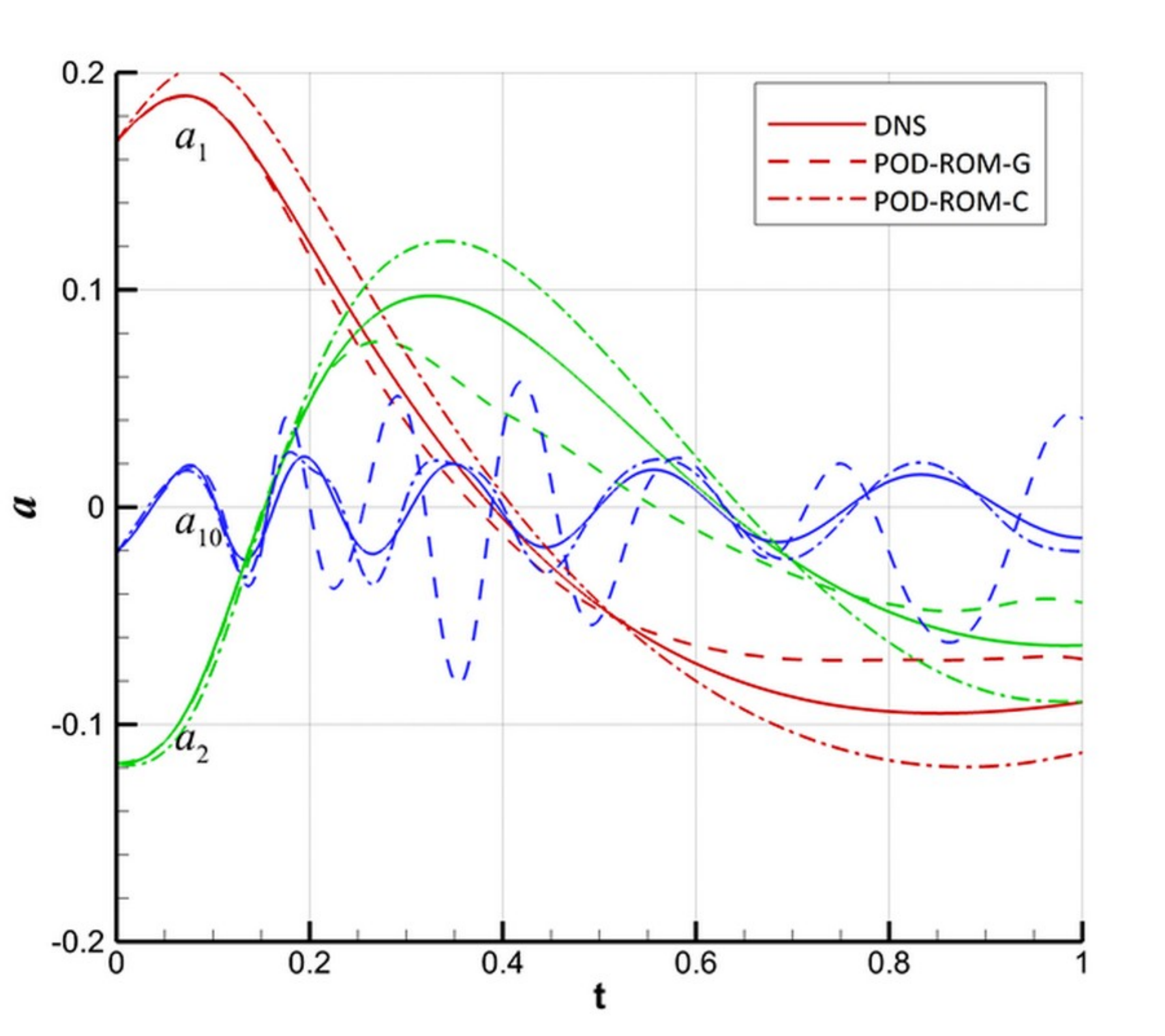}}
}
\caption{
Experiment 2:
Time evolutions of $a_1, a_2$ and $a_{10}$ coefficients of the POD-ROMs.
DNS and POD-ROM-G results are also included for comparison purposes.
}
\label{fig:9}
\end{figure}

Fig.~\ref{fig:8} and Fig.~\ref{fig:9} display the time evolutions of the coefficients $a_{1}$, $a_{2}$, and $a_{10}$ of all the POD-ROMs for Experiment 1 and Experiment 2, respectively.
All the POD-ROMs use $R=20$ POD modes and the optimal free parameter $\nu_e$, which is simply determined by taking the value of $\nu_e$ corresponding to the smallest RMS error value in Fig.~\ref{fig:7}.
The large numerical oscillations of the POD-ROM-G, displayed in particular by the time evolution of $a_{10}$, are significantly decreased by the POD-ROM closure models.
Fig.~\ref{fig:8} and Fig.~\ref{fig:9} also show that POD-ROM-H, POD-ROM-MK, POD-ROM-S and especially POD-ROM-C are not competitive.
All the other POD-ROMs, however, produce relatively accurate results.


\begin{figure}
\mbox{
\subfigure[DNS]{\includegraphics[width=0.35\textwidth]{s-dns.pdf}}
\subfigure[POD-ROM-G]{\includegraphics[width=0.35\textwidth]{s-gal-20.pdf}}
\subfigure[POD-ROM-SR ]{\includegraphics[width=0.35\textwidth]{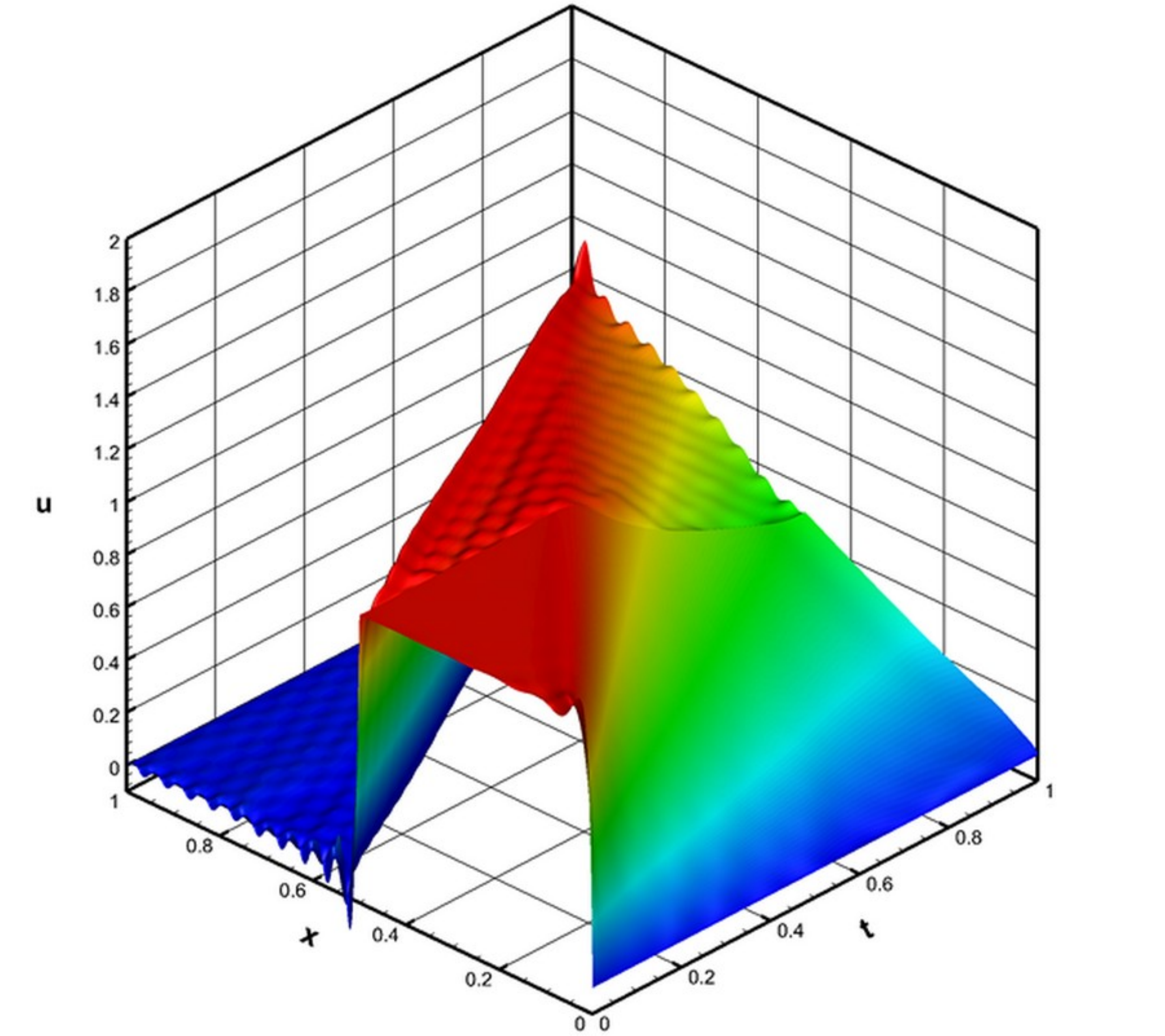}}
}\\
\mbox{
\subfigure[POD-ROM-H]{\includegraphics[width=0.35\textwidth]{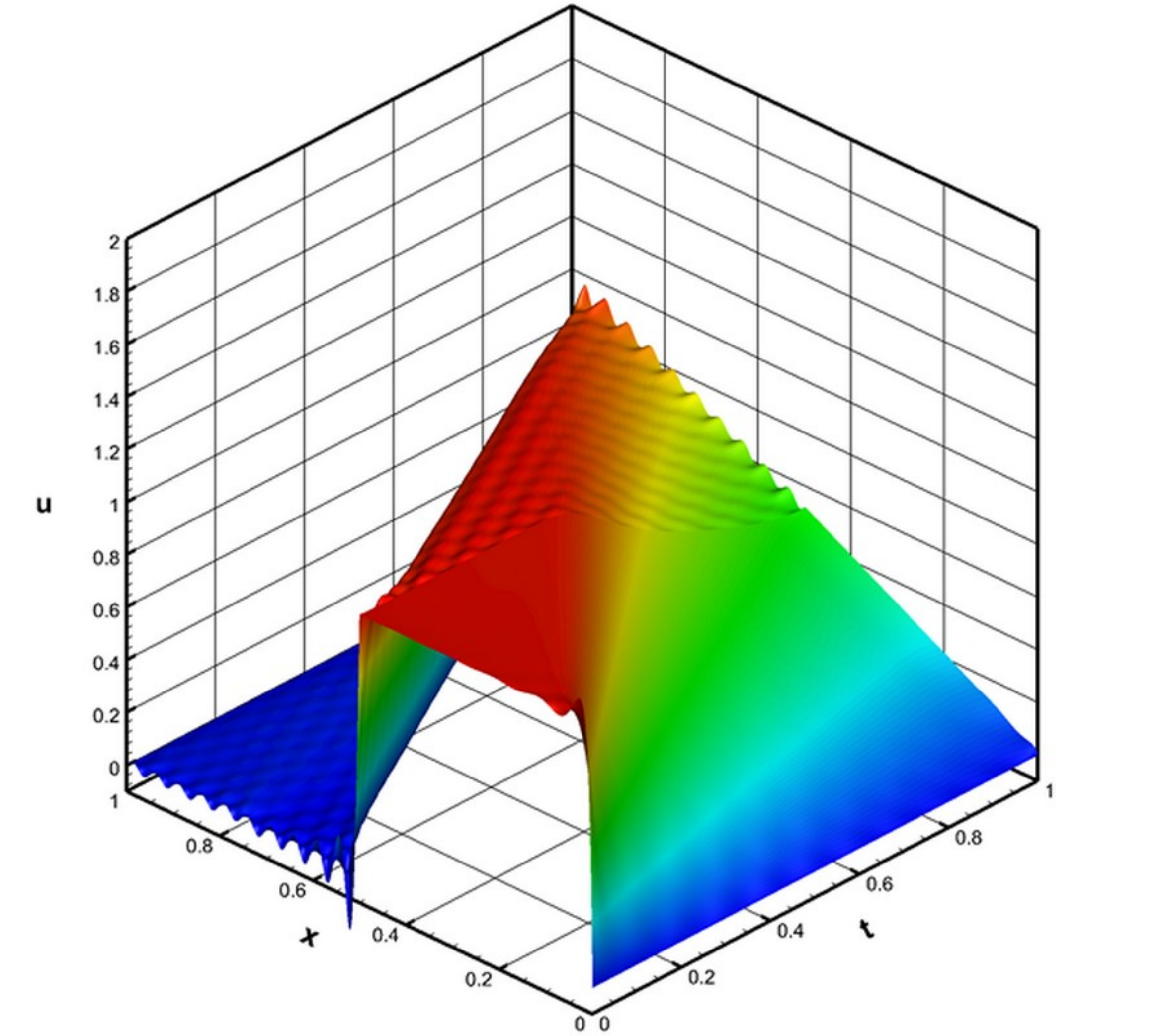}}
\subfigure[POD-ROM-R]{\includegraphics[width=0.35\textwidth]{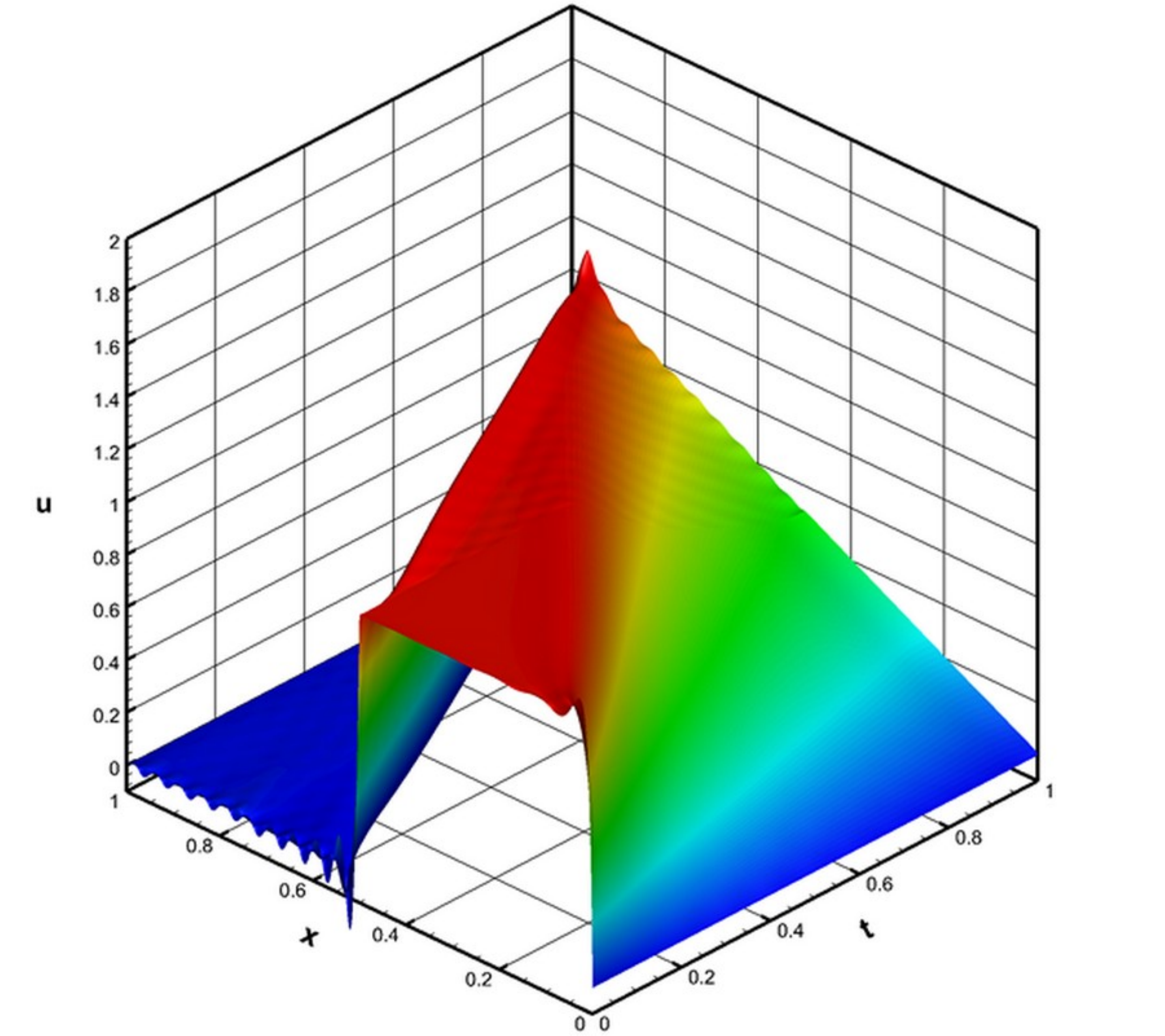}}
\subfigure[POD-ROM-RQ]{\includegraphics[width=0.35\textwidth]{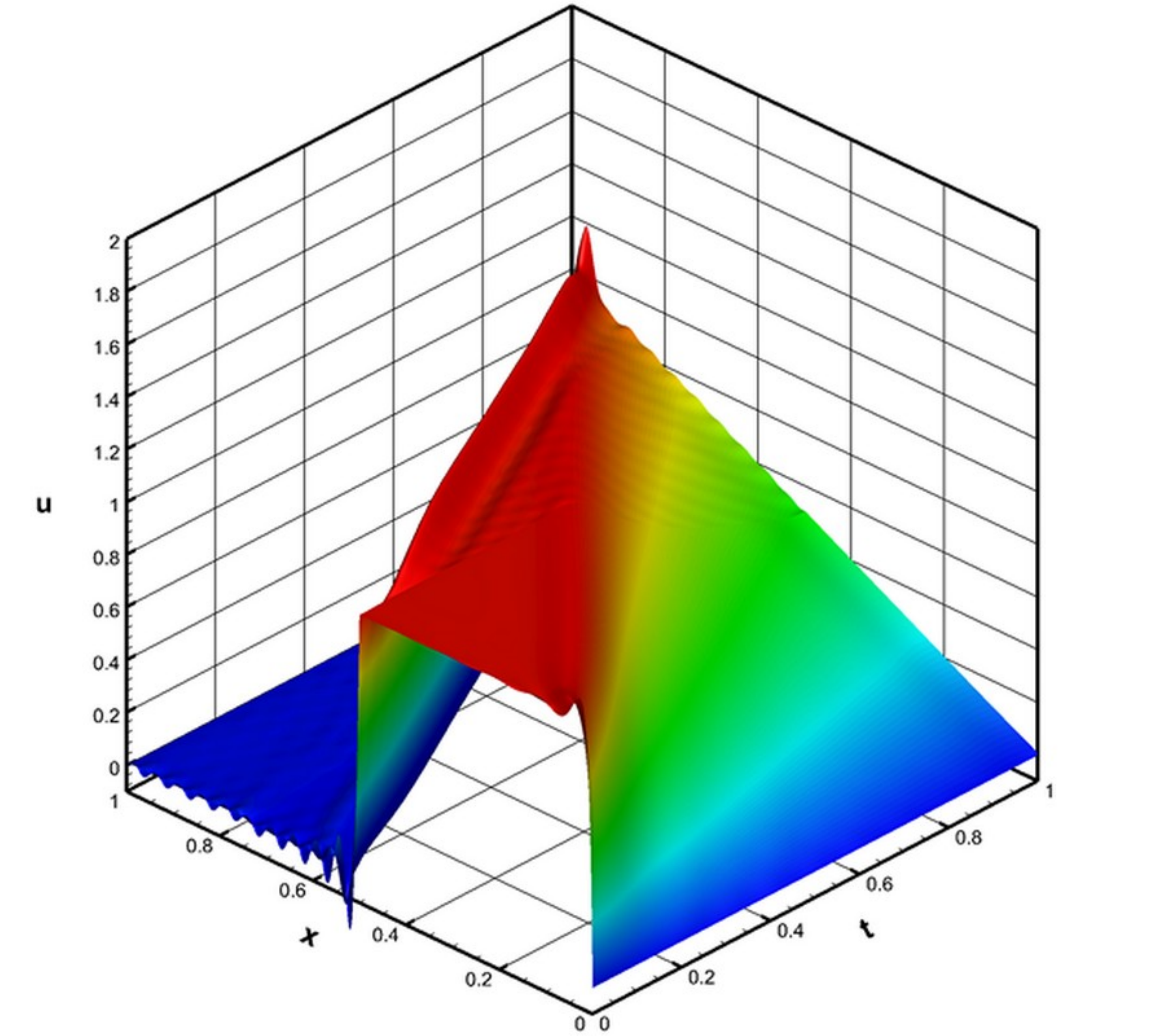}}
}\\
\mbox{
\subfigure[POD-ROM-RS]{\includegraphics[width=0.35\textwidth]{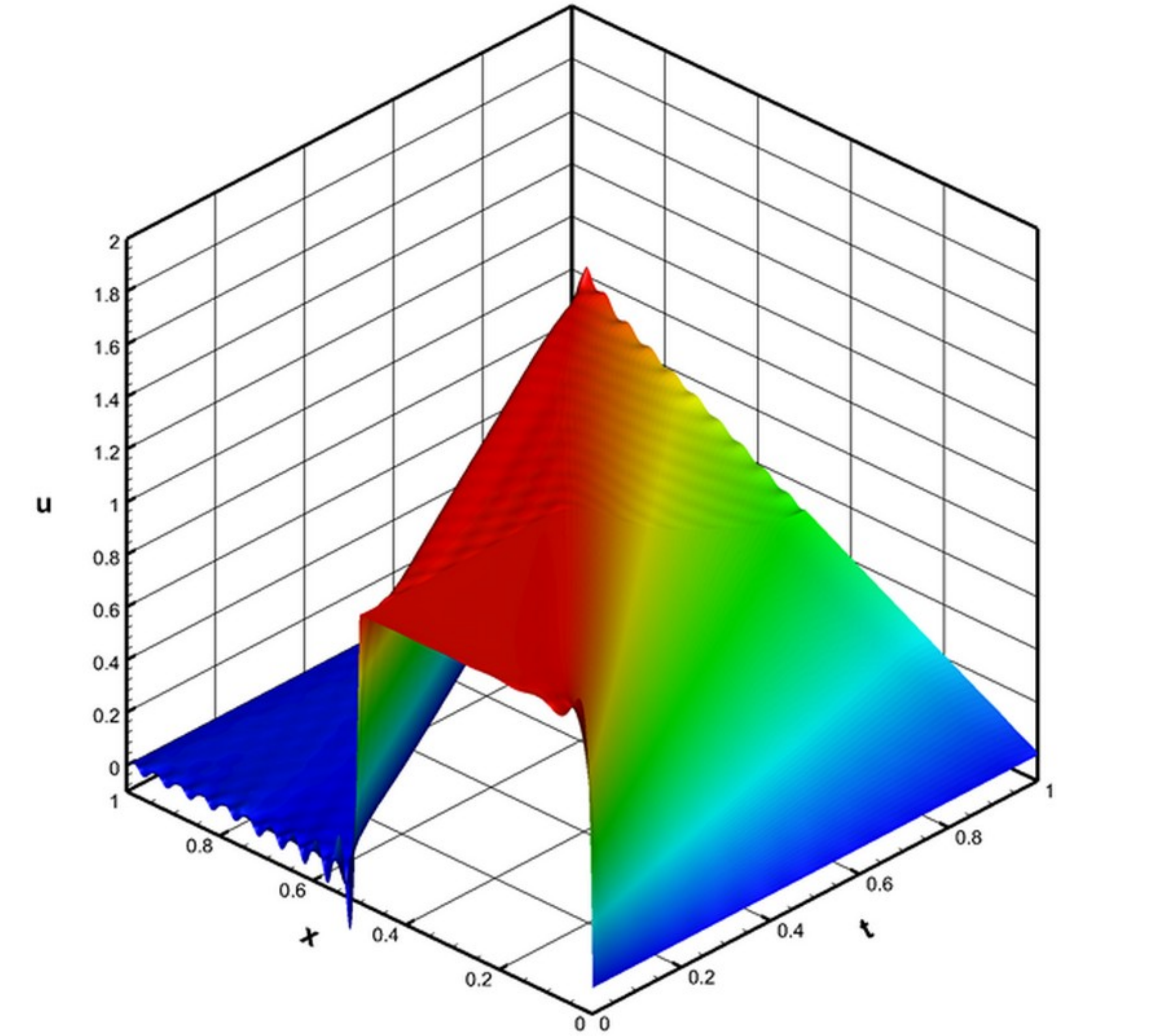}}
\subfigure[POD-ROM-T]{\includegraphics[width=0.35\textwidth]{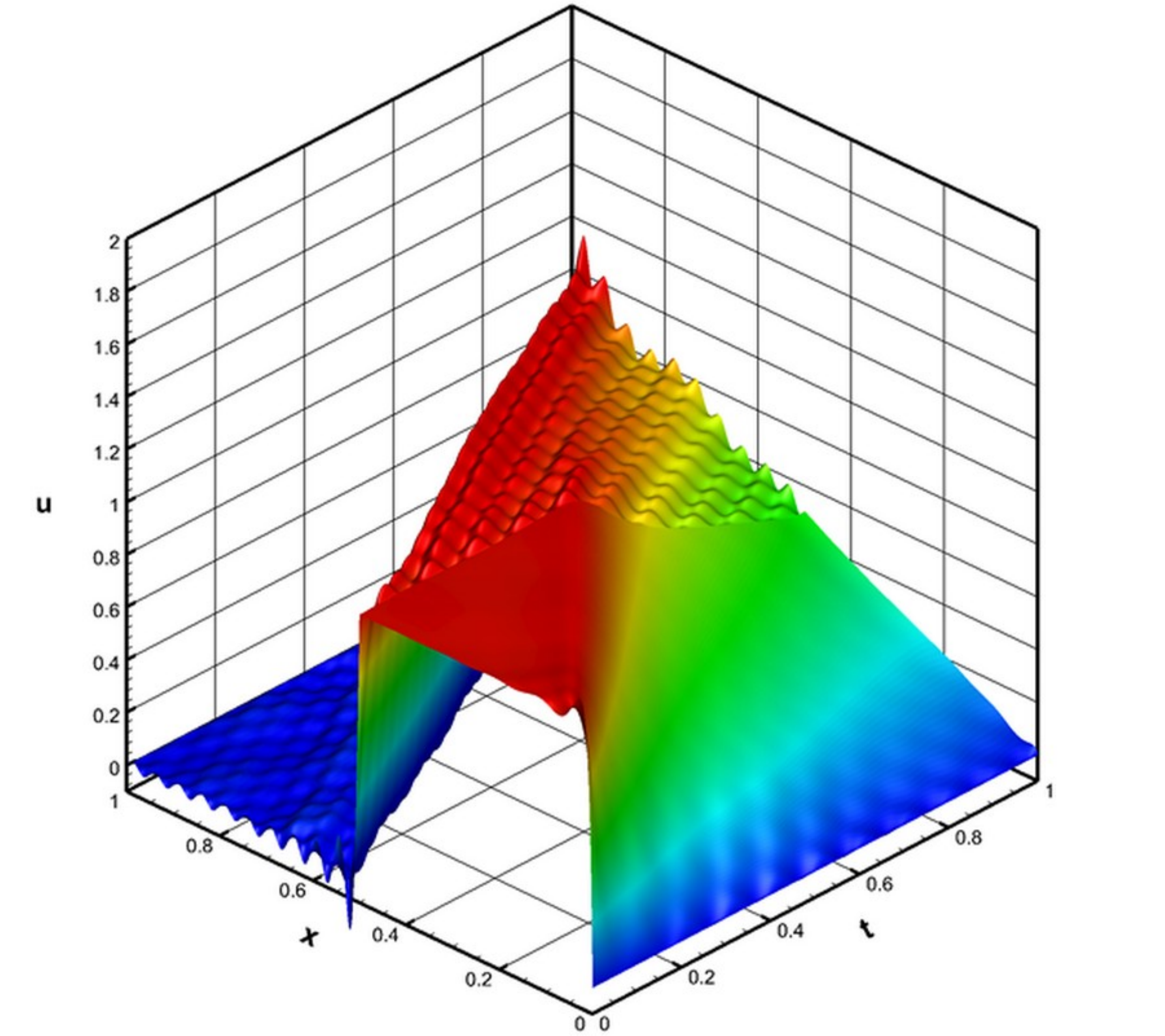}}
\subfigure[POD-ROM-MK]{\includegraphics[width=0.35\textwidth]{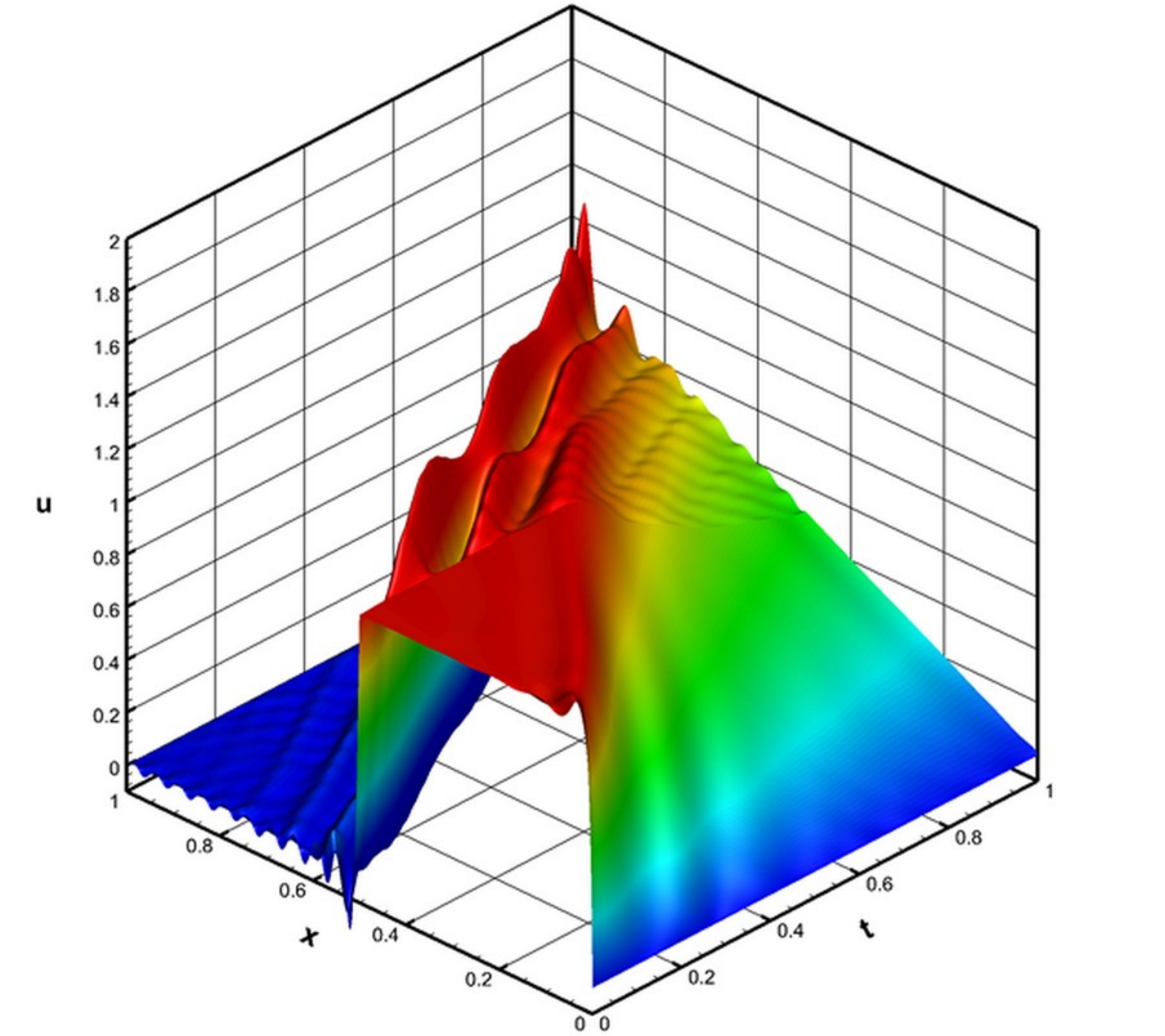}}
}\\
\mbox{
\subfigure[POD-ROM-CL]{\includegraphics[width=0.35\textwidth]{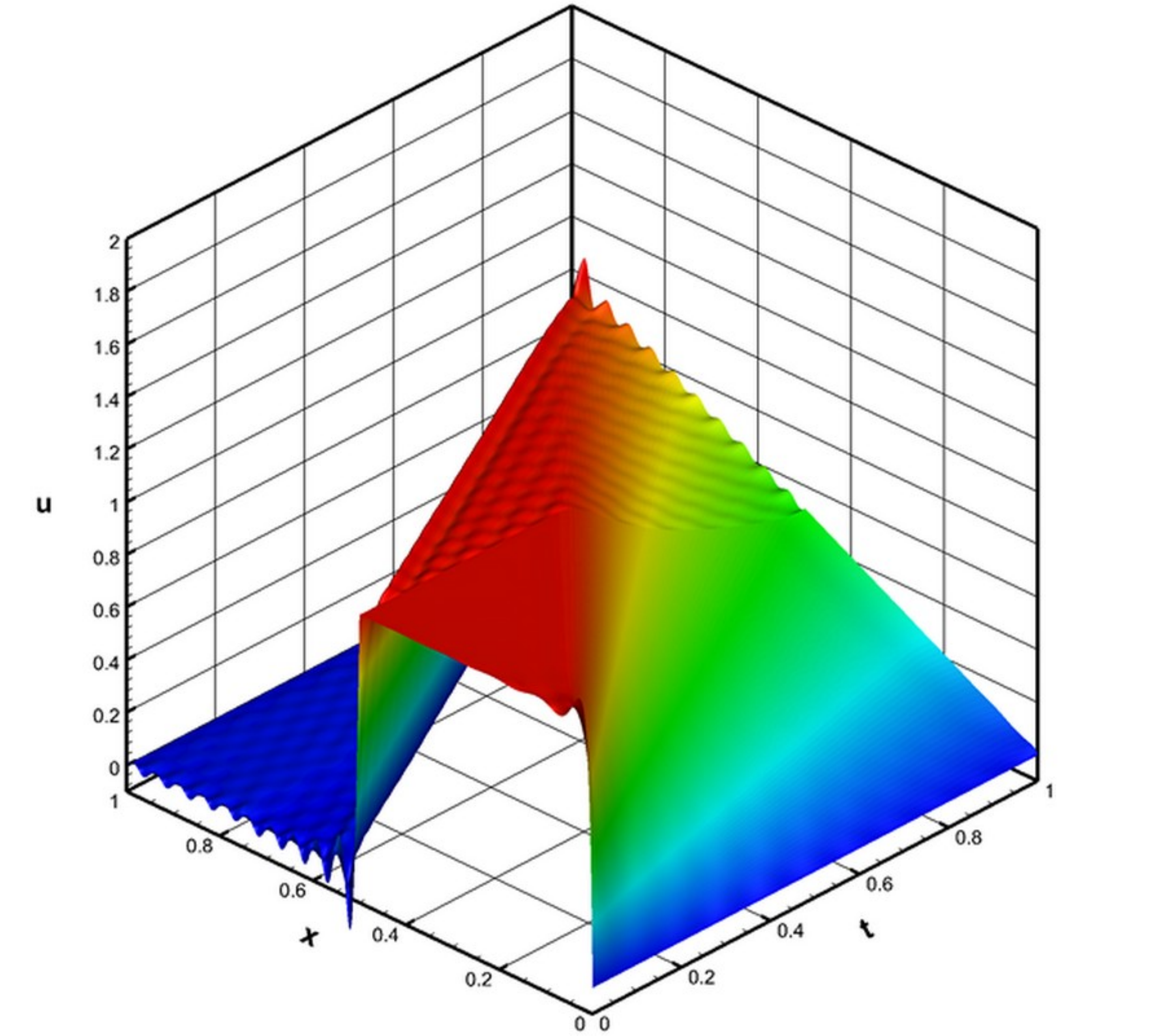}}
\subfigure[POD-ROM-S]{\includegraphics[width=0.35\textwidth]{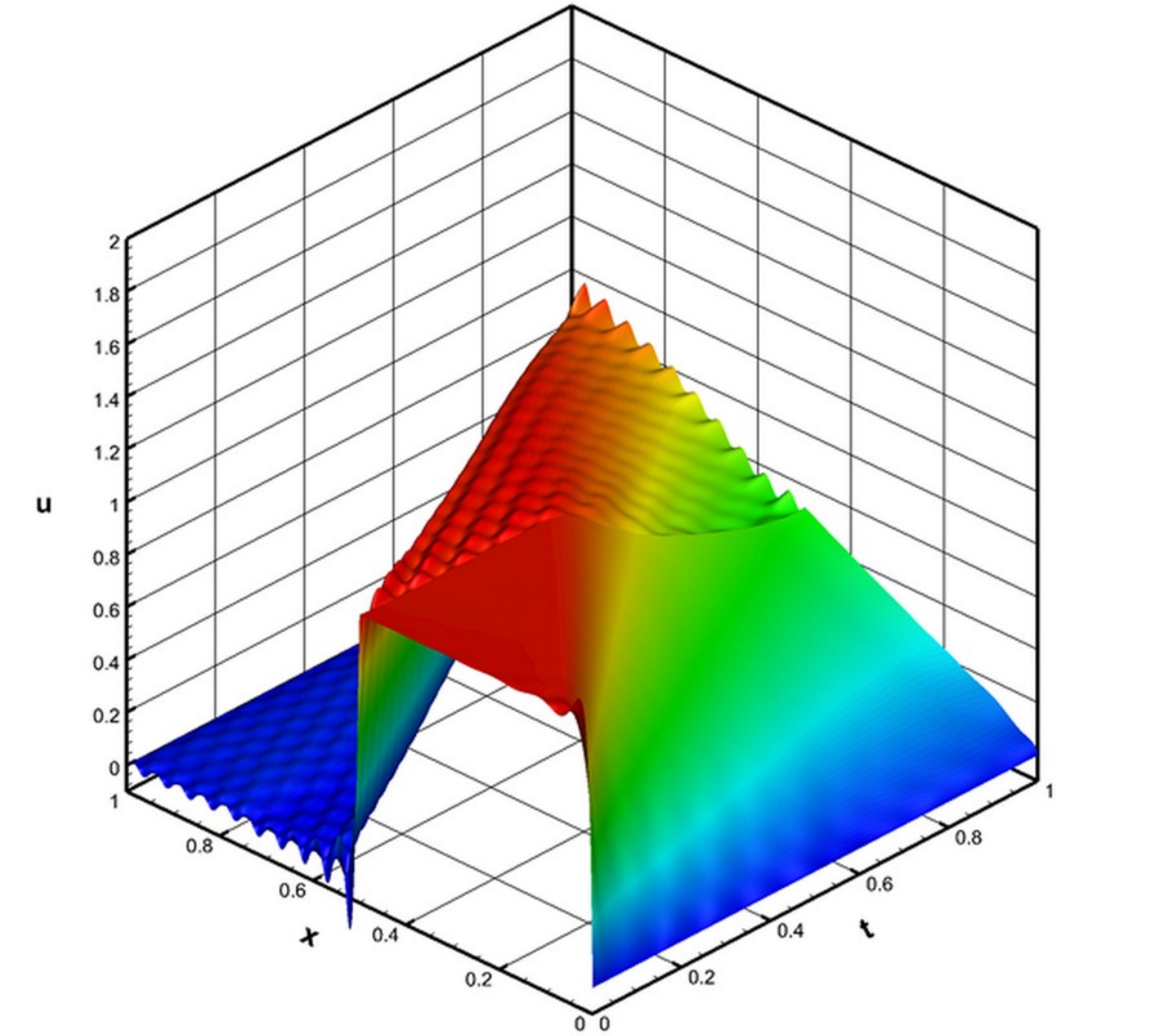}}
\subfigure[POD-ROM-C]{\includegraphics[width=0.35\textwidth]{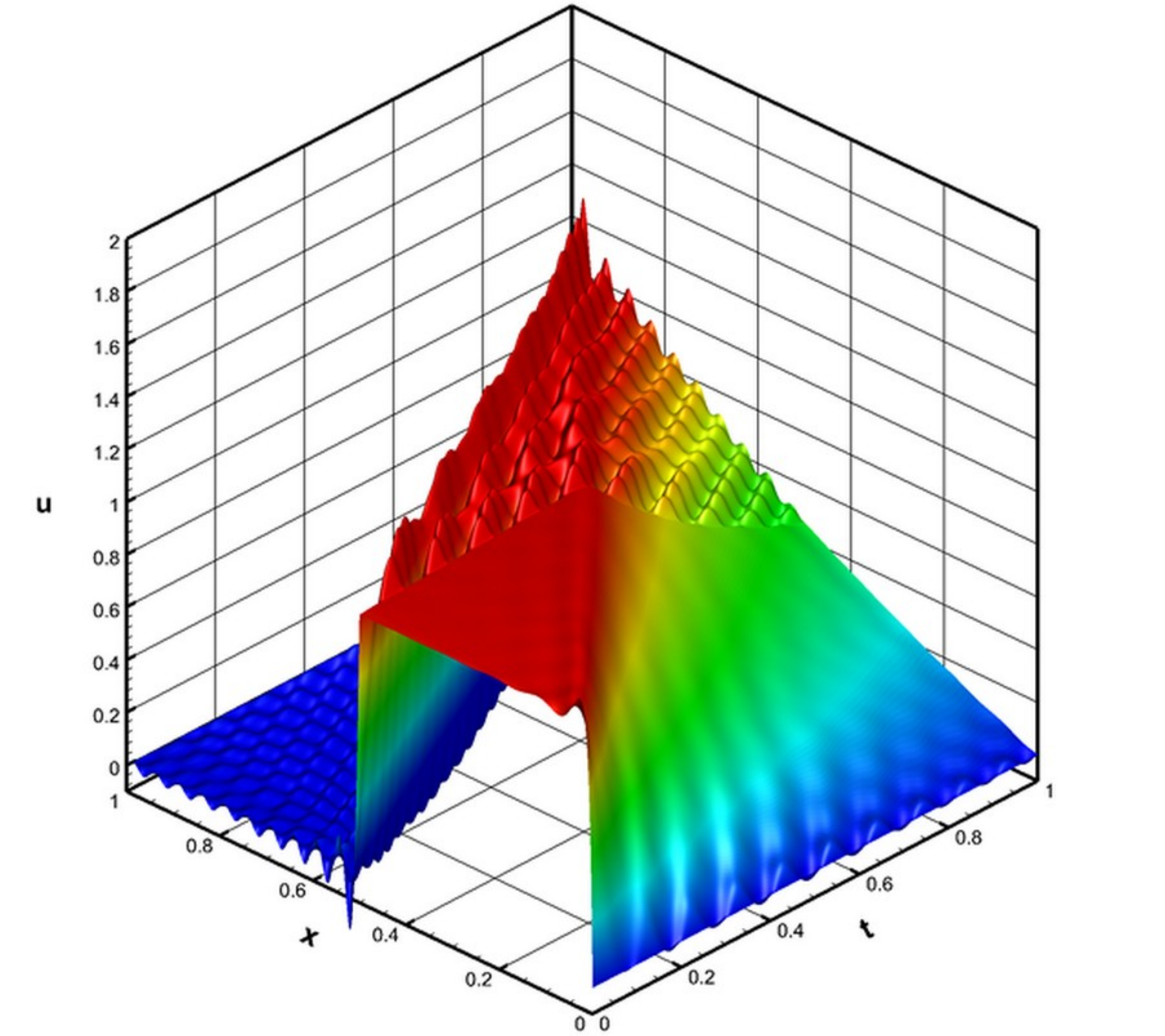}}
}
\caption{
Experiment 1:
POD-ROM results.
DNS and POD-ROM-G results are also included for comparison purposes.
}
\label{fig:10}
\end{figure}

\begin{figure}
\mbox{
\subfigure[DNS]{\includegraphics[width=0.35\textwidth]{g-dns.pdf}}
\subfigure[POD-ROM-G]{\includegraphics[width=0.35\textwidth]{g-gal-20.pdf}}
\subfigure[POD-ROM-SR]{\includegraphics[width=0.35\textwidth]{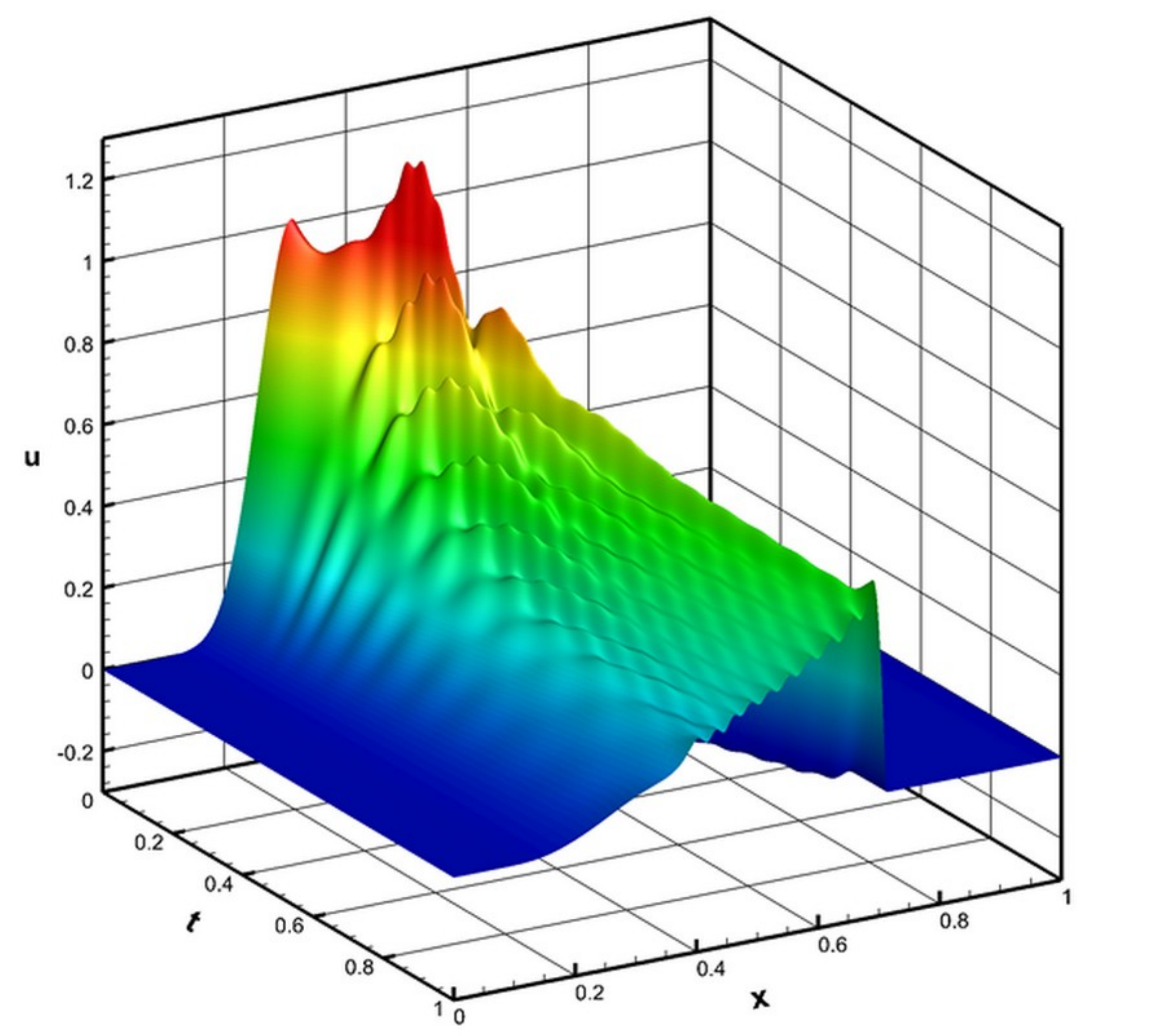}}
}\\
\mbox{
\subfigure[POD-ROM-H]{\includegraphics[width=0.35\textwidth]{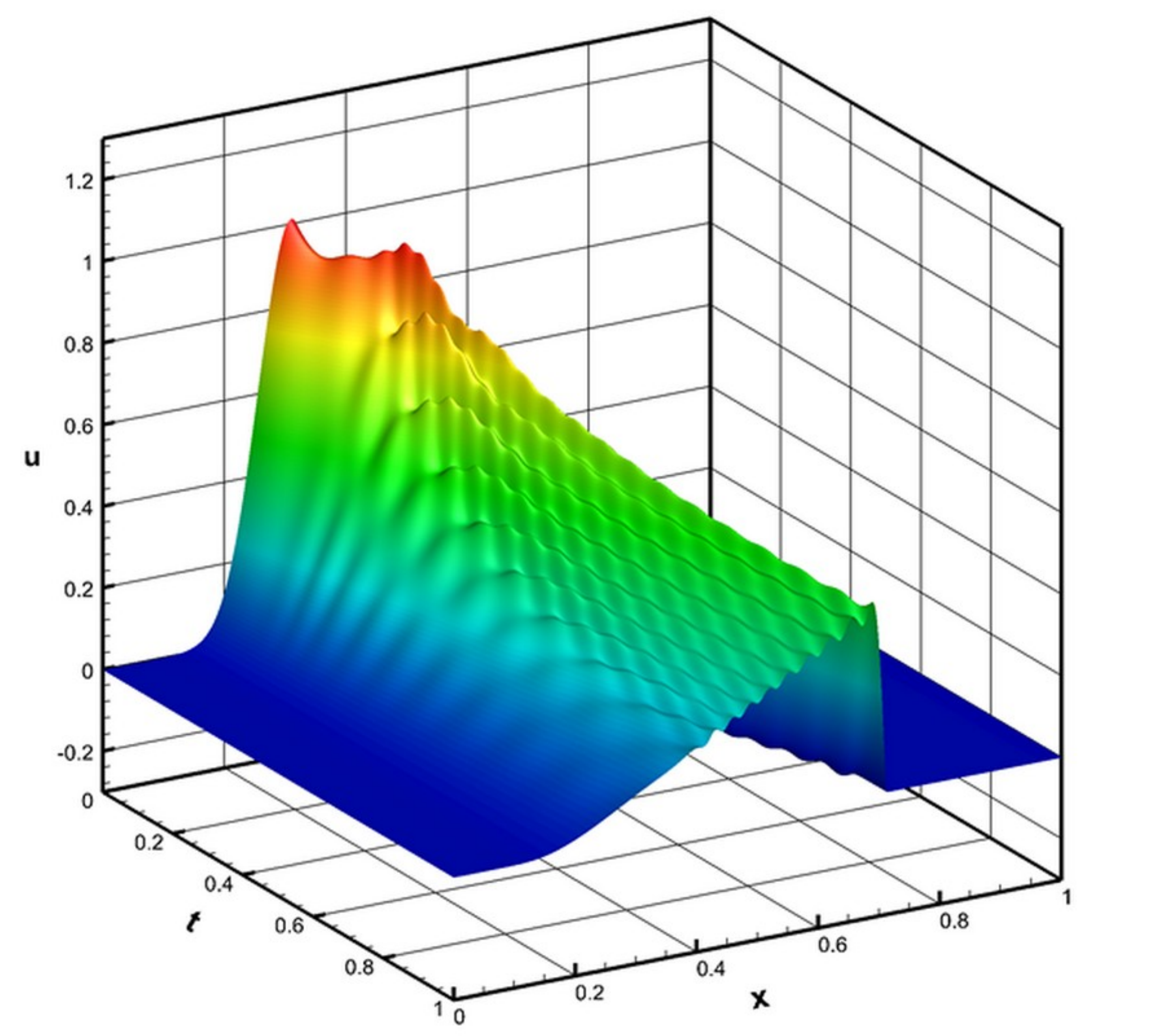}}
\subfigure[POD-ROM-R]{\includegraphics[width=0.35\textwidth]{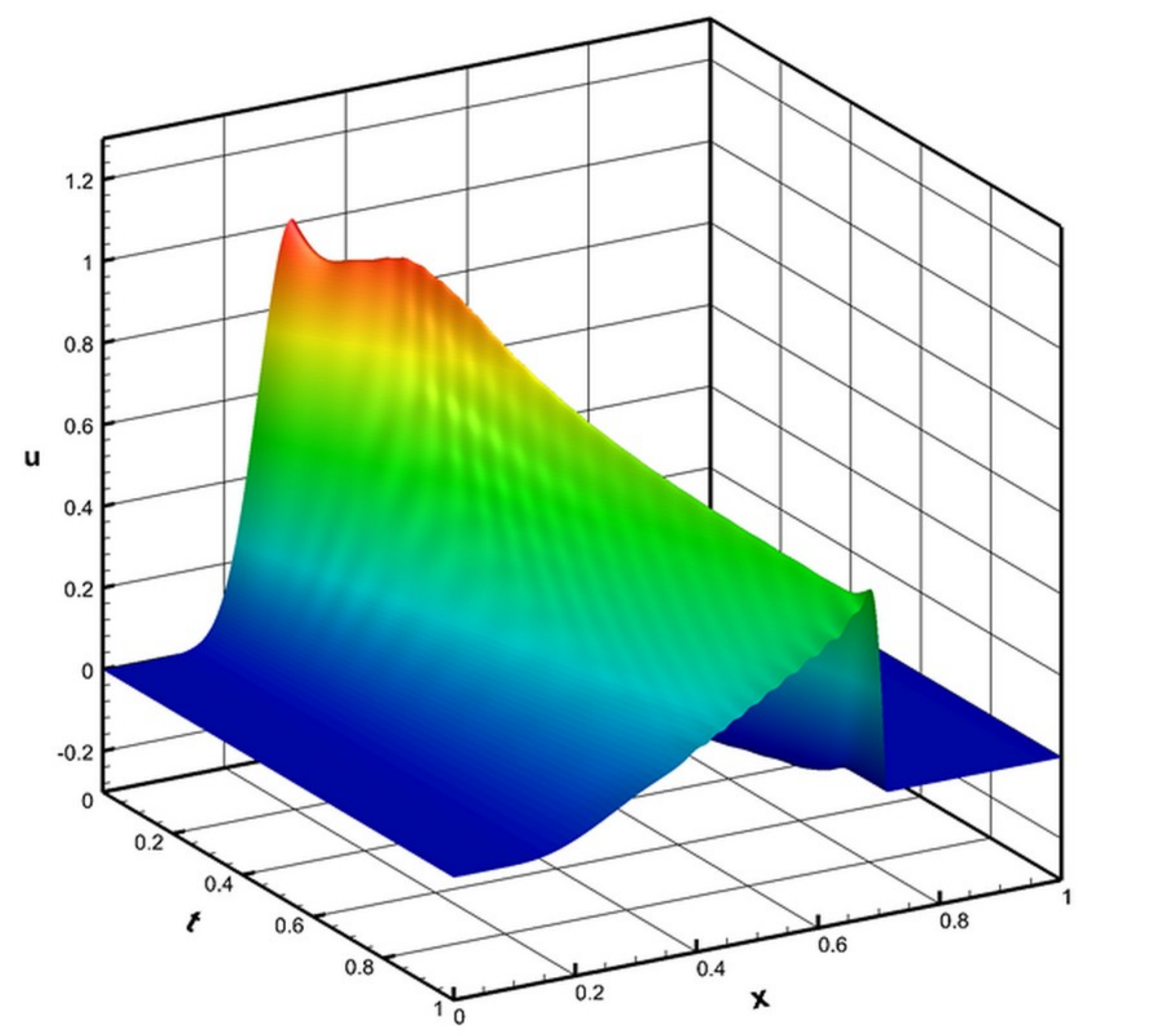}}
\subfigure[POD-ROM-RQ]{\includegraphics[width=0.35\textwidth]{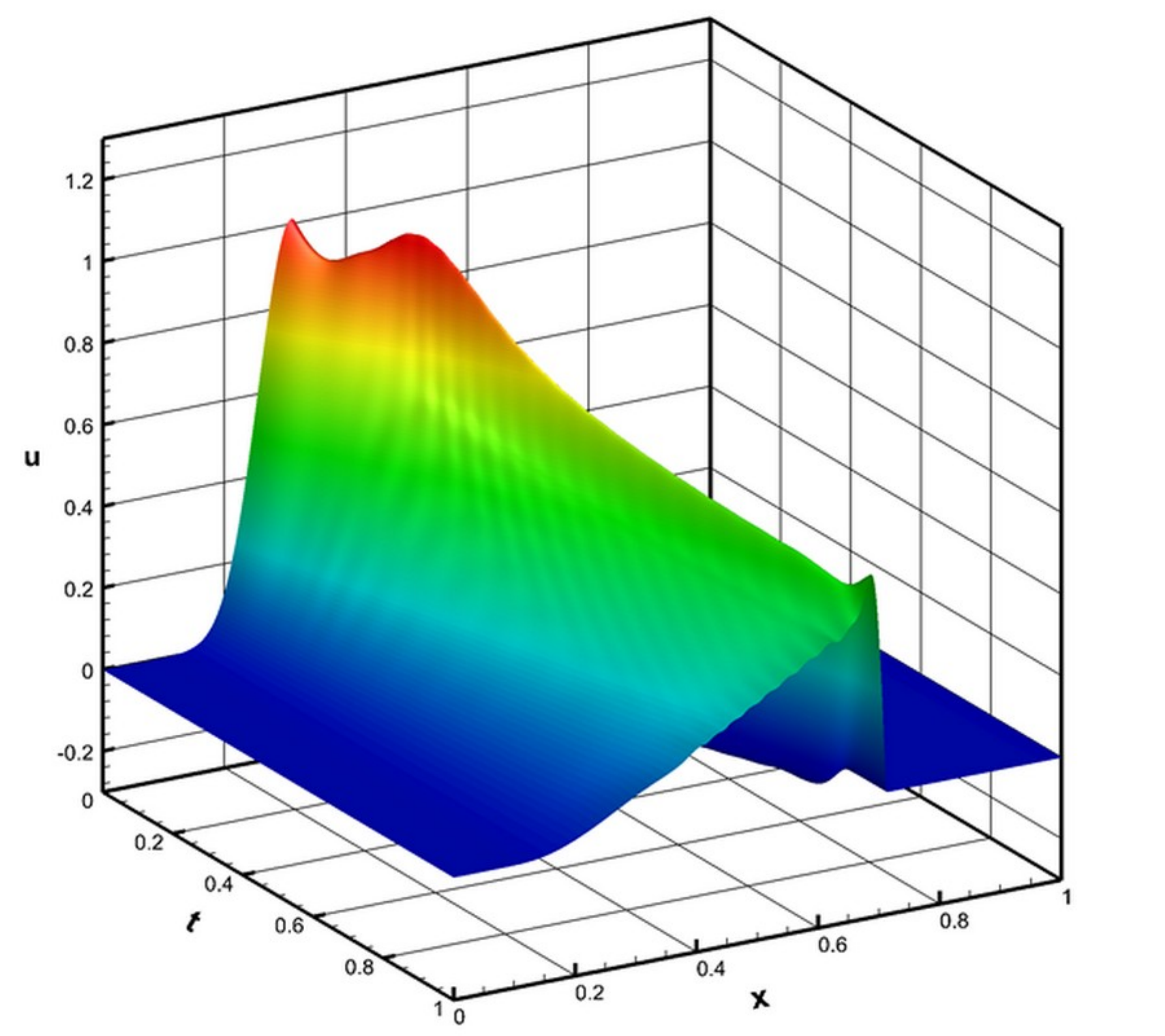}}
}\\
\mbox{
\subfigure[POD-ROM-RS]{\includegraphics[width=0.35\textwidth]{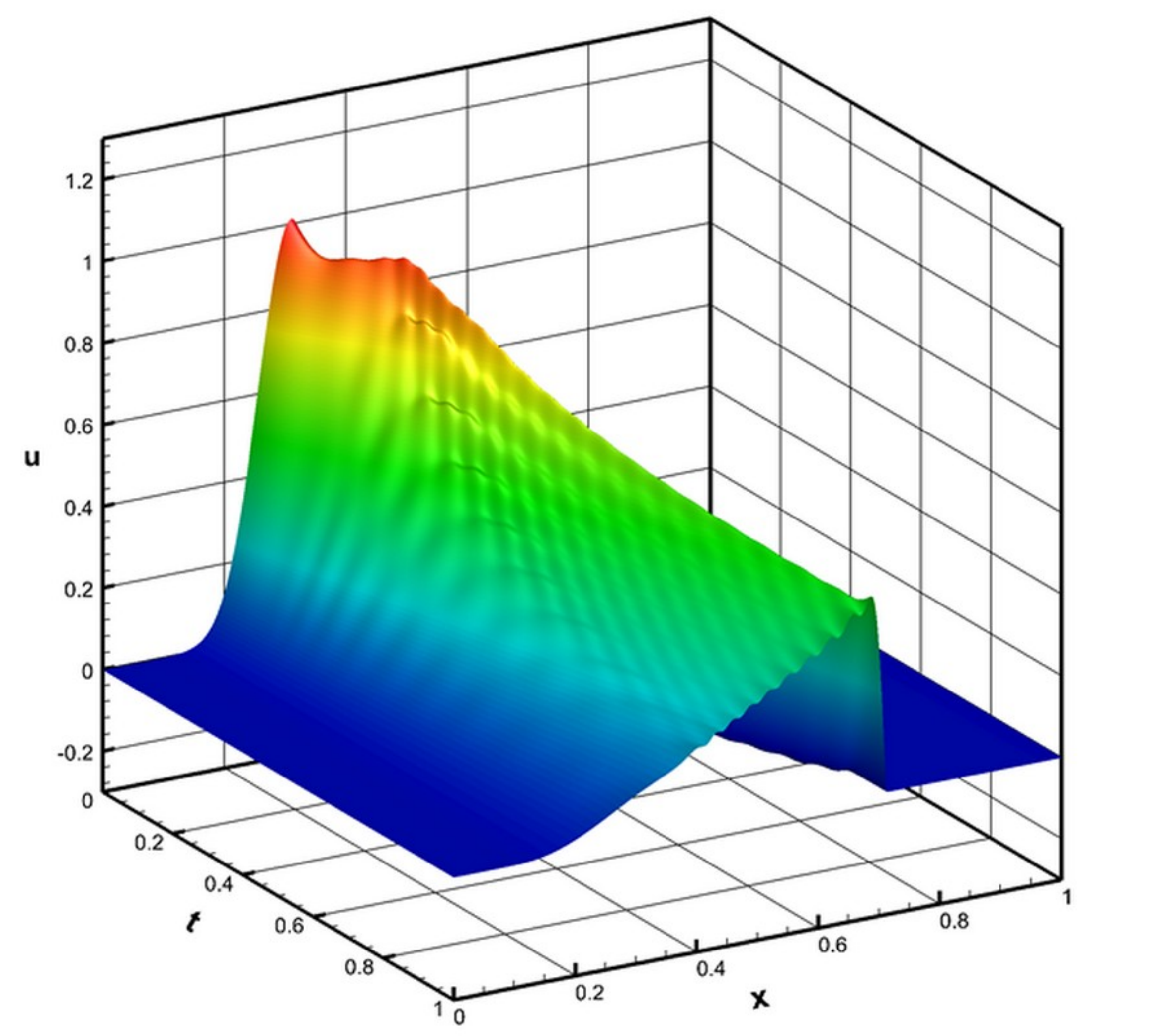}}
\subfigure[POD-ROM-T]{\includegraphics[width=0.35\textwidth]{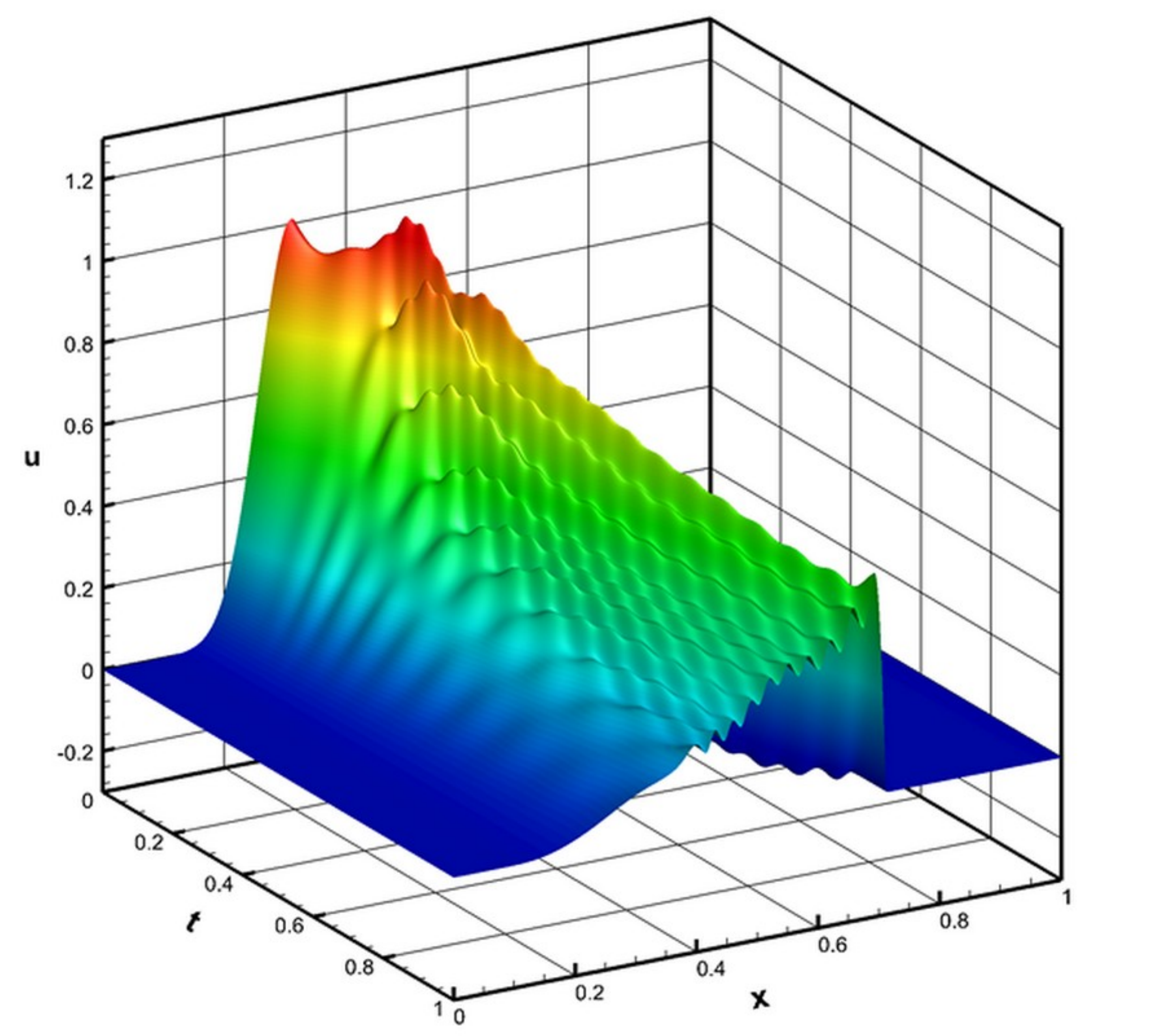}}
\subfigure[POD-ROM-MK]{\includegraphics[width=0.35\textwidth]{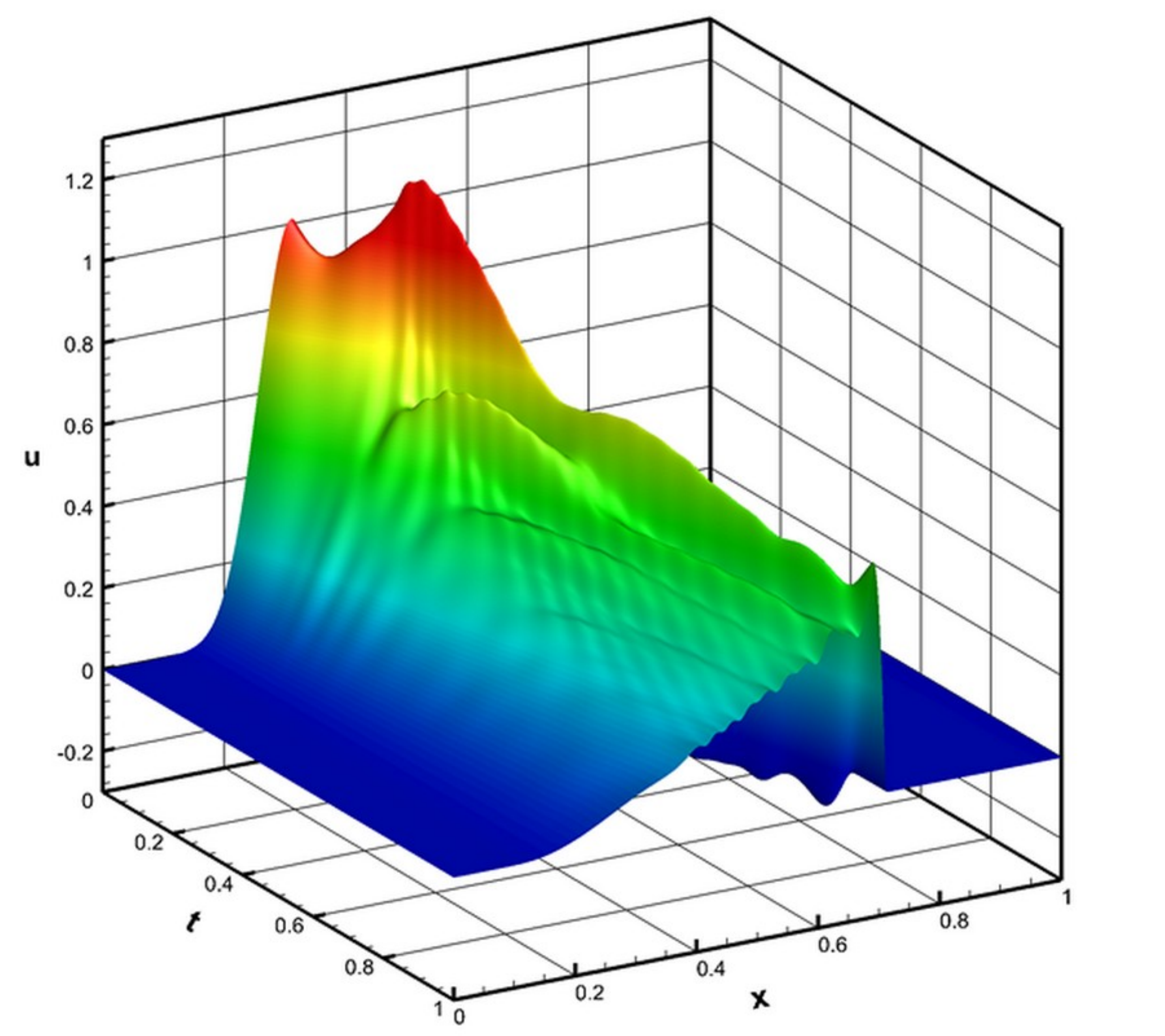}}
}\\
\mbox{
\subfigure[POD-ROM-CL]{\includegraphics[width=0.35\textwidth]{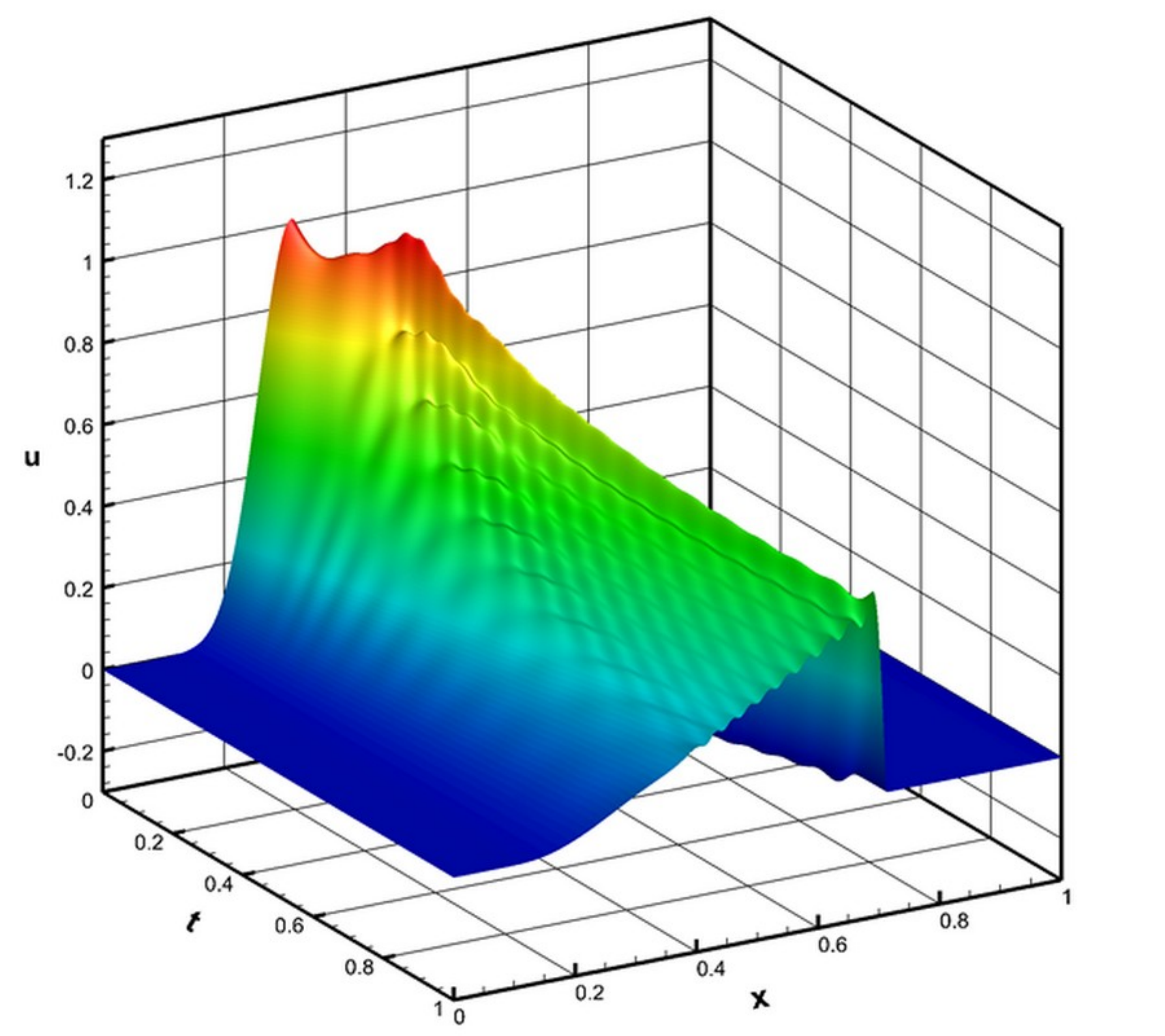}}
\subfigure[POD-ROM-S]{\includegraphics[width=0.35\textwidth]{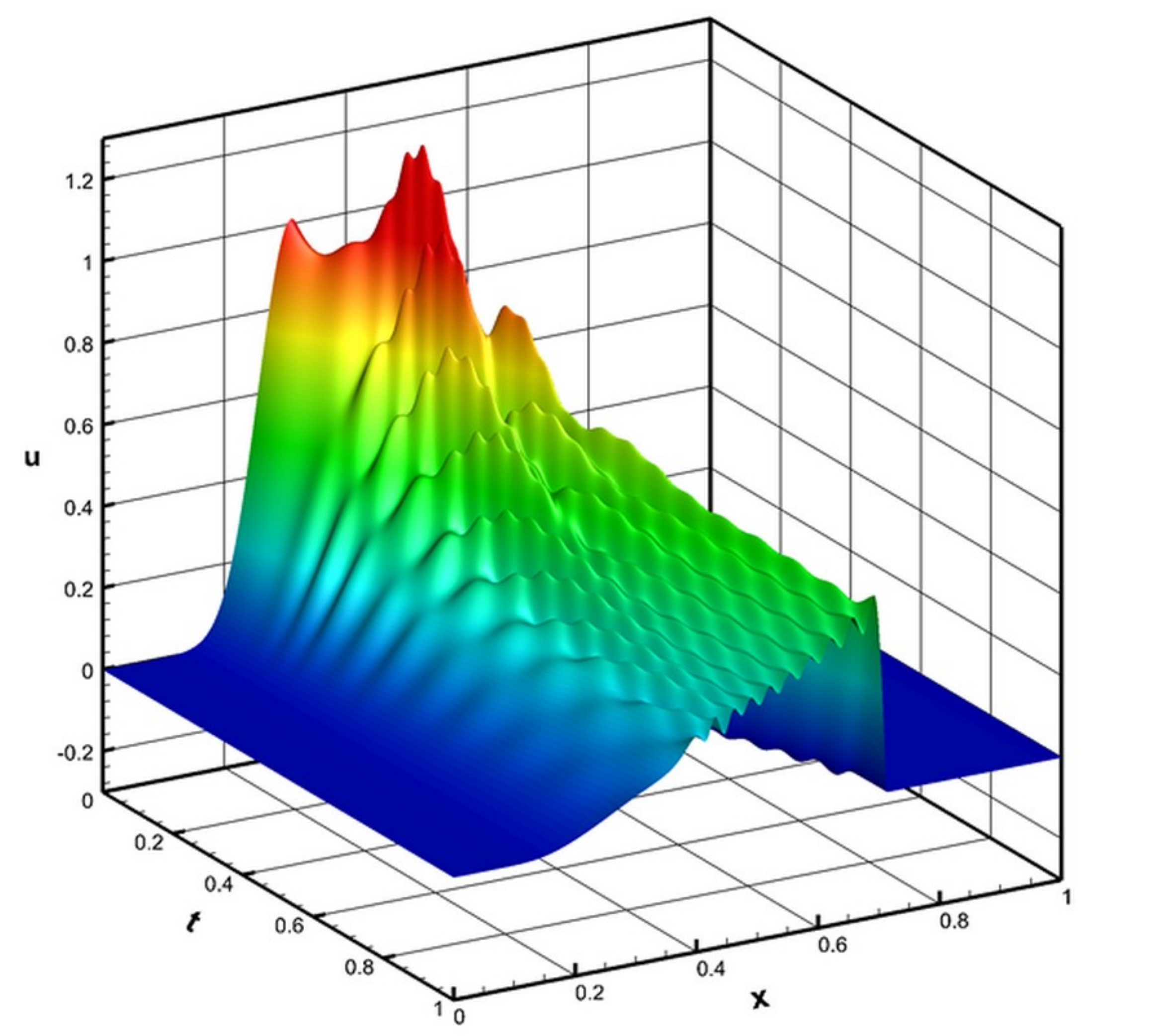}}
\subfigure[POD-ROM-C]{\includegraphics[width=0.35\textwidth]{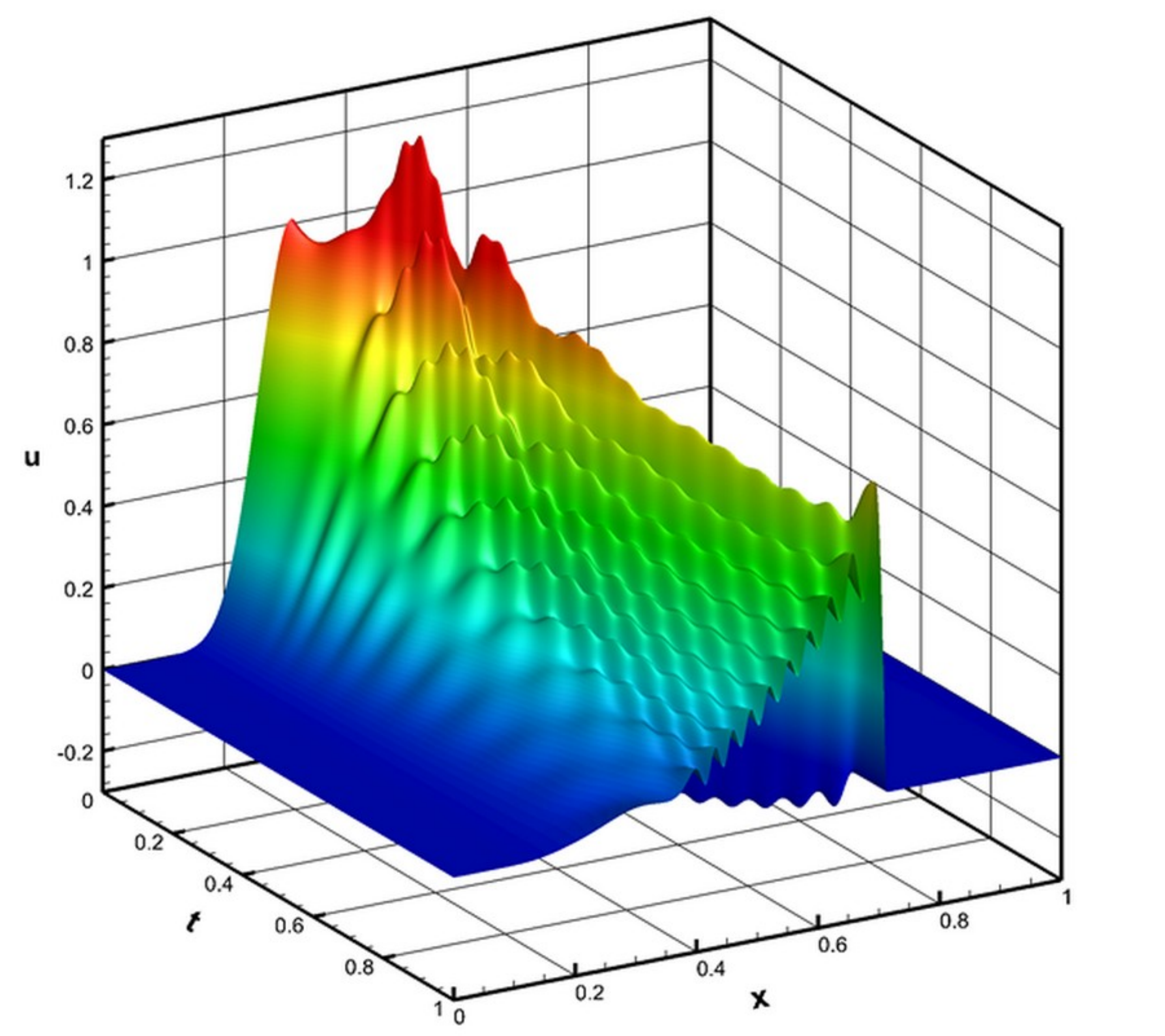}}
}
\caption{
Experiment 2:
POD-ROM results.
DNS and POD-ROM-G results are also included for comparison purposes.
}
\label{fig:11}
\end{figure}

Fig.~\ref{fig:10} and Fig.~\ref{fig:11} display the POD-ROM approximations for Experiment 1 and Experiment 2, respectively.
Results obtained by the DNS and the POD-ROM-G are also included for comparison.
All the POD-ROMs use $R=20$ modes and the optimal $\nu_e$.
The CPU time of the POD-ROMs (around 4 s for both Experiment 1 and Experiment 2) is dramatically lower than the CPU time of the DNS (around 130 s for Experiment 1 and 95 s for Experiment 2).
We also note that all the POD-ROMs yield results that are significantly more accurate than those obtained with the POD-ROM-G.
As in Fig.~\ref{fig:7} and Fig.~\ref{fig:8}, POD-ROM-MK and POD-ROM-C yield noncompetitive results and no clear overall ``winner" can be chosen from the remaining POD-ROMs.

\section{Conclusions}
\label{sec:cons}

Several new closure models for the POD-ROM of fluid flows were proposed.
These and other standard closure models were investigated in the numerical simulation of the Burgers equation.
DNS and POD-ROM-G results were also included for comparison purposes.
A detailed sensitivity analysis with respect to the modeling free parameters was also performed.
Two challenging test problems displaying moving shock waves were chosen as numerical tests.
Both numerical accuracy and computational efficiency were used to assess the performance of the POD-ROMs.

Two main conclusions can be drawn from the numerical investigation:
First, all the POD-ROMs showed a clear improvement in solution accuracy over the standard POD-ROM-G.
The POD-ROM-R and the POD-ROM-RQ outperformed the other POD-ROMs when the norm of the error was considered.
When the time evolution of the POD-ROM coefficients were considered, the POD-ROM-R and the POD-ROM-RQ performed well again, but other POD-ROMs were also competitive.
The second conclusion yielded by the numerical investigation is that all the POD-ROMs were computationally efficient, having a computational cost of the same order as that of the standard POD-ROM-G and much lower than that of the DNS.

From a practical point of view, the main conclusion of this study is that, when the model parameters are carefully chosen, closure models based on eddy viscosity terms that are constant in space and time, but POD mode dependent, are appropriate for POD-ROM of the Burgers equation.
More sophisticated closure models, such as POD-ROM-S and POD-ROM-C, which have a higher computational overhead, do not yield more accurate results.
Two caveats to this general conclusion should be included.
First, our numerical investigation has been centered exclusively around the one-dimensional Burgers equation.
This simplified setting was chosen as a first step in the investigation of the new closure models.
It allowed a thorough assessment of the performance of the new models, including a parameter sensitivity study.
We emphasize, however, that a similar investigation for realistic three-dimensional turbulent flows (which is the subject of a future study) could possibly yield different conclusions.
The second caveat to our general conclusion is that, as generally done in the evaluation of POD-ROMs (see, however, \cite{wang2012proper} for an exception), the {\it predictive} capabilities of the models were not investigated.
That is, the POD-ROMs were used on the same time interval on which the snapshots (employed in the POD basis generation) were collected.
Utilizing the POD-ROMs on longer time intervals might shed a different light on the more sophisticated closure models, such as the POD-ROM-S.

We intend to pursue several new research directions.
First, we will investigate the new POD-ROM closure models in the numerical investigation of realistic, three-dimensional turbulent flows (see \cite{wang2012proper} for such a numerical investigation of other closure models).
Second, we will develop and test closure models in which the parameters are computed dynamically (see \cite{wang2012proper} for the dynamic procedure applied to the POD-ROM-S).
This approach will eliminate the need of parameter optimization utilized in the present report.
Finally, we will investigate the predictive capabilities of the new POD-ROMs, by testing them on time intervals that are longer than that over which the snapshots were collected.



\bibliographystyle{plain}
\bibliography{ref}







\end{document}